\documentclass[amsmath,amssymb,11pt]{article}
\usepackage{jheppub2}
\pdfoutput=1
\usepackage{graphicx,epsfig}
\usepackage{float}
\usepackage{amsmath}
\usepackage{subfloat}
\usepackage[utf8]{inputenc}
\usepackage{hyperref}
\usepackage{caption}
\usepackage{subcaption} % For arranging multiple figures
\usepackage{color}
\usepackage{overpic}
\usepackage[dvipsnames]{xcolor}
\usepackage{physics}
\usepackage[multiple]{footmisc}
\usepackage{comment}
\usepackage{tensor}
\usepackage[normalem]{ulem}
\usepackage{orcidlink}

\usepackage{pifont}
\newcommand{\cmark}{\text{\ding{51}}}
\newcommand{\xmark}{\text{\ding{55}}}

\DeclareMathOperator\arctanh{arctanh}

\newcommand{\beq}{\begin{equation}}
\newcommand{\eeq}{\end{equation}}

\newcommand{\be}{\begin{equation}}
\newcommand{\ee}{\end{equation}}

\usepackage{tikz}
\usetikzlibrary{positioning}
\usetikzlibrary{intersections}
\usetikzlibrary{fadings} 
\usetikzlibrary{arrows.meta} 
\usetikzlibrary{arrows}

\tikzfading[name=fade out,
inner color=transparent!0,
outer color=transparent!100]

\definecolor{cherryblossompink}{rgb}{1.0, 0.72, 0.77}
\definecolor{lightblue}{rgb}{0.68, 0.85, 0.9}

\usetikzlibrary{decorations.pathmorphing}
\usetikzlibrary{decorations.pathreplacing,decorations.markings}

\usetikzlibrary{backgrounds,automata}

\begin{document}

\title{{\Large Thermodynamics of Regular Black Holes in Anti-de Sitter Space}}

\author[a]{Robie A.~Hennigar\orcidlink{0000-0002-9531-6440},}
\affiliation[a]{Centre for Particle Theory, Department of Mathematical Sciences, Durham University, Durham DH1 3LE, UK}

\author[b]{David Kubizňák\orcidlink{0000-0003-0683-4578},}
\affiliation[b]{ Institute of Theoretical Physics, Faculty of Mathematics and Physics, Charles University, V Holešovičkách 2, 180 00 Prague 8, Czech Republic}

\author[c]{Sebastian Murk\orcidlink{0000-0001-7296-0420},}
\affiliation[c]{Quantum Gravity Unit, Okinawa Institute of Science and Technology, 1919-1 Tancha, Onna-son, Okinawa 904-0495, Japan}

\author[d]{Ioannis Soranidis\orcidlink{0000-0002-8652-9874}}
\affiliation[d]{School of Mathematical and Physical Sciences, Macquarie University, Sydney, New South Wales 2109, Australia}

\emailAdd{robie.a.hennigar@durham.ac.uk}
\emailAdd{david.kubiznak@matfyz.cuni.cz}
\emailAdd{sebastian.murk@oist.jp}
\emailAdd{ioannis.soranidis@mq.edu.au}

\abstract{We construct regular black holes with anti-de Sitter asymptotics in theories incorporating infinite towers of higher-order curvature corrections in any dimension $D \ge 5$. We find that regular black branes are generically inner-extremal, potentially evading instabilities typically associated with inner horizons. Considering minimally coupled matter, we establish general criteria for the existence of singularity-free solutions. We analyze solutions coupled to Maxwell and nonlinear (Born--Infeld and RegMax) electrodynamics, demonstrating in the latter case the first examples of fully regular gravitational and electromagnetic fields for all parameter values. Here, we find that the ratio of the gravitational mass to the electrostatic self-energy determines whether the regular core is de Sitter or anti-de Sitter. We perform a detailed analysis of the black hole thermodynamics and show that the equation of state exhibits features akin to those of fluids with a finite molecular volume induced by the regularization parameter.}

\maketitle

\section{Introduction}

One of the central questions of theoretical physics is whether the {\em singularities} of general relativity are resolved --- and if so, how? It has long been known that singularities are generic within general relativity, occurring at the beginning of the universe and in gravitational collapse~\cite{Hawking:1973uf}. The physical reality of singularities would signal an irreconcilable incompleteness in our theories of nature, and it is therefore generally expected that they will somehow be cured. However, very little is known about singularity resolution in general. 

There has been a long history of studying the implications of singularity resolution on phenomenological grounds.  Dating back to the work of Bardeen, singularity-free black hole {\em metrics} have been postulated and their properties studied.\footnote{In fact, 30 years before Bardeen, the first regular black hole spacetime was constructed by Hoffmann and Infeld as a solution of their new theory of nonlinear electrodynamics \cite{hoffmann1937choice}. Unfortunately, this solution suffers from the `same drawbacks' as more recent regular electrically charged solutions in nonlinear electrodynamics (e.g., Ref.~\cite{Ayon-Beato:1998hmi}, discussed further below), and has since been largely forgotten.}\ Well-known examples include the Bardeen~\cite{1968qtr..conf...87B}, Hayward~\cite{Hayward:2005gi}, and Dymnikova~\cite{Dymnikova:1992ux} metrics, though there are many others, e.g., Refs.~\cite{Sakharov:1966aja,Borde:1994ai,Lemos:2011dq,Bambi:2013ufa,Simpson:2018tsi,Rodrigues:2018bdc}. These metrics are not \textit{a priori} solutions of any known theory, and therefore one is limited to {\em kinematical} rather than dynamical considerations. Nonetheless, the corresponding structure is rich and has been the subject of renewed and intensive investigation~\cite{Carballo-Rubio:2018pmi, Carballo-Rubio:2018jzw, Carballo-Rubio:2019nel, Carballo-Rubio:2019fnb, DiFilippo:2022qkl, Carballo-Rubio:2022kad} (see also \cite{Carballo-Rubio:2025fnc} for a recent review).

There has been considerable difficulty in realizing regular black holes as {\em exact solutions} of any realistic theory. One avenue that has received significant attention is embedding regular black hole metrics into theories of {\em nonlinear electrodynamics}~\cite{Ayon-Beato:1998hmi,Bronnikov:2000vy,Ayon-Beato:2000mjt,Bronnikov:2000yz,Ayon-Beato:2004ywd,Dymnikova:2004zc,Berej:2006cc,Balart:2014jia,Fan:2016rih,Bronnikov:2017sgg,Junior:2023ixh,Murk:2024nod}. In this approach, it is possible to obtain a large class of regular black hole metrics as exact solutions, but there are substantial drawbacks as well. For example, the required Lagrangians often do not have a Maxwell limit, meaning singularity resolution requires postulating exotic matter. Moreover, the required theories are often acausal or suffer from other dynamical problems. Perhaps worst of all is that embedding regular black holes into theories of nonlinear electrodynamics requires fine tuning of coupling and integration constants, indicating that the regular black hole metrics are a set of measure zero in the full solution space of the field equations. Singularity resolution in these models is therefore not generic, even for the simplest cases. While other proposals (e.g., Refs.~\cite{Cano:2020ezi, Li:2024rbw}) can resolve some of these issues, there remains the \textit{requirement} of adding matter to resolve singularities, and the fully singular Schwarzschild metric remains when the matter is switched off.

Recently, considerable progress has been made on this problem with the realization that the Schwarzschild singularity is resolved \textit{generically} and in any dimension $D \ge 5$ when infinite towers of higher-order curvature corrections are added to the Einstein--Hilbert action~\cite{Bueno:2024dgm}. The theories studied, known as quasi-topological gravities~\cite{Oliva:2010eb,Quasi,Dehghani:2011vu,Ahmed:2017jod,Cisterna:2017umf,Bueno:2019ycr, Bueno:2022res, Moreno:2023rfl}, are sufficiently general that they provide a basis for gravitational effective field theory in vacuum.\footnote{Strictly speaking, the regularization of the singularity takes one out of the effective field theory regime. Nonetheless, by treating the regularization parameter (the higher curvature coupling) perturbatively, one can match the corresponding expansion of the regular black hole metric to arbitrary accuracy \cite{Bueno:2024dgm}.}\ They also possess further desirable properties, such as second-order field equations in spherical symmetry and a Birkhoff theorem which guarantees that the static solution is the unique one. The fact that singularity resolution coincides with the resummation of an infinite tower of higher-order curvature corrections is appealing, as an infinite tower of such corrections is a rather ubiquitous prediction of different approaches to quantum gravity. However, perhaps the most appealing aspect is that the singularity is resolved without any additional fields beyond the metric. Overall, this has emerged as arguably the first comprehensive approach to the problem of regular black holes. It not only allows the problem of dynamics to be tackled~\cite{Bueno:2024zsx, Bueno:2024eig, Bueno:2025gjg}, but also permits the construction of regular black holes with inner-extremal horizons \cite{DiFilippo:2024mwm}. As such, the construction has attracted considerable recent attention~\cite{Konoplya:2024hfg, DiFilippo:2024mwm, Konoplya:2024kih, Ma:2024olw, Frolov:2024hhe, Wang:2024zlq, Fernandes:2025fnz, Fernandes:2025eoc}

Here, we extend the construction of Ref.~\cite{Bueno:2024dgm} in several directions, namely by considering a cosmological constant, constructing asymptotically anti-de Sitter (AdS) black holes, and including minimally coupled matter. The preliminary study of black hole thermodynamics conducted in Ref.~\cite{Bueno:2024dgm} revealed that regular black holes obtained in this manner avoid many of the problems and ambiguities associated with regular black holes embedded in nonlinear electrodynamics, such as degenerate potentials or a modified Hawking temperature~\cite{Simovic:2023yuv,Soranidis:2024}. We further  elaborate on the thermodynamic properties of regular black holes in these theories, uncovering some of their phase transitions on the way. Notably, we show that one can construct an equation of state exhibiting a `finite molecular volume subtraction', similar to what happens in van der Waals fluids.  

The remainder of this paper is organized as follows: In Section~\ref{sec:review}, we begin by reviewing quasi-topological gravities and their black hole solutions, emphasizing the inclusion of minimally coupled matter. We then provide a detailed analysis of the conditions required for regular black holes in quasi-topological theories incorporating an infinite number of terms in the action. In Section~\ref{sec:ads}, we analyze asymptotically AdS black holes in the resummed theories. We study spherical, planar, and hyperbolic black holes, finding in the planar case that the regular solutions are \textit{automatically} inner-extremal, and hence may evade mass inflation instabilities. In Section~\ref{sec:electro}, we study extensions of the theory incorporating minimally coupled matter, taking Maxwell electrodynamics and two theories of nonlinear electrodynamics (Born--Infeld and RegMax) as examples. In particular, we analyze how the inclusion of charged matter can alter the nature of the regular core, for example, necessitating an AdS core in the Maxwell case. In Section~\ref{sec:thermo}, we study the first law and Smarr relation for regular black holes in the extended thermodynamic phase space. We then go on in Sections~\ref{sec:bbthermo} -- \ref{sec:topothermo} to study in detail the thermodynamic properties of black holes with different horizon topologies and with different types of matter. Finally, in Section~\ref{sec:discussion}, we summarize the most important results of our work and indicate potential future directions. Quasi-topological gravities to quintic order are reviewed in Appendix ~\ref{app:lags}, and the ``effective energy conditions'' for various regular black hole models are summarized in Appendix~\ref{app:EnergyConditions}.

\paragraph{Note.} In Ref.~\cite{Julio} (appearing concurrently with our manuscript), thermodynamics of regular black holes is also studied, and there is some overlap with parts of our manuscript. Those authors also take the first steps toward understanding holographic implications of regular black holes in the `pure gravity' framework.

\section{Regular black holes in quasi-topological gravity}
\label{sec:review}

\subsection{Theory and equations of motion}

Consider a theory of gravity built from the metric and curvature with Lagrangian density $\mathcal{L}\left(g^{ab}, R_{abcd} \right)$. Allowing for minimally coupled matter, the corresponding equations of motion read
\begin{equation}
{\cal E}_{ab}\equiv P_a^{cde}R_{bcde}-\frac{1}{2}g_{ab}\mathcal{L}-2\nabla^c\nabla^d P_{acdb}=8 \pi G_{\rm N} T_{a b} \, ,  
\end{equation}
where $P^{abcd}\equiv \partial \mathcal{L}/\partial R_{abcd}$. Quasi-topological gravities are a particular class of this general set of theories defined by the requirement\footnote{Naively, it would seem that one could demand the weaker condition $\nabla^c\nabla^d P_{acdb} = 0$ to ensure second-order equations of motion on a particular background. However, this does not lead to any new theories of the polynomial type~\cite{Bueno:2019ycr}.}
\be 
\nabla^d P_{acdb}=0
\ee
for general spherically (planar, or hyperbolic) symmetric metrics.  Since it is this term that is responsible for higher-order terms in the equations of motion, quasi-topological gravities have by definition {\em second-order} equations of motion {\em on spherically symmetric} backgrounds.

Included in this family of theories is the well-known Lovelock theory of gravity~\cite{Lovelock1, Lovelock2} which, in addition to having second-order equations of motion on spherically symmetric backgrounds, has second-order equations of motion \textit{in general}. That the family of quasi-topological theories includes additional members beyond the Lovelock class came as somewhat of a surprise, with the first example being a cubic theory of gravity discovered almost simultaneously by two groups and for two different purposes~\cite{Oliva:2010eb,Quasi}. Subsequently, new examples were found at quartic~\cite{Dehghani:2011vu} and quintic orders~\cite{Cisterna:2017umf}, culminating with a construction of quasi-topological theories at all orders in curvature~\cite{Bueno:2019ycr, Bueno:2022res, Moreno:2023rfl}. 
Notably, such quasi-topological gravities  satisfy a Birkhoff theorem. Thus, the most general (vacuum) spherically symmetric solutions are also static~\cite{Bueno:2024zsx}.

Although explicit covariant Lagrangians are known for any curvature order --- see, for example, Eq.~(4.5) of Ref.~\cite{Moreno:2023rfl} --- the simplest way to construct quasi-topological theories at all orders is via a recurrence relation. It was proved in Ref.~\cite{Bueno:2019ycr} that, given the first five quasi-topological Lagrangian densities $\{\mathcal{Z}_{i}: i = 1, \dots, 5 \}$, all subsequent orders can be constructed via the formula
\begin{align}\label{recursive} 
\mathcal{Z}_{n+5} &= -\frac{3 (n+3)}{2 (n+1) D (D-1)} {\mathcal{Z}}_{1} {\mathcal{Z}}_{n+4} + \frac{3 (n+4)}{2 n D (D-1)} {\mathcal{Z}}_{2} {\mathcal{Z}}_{n+3} 
- \frac{(n+3)(n+4)}{2n (n+1)D(D-1)} {\mathcal{Z}}_{3} {\mathcal{Z}}_{n+2}  \, .  
\end{align}
Even the first five of these theories have rather complicated Lagrangians. We therefore do not present them explicitly in the main text, but instead dedicate Appendix~\ref{app:lags} to their form. However, we note that when using the above recurrence formula, the first-order term $\mathcal{Z}_1$ is simply the Einstein--Hilbert term, while the second-order term $\mathcal{Z}_2$ is the Gauss--Bonnet density. At higher orders, the overlap with Lovelock densities ceases. Crucially, while the Lovelock densities are limited to orders $n \le \left\lfloor D/2 \right\rfloor$ in a specific spacetime dimension $D$, quasi-topological theories of \textit{any order} can exist in dimensions $D \ge 5$. It is this crucial difference that allows one to consider a full, infinite tower of quasi-topological corrections. Importantly, as demonstrated in Ref.~\cite{Bueno:2024dgm}, it is only when an infinite number of corrections are included that black hole singularities are resolved.

Let us consider the following action:
\be 
I=I_{\rm QT}+I_{\rm matter}\,,\quad 
I_{\rm QT} = \int \frac{{\rm d}^D x \sqrt{|g|}}{16 \pi G_{\rm N}} \bigg[-2 \Lambda + R + \sum_{n=2}^{\infty} \tilde{\alpha}_n \mathcal{Z}_n \bigg]  \, ,
\ee
whose variation yields the field equations.
In our conventions (which match those of Ref.~\cite{Bueno:2024zsx}), 
this variation reads
\be 
\mathcal{E}_{ab} = 8 \pi G_{\rm N} T_{ab} \quad \text{with} \quad \mathcal{E}_{ab} = \frac{16 \pi G_{\rm N}}{\sqrt{|g|}} \frac{\delta I_{\rm QT}}{\delta g^{ab}} \quad \text{and} \quad T_{ab} = - \frac{2}{\sqrt{|g|}} \frac{\delta I_{\rm matter}}{\delta g^{ab}} \, .
\ee
We further assume a general static  and spherically/plane/hyperbolic symmetric metric of the form \be\label{eq:sss} 
{\rm d}s^2 = -N(r)^2 f(r) {\rm d} t^2 + \frac{{\rm d} r^2}{f(r)} + r^2 \Omega_{ij} {\rm d}x^i {\rm d} x^j \, ,
\ee
where $\Omega_{ij}$ is the metric on a $(D-2)$-dimensional space with curvature tensor $R_{ijkl} = k \left(\Omega_{ik} \Omega_{jl} - \Omega_{il} \Omega_{jk} \right)$, and $k = \{+1,0,-1 \}$ corresponds to spherical, planar, and hyperbolic geometries, respectively. The field equations for this metric ansatz can be readily obtained using reduced Lagrangian methods~\cite{Fels:2001rv, Deser:2003up, Frausto:2024egp}, evaluating the action on the ansatz~\eqref{eq:sss}, and varying with respect to the unknown functions. After integrating by parts and discarding total derivatives, the reduced action for the gravitational sector of all quasi-topological theories takes the form\footnote{We use $\Omega_{D-2}$ to denote the volume of the transverse space. In the case $k = 1$, this is the volume of a unit sphere $2 \pi^{(D-1)/2}/\Gamma \left[(D-1)/2 \right]$. In the cases $k = 0, -1$, this volume is, strictly speaking, infinite, but can be made finite by introducing identifications to the transverse geometry~\cite{Mann:1997iz}.}
\be 
I_{\rm QT}[N,f] = \frac{(D-2) \Omega_{D-2}}{16 \pi G_{\rm N}} \int {\rm d}t {\rm d}r N(r) \left[r^{D-1} h(\psi) \right]' \, ,
\ee
where the prime denotes the radial derivative and the function $h(\psi)$ will be explained below.

{Variation of the action with respect to $N$ and $f$ yields two components of the field equations $\mathcal{E}_{ab}$, 
\begin{align}
    \mathcal{E}_t^t &= - \frac{8 \pi G_{\rm N}}{r^{D-2} \Omega_{D-2}} \frac{\delta I_{\rm QT}[N, f]}{\delta N} = - \frac{(D-2)}{2 r^{D-2}} \frac{{\rm d}}{{\rm d}r} \left[ r^{D-1} h(\psi) \right] \, ,
    \\
    \mathcal{E}_r^r &= \mathcal{E}_t^t +  \frac{16 \pi G_{\rm N}} { \Omega_{D-2} N r^{D-2}} \frac{\delta I_{\rm QT}[N, f]}{\delta f} = \mathcal{E}_t^t + \frac{(D-2)f(r) N'(r)}{r N(r) } h'(\psi)
    \label{eq:Ett.Err} \, .
\end{align}
Due to the generalized Bianchi identity $\nabla_a \mathcal{E}^{ab} = 0$,
the angular components of the field equations are automatically satisfied provided Eq.~\eqref{eq:Ett.Err} is \cite{Bueno:2017sui}.} 

So far, the discussion has been quite general, but now let us assume that the minimally coupled matter inherits the spherical symmetry of the spacetime and satisfies $T_t^t = T_r^r$. From the reduced action and the definition of the EMT, we have
\be 
T_t^t = \frac{1}{\Omega_{D-2} r^{D-2}} \frac{\delta I_{\rm matter}}{\delta N} \, .
\ee

Then, performing the above calculations for the reduced action of quasi-topological theories, one obtains
\begin{align} 
\frac{\delta I[N, f]}{\delta f} \, \, &\Rightarrow \, \,  h'(\psi) N'(r) = 0 \, , 
\\
\frac{\delta I[N, f]}{\delta N} \, \, &\Rightarrow \, \,  \left[r^{D-1} h(\psi) \right]' = - \frac{16 \pi G_{\rm N} r^{D-2}}{D-2} T_t^t \, .
\end{align}

In the above, we have written the action and field equations in terms of a function $h(x)$ which depends on the details of the particular theory. Explicitly, we have 
\be 
h(x) \equiv \frac{-2 \Lambda}{(D-1)(D-2)} + h_{\rm f}(x) =   \frac{-2 \Lambda}{(D-1)(D-2)} + x + \sum_{n=2}^{\infty} \frac{(D-2n)}{D-2} \tilde{\alpha}_n x^n \,,
\ee
while 
\be 
\psi \equiv \frac{k-f}{r^2} \, .
\ee
Note that we have identified the contribution to $h(x)$ that arises in the asymptotically flat case as $h_{\rm f}(x)$. Henceforth, we will use the rescaling 
\be 
\alpha_n \equiv \left( \frac{D-2n}{D-2} \right) \tilde{\alpha}_n
\ee
of the higher-curvature coupling constants to simplify various expressions. The caveat is that if we are working in even dimensions with $D  = 2 n$, then the term with coupling $\alpha_n$ must be removed from the function $h(\psi)$. This is simply because quasi-topological theories of order $n$ do not contribute to the field equations in dimension $D = 2 n$ (the original motivation for the name `quasi-topological'). This has no effect on whether or not the singularity is resolved, as the resolution is due to the asymptotic properties of the series rather than its finite terms.\footnote{We refer the reader to the appendix of Ref.~\cite{Bueno:2024zsx} for further comments on this subtlety. It is from this property that certain restrictions on the applicable dimensions for certain resummations in Table~\ref{tab:metrics} arise.}

The field equation following from the variation $\delta I/\delta f$ is satisfied either if $N' = 0$ or if $h' = 0$. The second option is generically inconsistent with the remaining field equation, and thus we take $N' = 0$. We can therefore set $N = 1$ without loss of generality. Integrating the second field equation results in an algebraic equation that determines the metric function $f$:
\be \label{feq}
h_{\rm f}(\psi) = {\cal S}(r) \equiv \frac{2 \Lambda}{(D-1)(D-2)} +  \frac{m}{r^{D-1}} - \frac{16 \pi G_{\rm N}}{(D-2) r^{D-1}} \int_\infty^r r^{D-2} T_t^t {\rm d} r \, .
\ee
Here, $m$ is an integration constant that is directly related to the ADM or thermodynamic mass $M$,
\be\label{Mass} 
M = \frac{m (D-2)\Omega_{D-2}}{16 \pi  G_{\rm N}} \, .
\ee
Note that, when performing the integration for the EMT, the result should not include a constant of integration --- such a constant can always be reabsorbed into the definition of $m$. Here, we have assumed that the EMT falls off faster than $1/r^{D-2}$ at large $r$ to set this constant of integration to zero. 

\subsection{Conditions for regularity with minimally coupled matter}\label{sec:regularity conditions}

\begingroup
\setlength{\tabcolsep}{5pt} 
\renewcommand{\arraystretch}{3} 
\begin{table*}[t!]
	\footnotesize
	\centering
	\begin{tabular}{|c|c|c|c|}
	\hline
 Model & $\alpha_{n}$ & $ h_{\rm f}(\psi)$ & $f(r)$ 
 \\
 \hline \hline
 I & $\alpha^{n-1}$ & $\displaystyle \frac{\psi}{1-\alpha \psi}$ & $\displaystyle k - \frac{r^2 \mathcal{S}(r)}{1 + \alpha \mathcal{S}(r)}$ 
 \\ 
 \hline
  II & $\displaystyle\frac{\alpha^{n-1}}{n}$ & $\displaystyle -\frac{\log(1-\alpha\psi)}{\alpha}$ & $\displaystyle k - \frac{r^2}{\alpha } \left(1-e^{-\alpha \mathcal{S}(r)}\right)$  
    \\ 
    \hline
    III & $\displaystyle n\alpha^{n-1}$ & $\displaystyle \frac{\psi}{(1-\alpha \psi)^2}$ & \scriptsize  $\displaystyle k  - \frac{r^2 \left[1 + 2 \alpha \mathcal{S}(r) - \sqrt{1 + 4 \alpha \mathcal{S}(r)}\right]}{2 \alpha^2 \mathcal{S}(r)}$ 
    \\  
    \hline
    IV & $\displaystyle \frac{(1-(-1)^n)}{2}\alpha^{n-1}$ & 
    $\displaystyle \frac{\psi}{1-\alpha^2\psi^2}$
    & $\displaystyle k  + \frac{r^2 \left[1 - \sqrt{1 + 4 \alpha^2 \mathcal{S}^2(r)} \right]}{2 \alpha^2 \mathcal{S}(r)}$ 
    \\ 
    \hline 
   V &  $\displaystyle \frac{(1-(-1)^n)\Gamma \left(\frac{n}{2}\right)}{2\sqrt{\pi } \Gamma \left(\frac{n+1}{2}\right)}\alpha^{n-1}$ & $\displaystyle \frac{\psi}{\sqrt{1-\alpha^2\psi^2}}$ & $\displaystyle k - \frac{r^2 \mathcal{S}(r)}{\sqrt{1 + \alpha^2 \mathcal{S}^2(r)}}$  
   \\ \hline
    VI & $\displaystyle \frac{\left(1 + \left(-1\right)^{\left(n+1\right)} \right)}{2n} \alpha^{n-1}$& $\displaystyle \frac{\arctanh(\alpha \psi)}{\alpha}$ & $\displaystyle k - \frac{r^2}{\alpha} \tanh \left(\alpha \mathcal{S}(r) \right)$  
    \\ \hline
    VII & N/A & $\displaystyle \frac{\psi}{\left(1- \alpha^{\rm N} \psi^{\rm N} \right)^{1/{\rm N}}}$ & $\displaystyle k - \frac{r^2 \mathcal{S}(r)}{\left[1 + \alpha^{\rm N} \mathcal{S}^{\rm N}(r) \right]^{1/{\rm N}}}$ 
    \\ \hline
	\end{tabular}
	\caption{{\bf Models: summary.} Example resummations and their corresponding exact solutions to the equations of motion. Here, we write the metric function for general minimally coupled matter satisfying $T_t^t = T_r^r$ in terms of the function $\mathcal{S}(r)$ introduced in Eq.~\eqref{feq}. In the case of Model VII, the coefficients of the series can be obtained by expanding $h_{\rm f}(\psi)$ for a given value of ${\rm N}$ (a closed form expression is not simple for general ${\rm N}$). Note that this model includes Models I and V as special cases. Models I, II, and III are valid only in odd dimensions. Models IV, V and VI are valid in odd dimensions and also for $D = 4(k+1)$ with $k\in\mathbb{Z}^+$. Model VII is valid in odd dimensions for general ${\rm N}$, while particular cases of ${\rm N}$ also yield resummations valid in even dimensions. The simplest case for an even dimension $D$ is given by ${\rm N} = D/2$.}
	\label{tab:metrics}
\end{table*}
\endgroup

Let us now examine, as generally as possible, under what conditions the solutions of \eqref{feq} are regular.

The quantity $\mathcal{S}(r)$ introduced above is closely related to the quasi-local Misner--Sharp mass, suitably generalized to quasi-topological gravity,
\be 
M_{\rm MS}(r) \equiv  r^{D-1} \mathcal{S}(r) \, .
\ee
This makes the corresponding solutions much more natural. Effectively, the mass parameter appearing in the asymptotically flat solutions gets replaced by the corresponding Misner--Sharp energy in more general situations. In Table~\ref{tab:metrics}, we collect a number of examples of resummations and their corresponding exact solutions. Writing the results in terms of $\mathcal{S}(r)$ as introduced above, the problem of studying the solutions of a particular theory reduces to obtaining its corresponding $\mathcal{S}(r)$. 

In the general case, we can express the metric function as
\be 
f(r) = k - r^2 h_{\rm f}^{-1} \left(\mathcal{S}(r)\right) \, ,
\ee
where $h_{\rm f}^{-1}(x)$ is the inverse of the function $h_{\rm f}(x)$. This inverse will exist if and only if $h_{\rm f}(x)$ is bijective. The regularity of the spacetime can then be translated into requirements on the inverse function $h_{\rm f}^{-1}(x)$. Indeed, it is a simple exercise to show that the spacetime will be regular (in the sense that the Ricci and Kretschmann scalars are finite) provided the following two conditions are met: 
\begin{enumerate}
    \item The inverse function and its first two radial derivatives are \textit{finite} for all $r > 0$.
    \item The derivatives of $h_{\rm f}^{-1}$ have divergences \textit{no stronger} than $d \left(h_{\rm f}^{-1}(\mathcal{S}(r)) \right)/dr\sim 1/r$ and $d^2 \left(h_{\rm f}^{-1}(\mathcal{S}(r)) \right)/d r^2 \sim 1/r^2$ as $r \to 0$.  
\end{enumerate} 

Let us consider the first requirement. Using the definition of the inverse, we can readily obtain that
\be 
\label{eq:func_inv_rel}
\left(h_{\rm f}^{-1}(x)\right)'= \frac{1}{h'_{\rm f}\left(h_{\rm f}^{-1}(x)\right)}\quad \text{and} \quad \left(h_{\rm f}^{-1}(x) \right)'' = - \frac{h_{\rm f}''\left(h_{\rm f}^{-1}(x)\right)}{\left[h_{\rm f}'\left(h_{\rm f}^{-1}(x)\right)\right]^3} \, .
\ee
We then have 
\be 
\frac{d \left(h_{\rm f}^{-1}(\mathcal{S}(r)) \right)}{dr} = \frac{\mathcal{S}'(r)}{h_{\rm f}' \left[ \left(h_{\rm f}^{-1}(\mathcal{S}(r)) \right) \right]} \, .
\ee
This will be finite for $r > 0$ provided that $\mathcal{S}'(r)$ is finite and $h_{\rm f}'(x)$ has no zeroes. The latter property is equivalent to requiring that $h_{\rm f}(x)$ be monotonic on its domain. Examining the second derivative, we find
\be 
\frac{d^2 \left(h_{\rm f}^{-1}(\mathcal{S}(r)) \right)}{dr^2} = -\frac{h_{\rm f}''\left(h_{\rm f}^{-1}(\mathcal{S}(r)) \right) \mathcal{S}'(r)}{\left[ h_{\rm f}' \left( h_{\rm f}^{-1}(\mathcal{S}(r)) \right) \right]^3} 
 + \frac{\mathcal{S}''(r)}{h_{\rm f}' \left[ \left(h_{\rm f}^{-1}(\mathcal{S}(r)) \right) \right]}   \, .
\ee
Considering the information we have gained from the first derivative, the finiteness of the second derivative is guaranteed provided that $\mathcal{S}''(r)$ has no divergences for $r > 0$ and $h_{\rm f}''(x)$ is finite at all points in the domain except possibly at $r = 0$. 

Next, let us consider the limiting behavior at $r = 0$. For the metric to be regular at $r = 0$, we require that
\be 
f(r) = k + \psi_0 r^2 + \mathcal{O}(r^3) \, ,
\ee
for some constant $\psi_0$. By comparison with the above, this allows us to deduce that
\be 
\psi_0 = \lim_{r \to 0} h_{\rm f}^{-1} \left(\mathcal{S}(r)\right) \, .
\ee
In turn, this implies that $h_{\rm f}(x)$ must have a divergence at $x = \psi_0$ provided that $\mathcal{S}(r)$ diverges at $r=0$, which will generically be the case. It is easy to show that if near $x = \psi_0$ the function has the behavior $h_{\rm f}(x) \sim (x-\psi_0)^{-n}$ for some $n > 0$, then the derivatives of the inverse function are finite at $r = 0$. 

In summary: to guarantee a fully regular solution for the full range of $r \in [0, \infty)$, it is \textit{sufficient} for $h_{\rm f}(x)$ to be a monotonic function with a divergence, provided that $\mathcal{S}(r)$ and its derivatives have no divergences at finite $r$. As these conditions are only sufficient but not necessary, many other resummations with nonsingular black holes will exist.

Note that, for a given resummation $h_{\rm f}(\psi)$, the function $\mathcal{S}(r)$ determines the domain of $\psi$. Thus, it is entirely possible for a resummation that is nonsingular in vacuum to become singular when certain matter content is added. To see how this can be the case, consider the Hayward resummation (model I in Table~\ref{tab:metrics}):
\be 
h_{\rm f}(\psi) = \frac{\psi}{1 - \alpha \psi} \, . 
\ee
This function is monotonic only on the disjoint intervals $\psi \in (-\infty, 1/\alpha)$ and $\psi \in (1/\alpha, \infty)$. Restricted to each interval, the range of $h_{\rm f}(\psi)$ is limited. For $\psi \in (-\infty, 1/\alpha)$, the range is $(-1/\alpha, \infty)$, while for $\psi \in (1/\alpha, \infty)$, the range is $(-\infty, -1/\alpha)$. Hence, if the matter function $\mathcal{S}(r)$ is not of uniform sign, the Hayward model may not be able to accommodate it within a single domain for $\psi$. In such cases (which arise, for example, with a Maxwell field), the corresponding Hayward black hole may contain a singularity at finite $r$.

To ensure regular black holes exist regardless of the matter content, it is necessary to select a resummation $h_{\rm f}(\psi)$ with range $(-\infty, \infty)$ in a single domain on which it is a monotonic function. For example, model V of Table~\ref{tab:metrics},
\be 
h_{\rm f}(\psi) = \frac{\psi}{\sqrt{1-\alpha^2\psi^2}} \, ,
\ee
achieves this. Here, $h_{\rm f}(\psi)$ has range $(-\infty, \infty)$ for the domain of $\psi \in (-1/\alpha, 1/\alpha)$. Since for this model  $h_{\rm f}(\psi)$ is monotonic on the full interval, it can accommodate any $\mathcal{S}(r)$ and will always yield regular black holes.

\subsection{Asymptotically flat solutions}

Before beginning in earnest our study of the asymptotically AdS solutions, we pause to review the asymptotically flat vacuum solutions. These metrics were considered in detail in Refs.~\cite{Bueno:2024dgm, Bueno:2024eig, Bueno:2024zsx}, so we will be brief. In vacuum and with vanishing cosmological constant, we have simply 
\be\label{SrFlat} 
\mathcal{S}(r) = \frac{m}{r^{D-1}} \, .
\ee
As shown in Ref.~\cite{Bueno:2024dgm}, the sufficient conditions for regularity of asymptotically flat solutions are 
\be 
\alpha_n\geq 0 \quad \forall n\,,\quad \lim_{n\to \infty}(\alpha_n)^{\frac{1}{n}}=C>0\,.
\ee 
Consistency with these conditions requires $\psi$ to be positive. 
Then, the first condition ensures that $h_f(\psi)$ is monotonic, while the second condition 
implies that $h_f(\psi)$ has a radius of convergence of $\psi_0=1/C$, giving rise to a divergence at $\psi=\psi_0$. Thus, together with the  above expression for ${\cal S}(r)$, we obey the sufficient regularity criteria derived in the previous subsection.

In what follows, we shall consider the seven models summarized in Table~\ref{tab:metrics}. Substituting the expression \eqref{SrFlat} for ${\cal S}(r)$ in the last column of Table~\ref{tab:metrics} yields the corresponding exact solution. Importantly, for the solutions to satisfy asymptotically flat boundary conditions, we necessarily must have $k = +1$. At large distances, all models exhibit Schwarzschild-like asymptotics, 
\be 
f(r) = 1 - \frac{m}{r^{D-1}} + \cdots \quad \text{as $r\to\infty$}  \, ,
\ee
while the higher-curvature corrections become important in the interior leading to a regular core,
\be 
f(r) = 1 - \frac{r^2}{|\alpha|} + \cdots \quad \text{as $r\to 0$} \, .
\ee
Note that, irrespective of the sign of $\alpha$, the core is always of the de Sitter type for all models.

For all models, there exists a critical value of the mass $m_{\rm cr}$ which demarcates the possible solutions. This value of the mass is model-dependent and in some cases cannot be obtained in a simple closed-form expression. For example, for the Hayward black hole (model I), the critical mass parameter reads
\be 
m_{\rm cr} = \frac{(D-1)^{(D-1)/2}}{2 (D-3)^{(D-3)/2}} \alpha^{(D-3)/2} \, .
\ee
In terms of the critical mass parameter, there are three possible types of solutions:
\begin{enumerate}
    \item For $m > m_{\rm cr}$, there is an event and inner horizon. 
    \item For $m = m_{\rm cr}$, there is a single degenerate horizon representing an extremal black hole with vanishing Hawking temperature.
    \item For $0 < m < m_{\rm cr}$, there are no horizons. These represent gravitational solitons: Lorentzian, finite energy, geodesically complete metrics. 
\end{enumerate}

\section{Asymptotically anti-de Sitter solutions}
\label{sec:ads}
\subsection{Hayward-AdS example}

We will first consider the case of a negative cosmological constant, giving rise to asymptotically AdS solutions. Writing
\be 
\Lambda = - \frac{(D-1)(D-2)}{2 L^2} ,
\ee
we have
\be 
\mathcal{S}(r) = -\frac{1}{L^2} + \frac{m}{r^{D-1}} \, .
\ee
The corresponding solutions for various resummations can then be read off immediately from Table~\ref{tab:metrics}. 

Let us take the Hayward resummation --- model I from Table~\ref{tab:metrics} --- as an example to illustrate some of the general features of the asymptotically AdS metrics. The metric function reads
\be 
f(r) = k - \frac{r^2 \mathcal{S}(r)}{1 + \alpha \mathcal{S}(r)} \quad \text{with} \quad \mathcal{S}(r) = -\frac{1}{L^2} + \frac{m}{r^{D-1}} \, .
\ee
When $m = 0$, we have
\be 
f(r) = k + \frac{r^2}{L^2-\alpha} \, .
\ee
This describes an AdS geometry with effective cosmological constant 
\be 
L_{\rm eff} = \sqrt{L^2-\alpha} ,
\ee
provided that $\alpha < L^2$. For nonzero $m$, we have the same AdS asymptotics with the standard fall off now governed by the mass term,
\be 
f(r) = k + \frac{r^2}{L_{\rm eff}^2}  -  \frac{m L^4}{L_{\rm eff}^4 r^{D-3}} + \cdots \, .
\ee
The coefficient multiplying the mass term should be identified with the effective Newton constant of the theory. Specifically, in this case we have
\be 
G_{\rm eff} = \frac{L^4 G_{\rm N}}{L_{\rm eff}^4} = \frac{G_{\rm N}}{\left(1-a\right)^2}\,,
\ee 
where here as well as for other models in Table~\ref{tab:metrics}, we have defined
\be\label{aaa}
a \equiv \frac{\alpha}{L^2} \, .
\ee
In terms of $G_{\rm eff}$ and the thermodynamic mass $M$, the metric function then reads
\be 
f(r) = k + \frac{r^2}{L_{\rm eff}^2}  -  \frac{16 \pi G_{\rm eff} M}{(D-2) \Omega_{D-2} r^{D-3}} + \cdots \,, 
\ee
as required.

Near $r = 0$, the metric function admits the expansion
\be 
f(r) = k - \frac{r^2}{|\alpha|} + \cdots \, .
\ee
For the case $k = +1$, this is identical to the asymptotically flat (pure gravity) case. We will comment on the other values of $k$ below.

Since the function $\mathcal{S}(r)$ takes on both positive and negative values, we may be concerned about singularities at intermediate values of $r$. For the Hayward model, such singularities would arise where 
\be 
1 + \alpha \mathcal{S}(r) = 0 \quad \Rightarrow \quad L^2 - \alpha + \frac{L^2 \alpha m}{r^{D-1}} = 0 \, .
\ee
We require $\alpha > 0$ such that the asymptotically flat (pure gravity) limit is nonsingular. Then, clearly, provided that the conditions for the existence of AdS asymptotics (namely, $\alpha < L^2$) are met along with positivity of the mass, the Hayward-AdS black hole is completely regular.  

 \begingroup
\setlength{\tabcolsep}{5pt}
\renewcommand{\arraystretch}{3}
\begin{table*}[t!]
	\footnotesize
	\centering
	\begin{tabular}{|c|c|c|c|c|}
	\hline
        Model & $\displaystyle h(\psi)$ &$\displaystyle L^2_{\rm eff}/L^2$ & $\displaystyle G_{\rm eff}/G_{\rm N}$ & Parameter constraints
        \\ 
        \hline
        I & $\displaystyle \frac{\psi}{1-\alpha \psi}  $ & $\displaystyle 1-a$ & $\displaystyle \frac{1}{(1-a)^2}$ & $0< a < 1$ 
        \\
        \hline 
        II & $\displaystyle - \frac{\log \left(1-\alpha \psi \right)}{\alpha} $& $\displaystyle \frac{a}{e^a - 1}$ & $\displaystyle e^a$& $\displaystyle a > 0$
        \\
        \hline 
        III & $\displaystyle \frac{\psi}{\left(1-\alpha \psi \right)^2}$ & $\displaystyle \frac{2 a^2}{1 - 2 a - \sqrt{1-4a}}$ & 
        $\displaystyle \frac{2-8a+2\sqrt{1-4a}(-1+2a)}{4a^2(-1+4a)}$
        & $\displaystyle 0 < a < \frac{1}{4}$
        \\
        \hline 
        IV &  $\displaystyle \frac{\psi}{1-\alpha^2\psi^2}$ & $\displaystyle \frac{2 a^2}{-1 + \sqrt{1 + 4a^2}}$ & $\displaystyle \frac{1}{2a^2}\left[1 - \frac{1}{\sqrt{1+4a^2}} \right]$ &  $\displaystyle a \in \mathbb{R} $
        \\
        \hline
        V & $\displaystyle \frac{\psi}{\sqrt{1-\alpha^2\psi^2}}$ & $\displaystyle \sqrt{1+a^2} $ & $\displaystyle \left(1+a^2 \right)^{-3/2}$ &  $\displaystyle a \in \mathbb{R}$
        \\
        \hline 
        VI & $\displaystyle \frac{\arctanh(\alpha \psi)}{\alpha}$ & $\displaystyle a \coth a$ & $\displaystyle \sech^2 a$ & $a > 0$
        \\ 
        \hline 
        VII & $\displaystyle \frac{\psi}{\left(1- \alpha^{\rm N} \psi^{\rm N} \right)^{1/{\rm N}}}$ & $\displaystyle \left[1+\left(-a\right)^{\rm N}\right]^{1/{\rm N}} $& $\displaystyle  \left[1 + \left(-a\right)^{\rm N} \right]^{-\frac{({\rm N}+1)}{{\rm N}}} $& $\displaystyle \begin{cases} a \in \mathbb{R} \,\, &\text{if} \,\, {\rm N} \,\, \text{even} 
        \\
        0<a<1 \,\, &\text{if} \,\, {\rm N} \,\, \text{odd}
        \end{cases}$
        \\ 
        \hline 
	\end{tabular}
	\caption{{\bf Models: effective couplings and parameter constraints.} We tabulate the effective AdS length $L_{\rm eff}$ and the effective Newton constant $G_{\rm eff}$ for each of the resummations that appear in Table~\ref{tab:metrics}. We have defined $a = \alpha/L^2$ to ease the notation. The final column of the table gives the range of $a$ for which the metrics are fully regular and have AdS asymptotics. } 
   
	\label{tab:AdS_params}
\end{table*}
\endgroup
    
All we have said above generalizes straightforwardly to the other models. For an arbitrary resummation, the effective cosmological constant will be given by
\be 
L_{\rm eff}^2 = - \frac{1}{h^{-1}_{\rm f}\left(-1/L^2\right)} \, ,
\ee
while the effective Newton constant is\footnote{This can be obtained by taking the formal solution $f(r) = k - r^2 h^{-1}_{\rm f}(\mathcal{S}(r))$, linearizing it for small mass parameter $m$, and using the relationship between the derivative of a function and its inverse given in Eq.~\eqref{eq:func_inv_rel}. } 
\be 
G_{\rm eff} = \frac{G_{\rm N}}{h'_{\rm f} \left(-1/L_{\rm eff}^2 \right)}\, .
\ee
Whether or not the positivity of the effective AdS length, the effective Newton constant, or spacetime regularity places constraints on the domain of $\alpha$ will depend on the particular resummation. For convenience, we summarize the different possibilities for each of the models introduced earlier in Table~\ref{tab:AdS_params}. Note that for some of the models $\alpha<0$ also leads to regular black holes (including their asymptotically flat limits).

\subsection{Regular spherical black holes}

When the parameter $k = +1$, the transverse sections are $(D-2)$-dimensional spheres. These black holes behave in the same manner as the asymptotically flat ones. Near $r = 0$, for all of the resummations presented in Table~\ref{tab:metrics}, the metric function takes the form
\be 
f(r) = 1 - \frac{r^2}{|\alpha|} + \,\, \text{subleading}. 
\ee
Thus, in the immediate vicinity of $r = 0$, the spacetime metric takes the form
\be 
{\rm d}s^2 = - {\rm d}t^2 + {\rm d}r^2 + r^2 {\rm d}\Omega_{D-2}^2 \, .
\ee
We see that this is simply Minkowski space in spherical coordinates, with $r =0$ the origin of the coordinate system. Thus, as a particle worldline approaches $r = 0$, we can perform the standard {\em antipodal identification}. For example, in four-dimensions, a particle that approaches $r = 0$ at polar angle $\theta$ and azimuthal angle $\phi$ reemerges with $r$ now increasing, and with the angles shifted according to $\theta \to \pi - \theta$ and $\phi \to \phi + \pi$.

\begin{figure}[t]
    \centering
    \includegraphics[width=\linewidth]{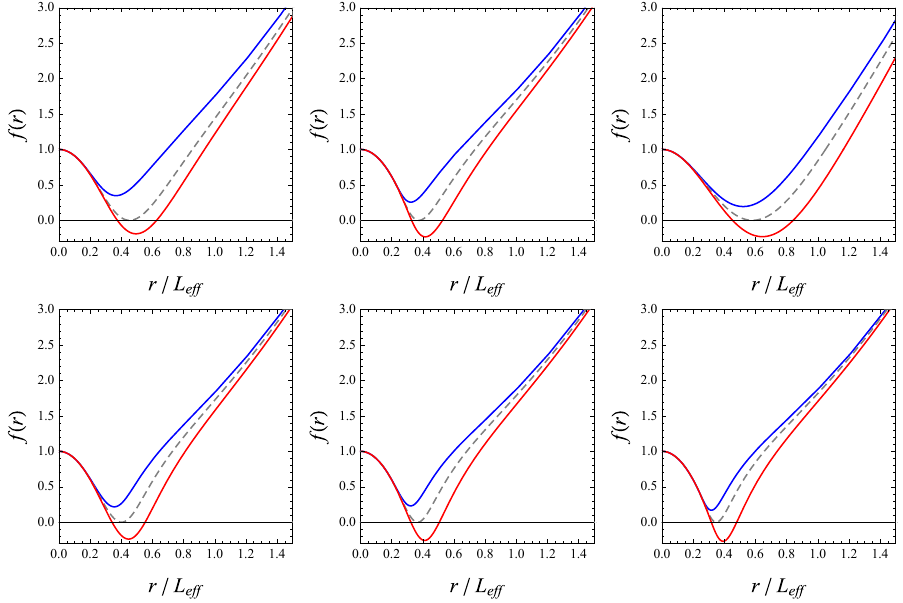}
    \caption{{\bf Metric functions: 
    AdS black holes.} Plots of the $k=1$ metric function for models I-VI for different values of the thermodynamic mass $M$. In each plot we have set $L_{\rm eff} = 1$, $G_{\rm eff}=1$, $a=0.1$, and $D = 5$. The top row represents models I–III from left to right, while the bottom row corresponds to models IV–VI in the same order. The gray dashed line represents the critical thermodynamic mass value associated with an extremal regular black hole. The values of the thermodynamic mass $M$ in the color order [blue, gray, red] are: $[0.3, 0.706, 1]$ (top left), $[0.2,0.362,0.55]$ (top center), $[1.2,1.849,2.8]$ (top right), $[0.2,0.326,0.5]$ (bottom left), $[0.15,0.253,0.4]$ (bottom center), and $[0.15,0.217,0.35]$ (bottom right). All models I-VI yield metric functions that have exactly the same qualitative structure.
    }
    \label{fig:f(r)_k1}
\end{figure}

Of course, the antipodal identification will only be an \textit{analytic} extension provided that no odd powers of $r$ appear in the expansion of the metric function near $r = 0$ --- see Ref.~\cite{Zhou:2022yio}. However, even in cases where odd powers of $r$ do occur, this results in an incredibly mild singularity that occurs at sufficiently high order such that it does not even act as a source in the field equations. In other words, the regularity of the metric is always $\mathcal{C}^3$ or better, ensuring finiteness of not only the Kretschmann and Ricci scalars, but also the action. We therefore do not view this as a physically important obstruction.

In Figure~\ref{fig:f(r)_k1}, we display plots of the metric function $f(r)$ for the first six models in Table~\ref{tab:metrics} for different choices of the thermodynamic mass $M$, fixing the effective cosmological constant $L_{\rm eff}$ and the parameter $a$. The structure of all six models is qualitatively identical. We summarize the main features in what follows.  At large values of the the thermodynamic mass, the metric function describes a black hole. There is an event horizon and inner horizon with the de Sitter core emerging near $r = 0$. As the thermodynamic mass is decreased, a critical value is reached for which the event and inner horizons merge, creating an extremal limit. Finally, for masses smaller than this critical value, the metric function describes nonsingular, horizonless objects. These can be thought of as completely regular gravitational solitons.

In general, expressions for the horizon radius in terms of the mass cannot be obtained in a useful closed form. However, there are simple parametric relationships that hold for any model. At any horizon (for the $k=+1$ metrics), we have
\be 
\mathcal{S}(r_+) = h_{\rm f}\left(\frac{1}{r_+^2}\right) \, . 
\ee
For a degenerate horizon, we have in addition to the above 
\be 
r_+^3 \mathcal{S}'(r_+) = 2 h'_{\rm f}\left(\frac{1}{r_+^2}\right) \, .
\ee
Relationships such as these are useful in obtaining a unified picture of the black hole thermodynamics, as we will discuss below. 

\subsection{Regular black branes}

Anti-de Sitter asymptotics allow for horizons that differ from spheres. Black branes, which correspond to $k = 0$, have $\mathbb{R}^{D-2}$ as the transverse geometries, or quotients thereof. 

In vacuum Einstein gravity, a black brane has metric function 
\be 
f(r) = \frac{r^2}{L^2} - \frac{m}{r^{D-3}} \, .
\ee
When $m > 0$, the black brane has a single horizon and a curvature singularity at $r = 0$, analogous to the spherical black hole. 

Let us consider the situation when the singularity of the black brane has been regularized by the infinite tower of higher-order curvature corrections. In all of the examples illustrated in Table~\ref{tab:metrics}, the metric function near $r = 0$ has the expansion
\be 
f(r) = - \frac{r^2}{|\alpha|} + \mathcal{O}(r^3) \, .
\ee
To understand the nature of this limiting geometry, let us  introduce a new coordinate $\tau = r/\sqrt{\alpha}$ and replace the coordinate $t\to \sqrt{\alpha}z$, to avoid confusion. Then, the metric near $r = 0$ reads
(dropping the overall constant conformal prefactor $\alpha$):
\be 
{\rm d} s^2 = - \frac{{\rm d} \tau^2}{\tau^2} + \tau^2 \left[{\rm d} z^2 + \delta_{ij} {\rm d}x^i {\rm d} x^j  \right] \,,
\ee
where $\delta_{ij}$ is just the flat Euclidean metric.

We immediately recognize the above as de Sitter space in flat slicing coordinates, which can be made manifest by switching to a proper time frame with $T = \log \tau$, upon which the metric reads

\be 
{\rm d} s^2 = -{\rm d} T^2 + e^{2 T} \left[{\rm d} z^2 + \delta_{ij} {\rm d}x^i {\rm d} x^j  \right]  \, . 
\ee

We then understand that the nature of $r=0$ is very different from the spherical case considered above. The surface $\tau = 0$ is null and in fact it is a Killing horizon generated by $\xi = \partial_z$. 
We should then interpret $r = 0$ of the regular black branes not as the origin of coordinates, but instead as  an inner horizon. The corresponding surface gravity reads

\be 
\kappa^2 = -\left(\nabla_a L\right) \left( \nabla^a L \right) = \tau^2  \quad \text{where} \quad L^2 = \xi_a \xi^a \, .
\ee
Therefore, we see that the inner horizon, being located at $\tau = 0$, is {automatically} an extremal one. In other words, the regular black branes are \textit{generically} {\em inner extremal}. No fine tuning of parameters is required.\footnote{A similar feature has recently been identified for charged, regular black holes in $D = 3$ --- see Ref.~\cite{Bueno:2025jgc} for details.}\ This is particularly interesting, as inner extremal regular black holes can avoid the problems typically associated with mass inflation at the inner horizon~\cite{Carballo-Rubio:2022kad}. Therefore, these black branes may in fact have stable interiors. 

An important issue is the extension of the metric across the inner horizon. Here, we shall not dwell on this in too much detail. The most straightforward method is to extend the chart across the $r = 0$ surface by allowing $r$ to take on negative values. If the metric function is not invariant under $r \to -r$, this extension may introduce singularities that arise in the negative $r$ region. It also appears reasonable to allow an extension that glues together two copies of the $r > 0$ geometry along the $r = 0$ surface. This extension would be very mildly singular at $r = 0$ if odd powers of $r$ appear in metric function. We leave a full analysis of the global structure to future work. 

As an example, the Hayward-AdS model solves the equations of motion in any odd dimension. The metric function is invariant under $r \to -r$, and therefore the Hayward black branes are completely nonsingular. The same feature is in fact shared by all of the models in Table~\ref{tab:metrics} in odd dimensions. In the even dimensions for which they exist, models IV, V, VI and VII with ${\rm N}$ even are completely regular under the extension to negative $r$ --- we will illustrate examples of this below in Figure~\ref{fig:maxwell_branes}.

\subsection{Regular topological black holes}

When $k = -1$, the transverse geometry is a hyperbolic space. These black objects are known as \textit{topological} black holes, since the hyperboloid can be identified to yield horizons that are Riemann surfaces of any genus $g \ge 2$~\cite{Mann:1997iz, Birmingham:1998nr, Emparan:1999gf}.  

In Einstein gravity, the vacuum metric function for topological black holes reads
\be 
f(r) = \frac{r^2}{L^2} - 1 - \frac{m}{r^{D-3}} \, .
\ee
Contrary to the cases we have considered so far, hyperbolic black holes are sensible even when the mass parameter is negative. Depending on the mass parameter, the geometry can have one or two horizons, while the surface $r = 0$ is a curvature singularity. 

More specifically, for nonzero $m$, the surface $r = 0$ is always a curvature singularity. If $m > 0$, there is a single horizon and the global structure is similar to the Schwarzschild-AdS black hole, with an event horizon and a spacelike singularity. On the other hand, if 
\be \label{eqn:topo-mass-bound}
- \frac{2 L^{D-3}}{(D-1)} \left(\frac{D-3}{D-1} \right)^{(D-3)/2} < m < 0 ,
\ee
then there are two horizons and the global structure is qualitatively the same as Reissner-Nordström-AdS, with event and inner horizons and a timelike singularity. For $m$ less than the lower bound given above, the spacetime is nakedly singular. 

When we consider topological black holes as solutions to the resummed quasi-topological theories, the nature of $r = 0$ becomes considerably more subtle. In all of the cases presented in Table~\ref{tab:metrics}, the metric function near $r=0$ has the expansion
\be 
f(r) = - 1 - \frac{r^2}{|\alpha|} + \,\, \text{subleading} \, .
\ee
Setting $\alpha = 1$ without loss of generality, and defining $r=\tau$ and $t = z$, the full metric has the limiting behavior
\be 
{\rm d}s^2 = - \frac{{\rm d} \tau^2}{1+\tau^2} + \left(1 + \tau^2 \right) {\rm d}z^2 + \tau^2 {\rm d}\Sigma_g^2 \, ,
\ee
where ${\rm d}\Sigma_g^2$ represents the metric on the hyperboloid. We have included the subscript $g$ as a reminder that it is possible (though not necessary) to perform topological identifications to this space. This is a de Sitter universe written in a somewhat unfamiliar Bianchi III slicing.\footnote{This can be made further manifest upon introducing a proper time frame 
\be 
{\rm d} \hat{\tau} = \frac{{\rm d}\tau}{\sqrt{1+\tau^2}} \, ,
\ee
in which the metric becomes
\be 
{\rm d}s^2 = - {\rm d} \hat{\tau}^2 + \cosh^2 \hat{\tau} \; {\rm d}z^2 + \sinh^2 \hat{\tau} \; {\rm d}\Sigma_g^2\, .
\ee
This can be compared with, for example, Eq.~(9.12) of Ref.~\cite{Eriksen:1995ws} for the case of $D = 4$.}  

To better understand what happens at $r = 0$, let us focus on the case of $D = 4$ and expand the above metric in the limit of small $\tau$ (recall $r = 0 \Leftrightarrow \tau = 0$). We then have
\be 
{\rm d}s^2 = - {\rm d}\tau^2 +  {\rm d} z^2 + \tau^2 \left({\rm d}\chi^2 + \sinh^2 \chi {\rm d}\phi^2 \right)\, ,
\ee
where we have now made the hyperboloid metric explicit. By zooming in to $\tau = 0$, we are examining the metric on scales much shorter than the de Sitter length scale, explaining the emergence of the line element for a {\em Milne universe} above. Performing the change of coordinates
\be 
T = \tau \cosh \chi \,, \quad X = \tau \sinh \chi \cos \phi \, , \quad Y = \tau \sinh \chi \sin \phi \, ,
\ee
the metric is brought into Minkowski form,
\be 
{\rm d} s^2  = - {\rm d} T^2 + {\rm d} X^2 + {\rm d} Y^2 + {\rm d} z^2 \, .
\ee
We have the relation $\tau^2 = T^2 - X^2 - Y^2$ and the region near to $\tau = 0$ is the portion of Minkowski space with $T > \sqrt{X^2 + Y^2}$ --- the {\em interior of the 
lightcone} in $(2+1)$-dimensions times $\mathbb{R}$. We therefore understand that the surface $r = 0$ in the topological regular black holes is null.  However, contrary to the case of the black branes considered above, this surface is \textit{not} a Killing horizon: No combination of boosts/translations of the Minkowski metric can be made to vanish at the surface $T^2 = X^2 + Y^2$.\footnote{In more than $(1+1)$ dimensions, the lightcone expands and cannot be a Killing horizon --- instead it is a {\em conformal Killing horizon}, see, for instance, Ref.~\cite{DeLorenzo:2017tgx}. However, in our case, we have a lightcone \textit{times} a line and it is more subtle. Hence, it remains to be seen whether $r = 0$ can be interpreted as a conformal Killing horizon.}

The continuation of the metric past $r = 0$ is also more subtle. Here, we will highlight the main subtleties and leave a detailed analysis of the global structure for future work. To illustrate the point, let us continue to work with the Milne/Minkowski charts introduced above. As explained above, ($\hat{\tau}, z, \chi, \phi$) covers the interior of the lightcone. To obtain an analogous form of the metric in the exterior of the lightcone, we can introduce coordinates
\be 
T = \rho \sinh \chi \, ,\quad X = \rho \cosh \chi \cos \phi \, , \quad Y = \rho \cosh \chi \sin \phi \, .
\ee
Then, the metric valid in the exterior reads
\be 
{\rm d}s^2 = {\rm d}\rho^2 + \rho^2 \left(-{\rm d}\chi^2 + \cosh^2 \chi {\rm d}\phi^2 \right) + {\rm d} z^2\, .
\ee
This metric describes the geometry just beyond $r = 0$ in the topological black holes on scales $r \ll \alpha$. The main subtlety should now be obvious. When we pass through $r = 0$ in the topological black hole, we emerge in a region where $\chi$ plays the role of a time coordinate. In the case of the black hole and the black brane, all of the interesting behavior takes place in the $(t,r)$ sector of the metric. However, this is no longer the case for the hyperbolic black hole, due to the intricate nature of the $r = 0$ surface.

The above makes clear an important fact: If we perform identifications to the hyperboloid, yielding horizon geometries which are higher-genus surfaces, then the surface $r = 0$ is a chronology horizon. That is because identifications of the hyperboloid become temporal identifications after we cross $r = 0$. A completely analogous situation arises in the interior of the BTZ black hole. There, the $r = 0$ surface is again null and is interpreted as a singularity in the causal structure --- see Ref.~\cite{Banados:1992gq} for further details. Therefore, while the infinite tower of corrections resolves the \textit{curvature} singularity of the topological black holes, it can leave in its place a singularity analogous to that of BTZ.

\section{Electrically charged black holes}
\label{sec:electro}

\subsection{Maxwell electrodynamics}

As the prototypical example of gravity minimally coupled to matter, we consider adding a Maxwell term to the action. The matter Lagrangian then reads
\be 
I_{\rm EM} = -\frac{1}{16 \pi G_{\rm N}} \int {\rm d}^D x \sqrt{|g|} F_{\mu \nu} F^{\mu \nu} \, ,
\ee
where the field strength is given in terms of a vector potential by the standard relation
\be 
F_{\mu \nu} = \partial_\mu A_\nu - \partial_\nu A_\mu \, .
\ee
Taking an electric ansatz for the vector potential,
\be \label{eq:gauge_ansatz}
A = \Phi(r) {\rm d} t \, ,
\ee
the reduced action for the Maxwell terms reads
\be 
I_{\rm EM} = \frac{\Omega_{D-2}}{8 \pi G_{\rm N}} \int {\rm d}r \frac{r^{D-2} \left(\Phi'\right)^2}{N} \, .
\ee
Varying the reduced action with respect to $\Phi(r)$ yields the Maxwell equation, which is solved by
\be 
\Phi(r) = \sqrt{\frac{D-2}{2(D-3)}} \frac{q}{r^{D-3}} \, ,
\ee
where $q$ is a constant of integration proportional to the electric charge. Varying with respect to $N(r)$ gives the relevant component of the EMT,
\be 
T_{t}^t = - \frac{\left(\Phi' \right)^2}{8 \pi G_{\rm N}} \, .
\ee
Plugging this into Eq.~\eqref{feq} allows us to determine $\mathcal{S}(r)$ for the Maxwell theory:
\be 
\mathcal{S}_{\rm EM}(r) = -\frac{1}{L^2} + \frac{m}{r^{D-1}} - \frac{q^2}{r^{2(D-2)}} \, .
\ee
Using this functional form for $\mathcal{S}(r)$ in the models of Table~\ref{tab:metrics} gives the corresponding metric function.

Let us remark on some of the properties of these black holes: First, we note that $G_{\rm eff}$ and $L_{\rm eff}$ derived from the asymptotic behavior of the metric function $r \to \infty$ are identical to before --- see Table~\ref{tab:AdS_params} for the different possibilities. This is because at large $r$ the charge is subleading. However, the behavior of the metric at smaller values of $r$ is different. We note that, restricting to $\alpha > 0$, models IV, V, VI, and VII with ${\rm N}$ even are regular, while the other models exhibit singularities at finite radius. This is due to the fact that, as discussed in Section~\ref{sec:regularity conditions}, %to the fact that 
the functions $h_{\rm f}(\psi)$ are not monotonic over the domain of $\psi$ required by $\mathcal{S}_{\rm EM}(r)$. If one wishes to allow for negative $\alpha$, the other models also admit completely regular Maxwell charged solutions. This is the case, for example, for the Hayward resummation. However, while negative $\alpha$ allows these models to be regular with charge, the $q \to 0$ limit results in singular solutions. We therefore will not dwell on these particular cases.  

Importantly, the charge-dependent contribution to $\mathcal{S}(r)$ \textit{dominates} at small $r$. This leads to a qualitative change in the behavior of the metric in that vicinity. We find that all regular configurations now take the form
\be 
f(r) = k + \frac{r^2}{|\alpha|} + \mathcal{O}(r^3) \, .
\ee
In other words, regular black holes with a Maxwell charge have \textit{anti}-de Sitter cores rather than de Sitter ones. This shift from a de Sitter core to an anti-de Sitter one is a consequence of the sign of $\mathcal{S}(r)$ near $r = 0$. In fact, it is easy to show that for models IV, V, VI, and VII with ${\rm N}$ even we have generally
\be 
\label{eq:core_nature}
f(r) = k - \frac{r^2}{|\alpha|} \lim_{r\to0}{\rm sign}\left( \mathcal{S}(r) \right)  + \cdots \, .
\ee
A physical explanation for this change in the core structure will become more transparent when we consider the theories of nonlinear electrodynamics below.  Despite the difference in the core, this does not change qualitatively what we discussed earlier for the continuation of the metric through $r = 0$ in the planar and hyperbolic cases. In each case, the de Sitter region is replaced by an anti-de Sitter one but the conclusion is still the same. Namely, that the metric function should be continued through to negative $r$ in the cases of $k = 0$ and $k=-1$. 

\begin{figure}[t]
    \centering
    \includegraphics[width=\linewidth]{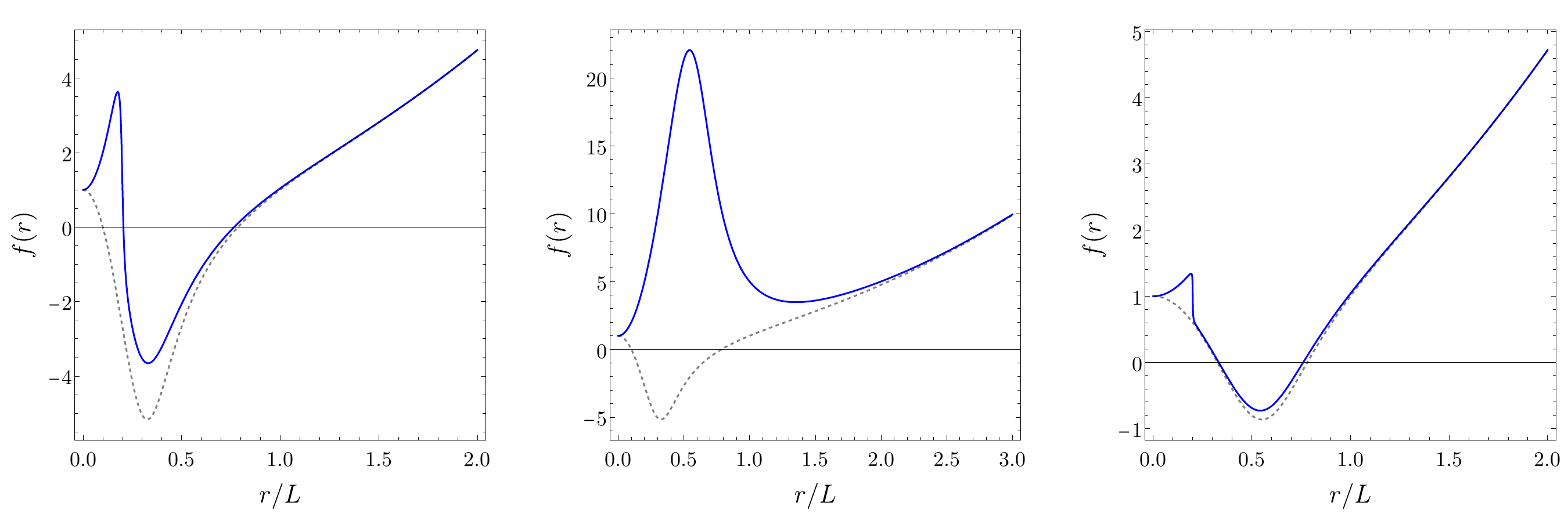}
    \caption{{\bf Metric functions:
    Maxwell charged black holes.} Plots of the $k=+1$ metric function for model IV for different values of the coupling and charge. In each plot, we have set $L = 1$, $m=1$, and $D = 5$. We have also included the corresponding metric function with $q=0$, shown in each panel as the dashed gray curve. The parameter values are: $\alpha = 1/100, q = 1/20$ (left), $\alpha = 1/100, q = 2$ (center) and $\alpha = 1/10, q = 1/20$ (right). Models V, VI and VII with ${\rm N}$ even yield metric functions that have exactly the same qualitative structure.}
    \label{fig:maxwell_bhs}
\end{figure}

\begin{figure}[t]
    \centering
    \includegraphics[width=\linewidth]{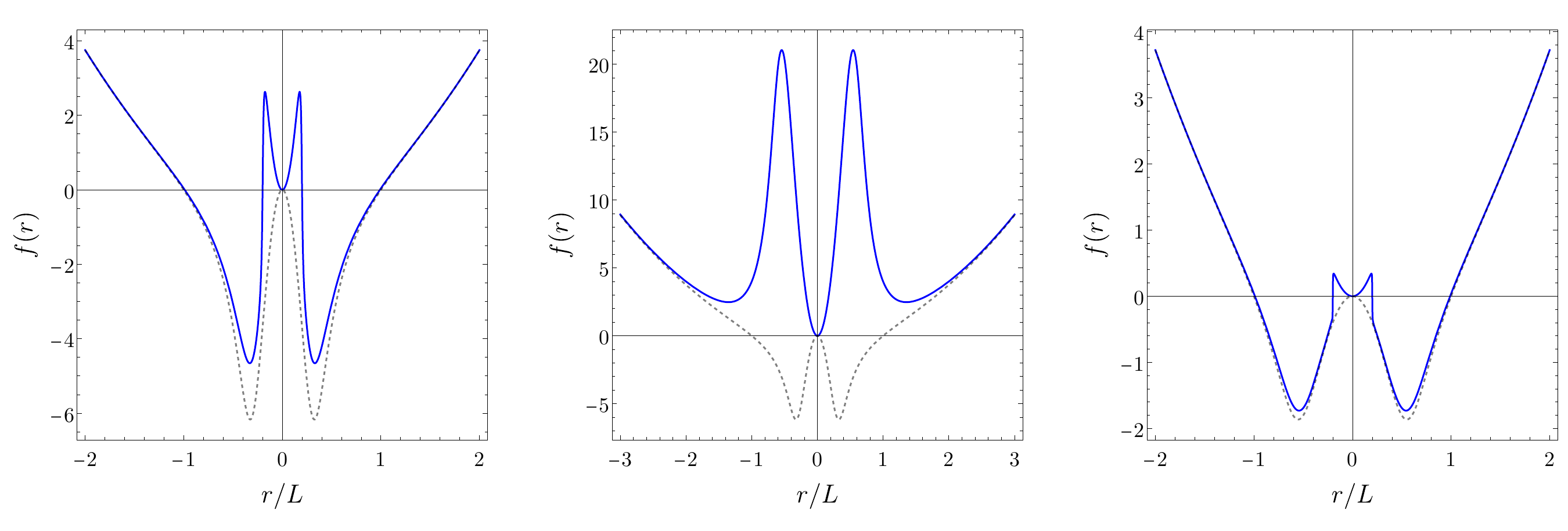}
    \caption{{\bf Metric functions: black branes.} Plots of the $k=0$ metric function for model IV for different values of the coupling and charge. In each plot, we have set $L = 1$, $m=1$, and $D = 5$. We have also included the corresponding metric function with $q=0$, shown in each panel as the dashed gray curve. The parameter values are: $\alpha = 1/100, q = 1/20$ (left), $\alpha = 1/100, q = 2$ (center) and $\alpha = 1/10, q = 1/20$ (right). Models V, VI and VII with ${\rm N}$ even yield metric functions that have exactly the same qualitative structure.}
    \label{fig:maxwell_branes}
\end{figure}

In Figure~\ref{fig:maxwell_bhs} we plot the metric function for the $k = +1$ black holes in five dimensions. The structure of the metric function is qualitatively similar for each of the regular models, and thus we present only model IV. Depending on the particular parameters, there will be either zero, two, or one degenerate horizon. When there is no horizon, the metric describes a self-gravitating charged ``soliton''. When horizons exist, there is typically an event and inner Cauchy horizon. These horizons merge in an extremal/critical limit. For large $r$, the metric functions are qualitatively similar to the uncharged case considered in the previous section. The effect of charge is most pronounced in the deep interior. There, we observe a relatively sharp transition where the metric function takes on the necessary behavior to exhibit the AdS core. Here, the metric function almost appears to vertically ``jump''.  However, even though the gradients can be large in this region, the curvature can be large but nonetheless \textit{always} finite. In Figure~\ref{fig:maxwell_branes}, we consider the Maxwell charged black branes, to which similar comments apply. The topological black holes have identical metric functions, just shifted vertically by $1$. 

\subsection{Regular black holes with regular electric field}

So far, we have constructed regular black holes coupled to a Maxwell field. However, while the spacetime metric could be made regular, there remained a divergence of the electric field at the origin. To resolve such divergences, nonlinear electrodynamics (NLE) was introduced as a modification of Maxwell's theory, originally aimed at addressing the self-energy problem of charged particles. Among various NLE models, the Born--Infeld theory, introduced by M. Born and L. Infeld in Refs.~\cite{born1933electromagnetic, %Born--Infeld:1934,
Born:1934gh}, remains one of the most influential. It ensures a finite self-energy for point charges,  
exhibits electromagnetic duality \cite{Gibbons-Rasheed:1995}, and recovers Maxwell’s electrodynamics in the weak-field limit, while 
being at the same time a unique NLE (apart from Maxwell) not suffering from 
birefringence \cite{plebanski1970lectures, Russo:2022qvz}.
Another much more recent model of NLE is the Regularized Maxwell (RegMax) theory \cite{Tahamtan:2021, Hale-Kubizňák-Svítek-Tahamtan:2023}, which introduces a minimal
regularization to the field strength of a point charge. Despite this modification, many of its self-gravitating solutions retain a strikingly Maxwell-like character \cite{Tahamtan:2021, Hale-Kubizňák-Svítek-Tahamtan:2023,
Kubiznak-Tahamtan-Svitek:2022,
Hale:2025veb}.

We thus explore the Lagrangians of these matter fields as a means to construct black holes --- and their horizonless counterparts --- featuring \textit{both finite} gravitational and electromagnetic fields.  The contribution of NLE fields to the matter sector of the total action, assuming dependence solely on the field strength $\mathcal{F}$, as is the case in the original
Born--Infeld and RegMax NLEs, is given by\footnote{The shift from linear Maxwell theory to NLE is very natural in our setting. Namely, apart from considering infinite series in the gravity sector, we now consider an infinite series, $\sum_n b_n {\cal F}^n$ in the electromagnetic sector as well.}
\begin{align}
    I_{\mathrm{matter}}=\frac{1}{16\pi G_{\rm N}}\int d^{D}x\sqrt{|g|}4\mathcal{L}(\mathcal{F})\,, \quad \mathcal{F}=\tensor{F}{^\mu^\nu}\tensor{F}{_\mu_\nu}\,.
\end{align}
Varying this action with respect to the metric tensor leads to the EMT for such a theory:
\begin{align}
    T^{\mu\nu}=-\frac{1}{4\pi}\bigg(4 \tensor{F}{^\mu^\sigma}\tensor{F}{^\nu_\sigma}\mathcal{L}_{\mathcal{F}}-\mathcal{L}\tensor{g}{^\mu^\nu}\bigg)\,.\label{eq:EMT-NLE}
\end{align}
Using Eq.~\eqref{feq} 
along with the corresponding components of the EMT from Eq.~\eqref{eq:EMT-NLE}, we now analyze these two NLE models in detail.

\subsubsection{Born--Infeld black holes}

We begin by analyzing the Born--Infeld model, which has been extensively studied to obtain gravitational solutions in both general relativity and modified gravity theories across arbitrary spacetime dimensions. The Lagrangian density is given by \cite{Dey:2004}:\footnote{Note that this form of the action, which coincides with the original formulation of the theory in Ref.~\cite{born1933electromagnetic}, is not the same as the (more distinguished and popular) later version of Born--Infeld electrodynamics written in Ref.~\cite{Born:1934gh}, which reads
\be 
\mathcal{L} = -b^2 \sqrt{-{\rm det} \left( g_{\mu\nu} + \frac{F_{\mu\nu}}{b} \right)} + b^2 \, . 
\ee 
However, for static, purely electric solutions as the ones we consider here, the two formulations are identical~\cite{Li:2016nll}. }
\begin{align}
   \mathcal{L}_{\mathrm{BI}}(\mathcal{F}) = b^2 \left( 1 - \sqrt{1 + \frac{\mathcal{F}}{2 b^2}} \right)\,,
\end{align}
where $b$ is a dimensionful parameter introduced to regularize the electric field, and in the limit $b\rightarrow +\infty$, we have
\begin{align}
    \lim_{b\rightarrow +\infty}\mathcal{L}_{\mathrm{BI}}(\mathcal{F})=-\frac{1}{4}\mathcal{F}\,,
\end{align}
which corresponds to the Maxwell theory. 

The reduced action for the Born--Infeld theory takes the form
\be 
I_{\rm BI} = \frac{\Omega b^2 N}{4\pi G_{\rm N}}\left[1- \sqrt{1- \frac{\left(\Phi'\right)^2}{b^2 N^2}} \right] \, ,
\ee
where we have taken the same electric ansatz for the field strength as given in~\eqref{eq:gauge_ansatz}. As before, varying the action with respect to $\Phi(r)$ yields the Maxwell equation, which can be solved explicitly to give the electric field
\begin{align}
    \label{eq:BI_Efield}
    E_{\mathrm{BI}}(r) \equiv \frac{d\Phi}{dr} =\frac{\sqrt{(D - 2) (D - 3)} \, b \, q}{\sqrt{2 b^2 r^{2 (D - 2)} + (D - 2) (D - 3) q^2}}\,,
\end{align}
where $q$ is an integration constant related to the electric charge. The Born--Infeld modification of Maxwell's electrodynamics becomes important at short distances and results in a finite (as opposed to divergent) electric field at $r = 0$. 

Varying the action with respect to $N$ (and subsequently fixing $N = 1$) gives the relevant stress-tensor component,\footnote{The same result can be obtained working directly with the EMT as given in Eq.~\eqref{eq:EMT-NLE}.} 

\be 
T_t^t = \frac{b^2}{4 \pi G_{\rm N}}  \left[1 - \left(1-\frac{E_{\rm BI}^2}{b^2} \right)^{-1/2} \right] \, .
\ee
The required integral from Eq.~\eqref{feq} has a closed-form but not particularly illuminating expression,
\begin{align}
    \mathcal{S}_{\mathrm{BI}}(r) &=  -\frac{1}{L^2} + \frac{m}{r^{D-1}} - \frac{4 b^2}{(D-2)(D-1)}
    \nonumber
    \\
    &+  \frac{4 b^2}{(D-2)(D-1)} \,\, {}_2F_1 \!\left(-\frac{1}{2}, - \frac{(D-1)}{2(D-2)}; \frac{(D-3)}{2(D-2)}; - \frac{(D-3)(D-2) q^2}{2 b^2 r^{2(D-2)}} \right)\,,
\end{align}

where ${}_2F_1$ is the hypergeometric function. The integration constant and conventions in the above have been carefully chosen such that the Maxwell limit ($b\to \infty)$ is reproduced in the same conventions as in the previous section. 

Let us make some general remarks about the behavior of the solutions with Born--Infeld electrodynamics: First, note that the large $r$ asymptotics have the form
\be 
\mathcal{S}_{\mathrm{BI}}(r) = -\frac{1}{L^2} + \frac{m}{r^{D-1}} - \frac{q^2}{r^{2(D-2)}} + \mathcal{O}\left( \frac{1}{r^{4(D-2)}}\right) \, .
\ee
Thus, all of the asymptotic information such as the effective cosmological length scale and the effective Newton constant are the same as considered earlier and summarized in Table~\ref{tab:AdS_params}. On the other hand, the behavior of $\mathcal{S}_{\rm BI}(r)$ at small radius is much more interesting. We have
\be 
\mathcal{S}_{\mathrm{BI}}(r) = \frac{m - m_\star}{r^{D-1}} + \sqrt{\frac{8(D-3)}{D-2}} \frac{b |q|}{r^{D-2}} +  \cdots\,, 
\ee
where we have included only the two most dominant terms in the limit of small $r$. To simplify the notation, we have introduced the self-energy quantity\footnote{Note that, since the energy density is just $\rho = -T_t^t$, $m_\star$ is simply related to the electrostatic self-energy $M_\star$ of a Born--Infeld point particle,
\be 
M_\star = \Omega_{D-2} \int_0^\infty r^{D-2} \rho \, {\rm d} r \quad \Rightarrow \quad M_\star = \frac{m_\star (D-2) \Omega_{D-2}}{16 \pi G_{\rm N}} \,  .
\ee
}
\be 
m_\star = \frac{2^\frac{D-3}{2(D-2)} b^\frac{D-3}{D-2} |q|^\frac{D-1}{D-2} \left[(D-3)(D-2) \right]^\frac{D-1}{2(D-2)} }{\sqrt{\pi} (D-2)(D-1)}  \Gamma\left[\frac{1}{2(D-2)} \right] \Gamma\left[\frac{(D-3)}{2(D-2)} \right]\,.
\ee

The important point here is that in Born--Infeld theory, at small distances, the charge term $m_\star$ contributes
to $\mathcal{S}(r)$ at the same order as the mass. This means, returning to Eq.~\eqref{eq:core_nature}, whether the Born--Infeld black holes have a de Sitter or anti-de Sitter core depends on the ratio of $m_\star/m$. Specifically, when $m_\star > m$, the core is anti-de Sitter, while if $m_\star < m$, the core is de Sitter. In the marginal case where $m = m_\star$, the nature of the core is determined by the subleading term in $\mathcal{S}_{\rm BI}(r)$. Since this term is manifestly positive, the result is a de Sitter core in this case. 

Since $\mathcal{S}_{\rm BI}(r)$ takes on both positive and negative values, not all of the resummations in Table~\ref{tab:metrics} will yield fully regular black holes for all parameter values. Just as in the Maxwell case, models IV, V, VI, and VII with ${\rm N}$ even are fully regular for all values of the charge. However, due to the appearance of the parameter $m_\star$, the situation is more intricate. We find that when $m > m_\star$, then all models are regular provided the constraints on the cosmological length scale described in Table~\ref{tab:AdS_params} are met. This is a result of the fact that $\mathcal{S}(r)$ cannot become arbitrarily negative when $m > m_\star$. 

\begin{figure}[t]
    \centering
    \includegraphics[width=\linewidth]{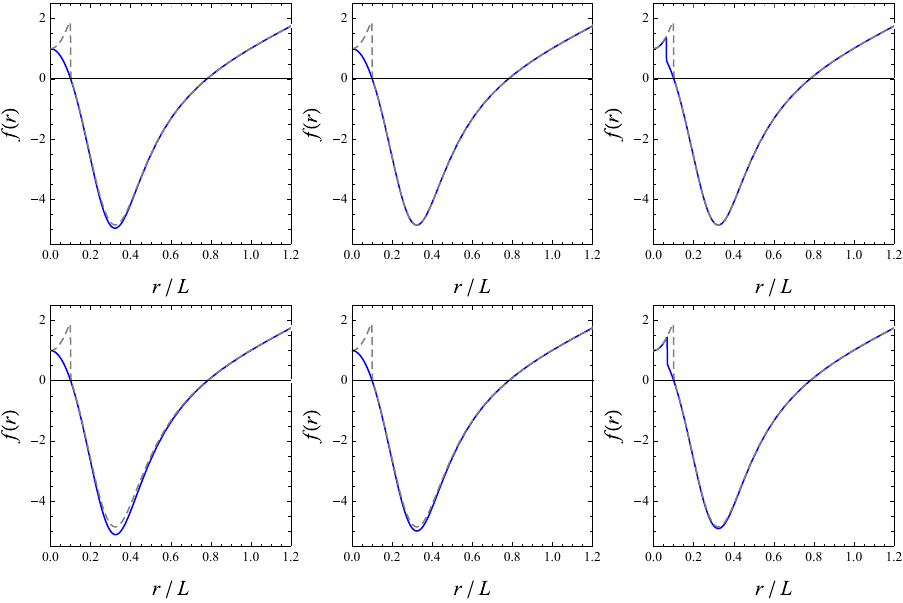}
    \caption{{\bf Metric functions: charged 
    black holes in nonlinear electrodynamics.
    } Plots of the $k=1$ metric function for model IV, where we have set $m=1$, $L=1$, $\alpha=0.01$, $q=0.1$, and $D=5$ for two different NLE Lagrangians illustrated by the blue solid line. The Maxwell case is shown as a gray dashed line. \textbf{Top:} Born---Infeld case for $m>m_{\star}$ (left), $m=m_\star$ (middle), and $m<m_\star$ (right), achieved by choosing $b=1$, $b\simeq 20.07$, and $b=40$, respectively; \textbf{Bottom:} RegMax case for $m>m_{\star}$ (left), $m=m_\star$ (middle), and $m<m_\star$ (right), achieved by choosing $\beta=1$, $\beta\simeq 4.46$, and $\beta=15$, respectively. Both Born--Infeld and RegMax metric functions have exactly the same qualitative structure.}
    \label{fig:BI-RegMax}
\end{figure}

We show in Figure~\ref{fig:BI-RegMax} (top row) plots of the metric function for the spherical Born--Infeld black holes. We focus on one particular model, in this case model IV, as all models produce qualitatively similar metric functions. The plots show three cases, one with $m > m_\star$, one with $m = m_\star$, and one with $m < m_\star$, from left to right. In the same plots we also include the metric function in the Maxwell case for comparison. The effects of NLE become most pronounced at small radius, with the large $r$ behavior essentially indistinguishable from the Maxwell case. Focusing on small $r$, we see explicitly the feature discussed above concerning how the ratio of $m/m_\star$ determines whether the core is a de Sitter or anti-de Sitter one. 

\subsubsection{RegMax charged black holes}

Regularization of Maxwell’s theory is possible in any number of dimensions, with a detailed method provided in Appendix C of Ref.~\cite{Kubizňák-Svítek-Tahamtan:2024}. The Lagrangian density is expressed in terms of a hypergeometric function, given by
\begin{align}
\mathcal{L}_{\mathrm{RegMax}}(s)=\frac{1}{2}\beta^{2(D-2)}s^{2(D-2)} \;{}_2F_1\Big(D-2, 2(D-2); 2(D-2)+1;s \Big)\,,
\end{align}
where $\beta$ is a dimensionful regularization parameter for the electric field, analogous to the parameter $b$ in the Born--Infeld theory, and $s$ is defined as
\begin{align}
    s=\frac{1}{\beta}\left(-\frac{\mathcal{F}}{2}\right)^{\frac{1}{2(D-2)}}, \qquad s\in(0,1)\,.
\end{align}
As before, we can obtain the reduced action
\be 
I_{\rm RegMax} = \frac{\Omega_{D-2}}{8 \pi G_{\rm N}} \int r^{D-2} \frac{\Phi'^2}{N} \,\, {}_2F_1 \!\bigg(D-2, 2(D-2); 2D-3; \frac{1}{\beta} \left(\frac{\Phi'^2}{N^2} \right)^\frac{1}{2(D-2)}  \bigg) \, {\rm d} r\,.
\ee
Varying with respect to $\Phi(r)$ yields the Maxwell equations, which in this case are solved by the electric field
\be \label{eq:RM_Efield}
E_{\rm RegMax} \equiv \frac{d \Phi}{dr} = \frac{1}{\sqrt{2}} \frac{ \sqrt{(D-2)(D-3)} q}{\left[r + \frac{1}{\beta}  \left(\sqrt{\frac{(D-2)(D-3)}{2}} q \right)^{1/(D-2)} \right]^{D-2}} \, . 
\ee
Here, $q$ is a constant of integration, and we have chosen the conventions such that when $\beta \to \infty$ we reproduce the Maxwell limit with the same conventions introduced above. Varying the action with respect to $N$ allows us to isolate the relevant component of the EMT,

\begin{align}\label{eq:Ttt_regmax}
T_t^t &= \frac{\beta^{D-3} E_{\rm RegMax}^2}{8 \pi G_{\rm N} }\left[\beta - \left(E_{\rm RegMax}^2\right)^\frac{1}{2(D-2)}\right]^{2-D} \bigg[-2 \beta 
\nonumber
\\
&\quad + \left(\beta - \left(E_{\rm RegMax}^2\right)^\frac{1}{2(D-2)} \right) \;\! {}_2F_1 \!\left(1, D-1; 2D-3; \frac{1}{\beta} \left(E_{\rm RegMax}^2\right)^\frac{1}{2(D-2)} \right) \bigg]\,.
\end{align}

Unfortunately, the required integral of the EMT does not admit a simple closed form in general dimensions. Nonetheless, in a dimension-by-dimension case, the integral can be evaluated. For example, in five dimensions, we have

\begin{align}
    \mathcal{S}^{D=5}_{\rm RegMax}(r)= & -\frac{1}{L^2} + \frac{m}{r^4} - \frac{3^{2/3} q^{4/3} \beta^2}{2 r^4} + 
\frac{4 q \beta^3}{\sqrt{3} r^3} - \frac{3^{4/3} q^{2/3} \beta^4}{r^2} \notag \\
& + \frac{8 \cdot 3^{1/6} q^{1/3} \beta^5}{r} + 2 \beta^6 
- \frac{2 r \beta^7}{3^{1/6} q^{1/3} + r \beta} 
+ 10 \beta^6 \log \left(\frac{r \beta}{3^{1/6} q^{1/3} + r \beta} \right)\,.
\end{align}

The geometry is fully regular for models IV, V, VI and VII with ${\rm N}$ even, just as for the Maxwell case. 

The small $r$ behavior of $\mathcal{S}(r)$ differs strongly from the Maxwell case. For RegMax, we find
\be 
\mathcal{S}_{\rm RegMax}(r) = \frac{m-m_\star}{r^{D-1}} + \cdots  \, ,
\ee
which closely mimics the Born--Infeld case we studied above. Here, the quantity $m_\star$ can be obtained in a remarkably simple closed form, 
\be 
m_\star = \frac{4 \beta^{D-3}}{(D-1)(D-3)} \left(\sqrt{\frac{(D-2)(D-3)}{2}} q\right)^\frac{(D-1)}{(D-2)} \, .
\ee
When $m > m_\star$, there is a de Sitter core, while there is an anti-de Sitter core for $m < m_\star$. When $m > m_\star$, it turns out that all models in Table~\ref{tab:metrics} are regular, just as in the Born--Infeld case.

The bottom row of Figure~\ref{fig:BI-RegMax} shows the metric function for RegMax black holes. As for the Born--Infeld case, we display
$m> m_\star$ (left), $m = m_\star$ (middle), and $m < m_\star$ (right). The large $r$ behavior of the metric function is qualitatively identical to the Maxwell case, with differences appearing at small $r$.

\section{Thermodynamics of regular black holes}

\subsection{First law and Smarr relation}
\label{sec:thermo}

The thermodynamics of asymptotically flat regular black holes was briefly considered in Ref.~\cite{Bueno:2024dgm}, with the expressions for the mass, temperature and black hole entropy provided. This analysis makes use of the general considerations of quasi-topological black hole thermodynamics outlined in Refs.~\cite{Bueno:2019ycr, Bueno:2022res} --- in particular, see Section 4 of Ref.~\cite{Bueno:2022res}. Here, we extend this analysis, constructing also the Smarr formula and allowing for the inclusion of minimally coupled matter.

For $k = \pm 1$, the thermodynamic potentials read:
\begin{align}\label{eq:thermo_eqns}
M &= \frac{(D-2) \Omega_{D-2} r_+^{D-1}}{16 \pi G_{\rm N}} h(\psi_+) + \frac{(D-2) \Omega_{D-2}}{16 \pi G_{\rm N}} \frac{r_+^{D-1}}{L^2} + \Omega_{D-2} \int_\infty^{r_+} r^{D-2} T_t^t {\rm d} r \, , 
\nonumber\\
T &= -\frac{1}{4 \pi r_+ }\left[ \frac{r_+^3 \mathcal{S}'(r_+)}{h'(\psi_+)} + 2 k \right] \, ,
\quad
S = \frac{(D-2) \Omega_{D-2}}{4 G_{\rm N}} \int r_+^{D-3} h_{\rm f}'(k/r_+^2) {\rm d} r_+ \, ,
\nonumber\\
V &= \frac{\Omega_{D-2} r_+^{D-1}}{D-1} \, ,
\quad 
P = -\frac{\Lambda}{8 \pi G_{\rm N}} = \frac{(D-1)(D-2)}{16 \pi G_{\rm N} L^2} \, ,
\nonumber \\
\Psi &= \frac{1}{2\alpha} \left[(D-3)M - (D-2) TS + 2 PV \right] \, .
\end{align}
Here, $M$ is the ADM mass, $T$ is the Hawking temperature computed via the ordinary Euclidean trick, and $S$ is the Wald entropy. Note that the temperature differs from the expression in Ref.~\cite{Bueno:2024dgm} because that work made use of some simplifications that hold only in the vacuum asymptotically flat case. The entropy on the other hand is identical --- since the matter is minimally coupled, the Wald entropy receives no modifications. In the above, we have introduced the short-hand $\psi_+ \equiv k/r_+^2$. When $k = 0$, the only difference is that the entropy is given by the same area law that holds in general relativity,
\be 
S = \frac{A}{4 G_{\rm N}}\,, \quad \text{when $k=0$} \, .
\ee

When the EMT vanishes, it can be verified in general that the first law and Smarr formula hold with the thermodynamic parameters defined as above, 
\begin{align}
{\rm d}M &= T {\rm d}S + V {\rm d P} + \Psi {\rm d} \alpha \, ,
\\
(D-3) M &= (D-2) T S - 2 V P + 2 \alpha \Psi  \, .
\end{align}
In the above, $\Psi$ is a potential conjugate to the coupling constant $\alpha$.  Since this parameter introduces a new scale, the Smarr formula necessitates its inclusion. Similar considerations hold for any higher-order theory of gravity~\cite{Kastor:2011qp, Hennigar:2015esa}. (Note that, in fact, such a potential represents a resummation of an infinite sum of potentials corresponding to the couplings $\alpha_n$.)

When minimally coupled matter is included, there may also be additional conserved charges that need to be considered. In the case of the Maxwell and nonlinear theories of electrodynamics we have studied in this work, there is a conserved $U(1)$ charge that takes the form
\be 
\mathcal{Q} = \frac{\sqrt{2(D-2)(D-3)}}{8 \pi G_{\rm N}} \Omega_{D-2} q
\,,
\ee
where $q$ is the charge parameter we introduced in the Maxwell, Born--Infeld and RegMax examples. In terms of this charge, the electric field in all of the examples has the following asymptotic form at large $r$:
\be 
E = - \frac{4 \pi G_{\rm N}}{\Omega_{D-2}} \frac{\mathcal{Q}}{r^{D-2}} + \cdots  \, .
\ee
For the Maxwell theory, this electric field is exact, while for the Born--Infeld and RegMax models there are subleading corrections --- see Eqs.~\eqref{eq:BI_Efield} and~\eqref{eq:RM_Efield}.

While the formal relationship between the physical charge $\mathcal{Q}$ and the integration constant $q$ is the same for all of the models we have studied, its conjugate potential depends on the particular model. In general, it is the difference between the electric potential at infinity and at the horizon that enters into the thermodynamic description, and we can express this in terms of the electric field as
\be 
\Phi = \int_\infty^{r_+} E \, {\rm d} r \, .
\ee
Performing the integration, we have the following potentials: 

\begin{align}
    \Phi_{\rm EM} &= \sqrt{\frac{(D-2)}{2(D-3)}} \frac{q}{r_+^{D-3}} \, ,
    \\
    \Phi_{\rm BI} &= \sqrt{\frac{D-2}{2(D-3)}}\frac{q}{r_+^{D-3}} \,\,{}_2F_1\!\bigg(\frac{1}{2},\frac{D-3}{2
   (D-2)};\frac{3D-7}{2 (D-2)};-\frac{(D-3) (D-2) q^2}{2
   b^2 r_+^{2(D-2)}}\bigg)\,,
   \quad\\
   \Phi_{\rm RegMax}&=\sqrt{\frac{D-2}{2(D-3)}}\frac{q}{r_+^{D-3}} \left(1+\frac{1}{\beta r_+}\left(\sqrt{\frac{(D-2)(D-3)}{2}}q\right)^{\frac{1}{D-2}}\right)^{3-D}\,.
\end{align}
It is then straightforward to show that the first law holds,
\be 
{\rm d}M = T {\rm d}S + \Phi {\rm d} \mathcal{Q} +  V {\rm d P} + \Psi {\rm d} \alpha \, .
\ee
It is easy to construct a Smarr formula as well. In the Maxwell theory, it simply reads
\be 
(D-3) M = (D-2) T S  + (D-3) \Phi_{\rm EM} \mathcal{Q} - 2 V P + 2 \alpha \Psi  \, .
\ee
Theories of nonlinear electrodynamics introduce an additional length scale associated with their coupling constants. By the standard Eulerian scaling arguments, this must be included to obtain a valid Smarr formula,
\be 
(D-3) M = (D-2) T S  + (D-3) \Phi_{ i} \mathcal{Q} - 2 V P + 2 \alpha \Psi +[b_i] b_i {\cal B}_i\,.
\ee
Here, we have denoted the electromagnetic couplings as $b_i$, denoted their length dimensions by $[b_i]$ (for Born--Infeld $b_i=b$ has dimensions $-1$, while for RegMax $b_i=\beta$ has dimensions $-1/(D-2)$), and introduced the corresponding potential
\be 
\mathcal{B}_i \equiv \left(\frac{\partial M}{\partial b_i}\right)_{S, \mathcal{Q}, P, \alpha} \, .
\ee
The implications of such terms have been studied before in Einstein gravity coupled to nonlinear electrodynamics --- see, e.g., Ref.~\cite{Gunasekaran:2012dq}.

\subsection{Thermodynamics of regular black branes}
\label{sec:bbthermo}

Let us first focus on the thermodynamics of black branes, i.e., metrics with $k = 0$. In this case, the thermodynamics simplifies considerably and permits a {\em universal description}. This is because many of the expressions depend on functions $h(\psi_+)$ and/or $h'(\psi_+)$, which are respectively $0$ and $1$ for black branes.\footnote{The universality of black brane thermodynamics for Lovelock, quasi-topological gravities (with a finite number of corrections in the action), and several other models has been previously reported in Refs.~\cite{Cadoni:2016hhd, Hennigar:2017umz, Hennigar:2020fkv, Fernandes:2025eoc}.}\ (This is just a manifestation of the fact that all of the resummations we consider include the Einstein--Hilbert term.) Focusing first on the vacuum AdS case, the thermodynamics of black branes in \textit{any} resummation takes the form
\begin{align}\label{eq:thermo_eqns_bb_vac}
M &= \frac{(D-2) \Omega_{D-2}}{16 \pi G_{\rm N}} \frac{r_+^{D-1}}{L^2} \,,  \quad 
T = \frac{(D-1) r_+}{4 \pi L^2}  \, ,
\quad S = \frac{\Omega_{D-2} r_+^{D-2}}{4 G_{\rm N}} 
\,, \quad 
\nonumber\\
V &= \frac{\Omega_{D-2} r_+^{D-1}}{D-1} \, ,
\quad 
P =  \frac{(D-1)(D-2)}{16 \pi G_{\rm N} L^2} \, ,
\end{align}
while the potential $\Psi$ vanishes. It is easy to recognize that the  above thermodynamic quantities are \textit{identical} to those for black branes in Einstein gravity. Therefore, the thermodynamics and phase structure for regular black branes in \textit{any} resummation can be immediately deduced from  the corresponding result for Einstein gravity.
For example, in the vacuum case presented above, there are no Hawking--Page transitions, and the black brane generically has negative free energy,
\be 
F = M - T S = - \frac{\Omega_{D-2} r_+^{D-1}}{16 \pi G_{\rm N} L^2 } \, .
\ee
The equation of state --- found by re-arranging the equation defining the temperature for the pressure --- provides a different perspective on the lack of interesting phase behavior. We have
\be 
P = \frac{(D-2) T}{4 G_{\rm N} r_+}\,,
\ee
which is seen to be the equation for an ideal gas provided the  specific fluid volume is identified as
\be 
v = \frac{4 G_{\rm N} r_+}{D-2} \, .
\ee
Being an ideal gas, the corresponding ``molecules'' have neither finite volume nor are they subject to any intermolecular forces. Therefore, criticality is not possible.

In fact, even when minimally coupled matter is included, the equivalence with Einstein gravity thermodynamics remains true. For example, let us consider the addition of a Maxwell field. Then, the above thermodynamic expressions are modified to the following:
\begin{align}\label{eq:thermo_eqns_bb_max}
M &= \frac{(D-2) \Omega_{D-2}}{16 \pi G_{\rm N}} \left[ \frac{r_+^{D-1}}{L^2} + \frac{q^2}{r_+^{D-3}}  \right]\,,  \quad 
T = \frac{(D-1) r_+}{4 \pi L^2} - \frac{(D-3) q^2}{4 \pi r_+^{2D-5}}  \, ,
\quad S = \frac{\Omega_{D-2} r_+^{D-2}}{4 G_{\rm N}} 
\,, \quad 
\nonumber\\
V &= \frac{\Omega_{D-2} r_+^{D-1}}{D-1} \, ,
\quad 
P =  \frac{(D-1)(D-2)}{16 \pi G_{\rm N} L^2} \, ,
\end{align}
while the free energy of Maxwell charged branes in the canonical ensemble reads
\be 
F  = M - TS =  - \frac{\Omega_{D-2} r_+^{D-1}}{16 \pi G_{\rm N} L^2} \left[1 - \frac{(2D-5)L^2 q^2}{r_+^{2(D-2)}} \right] \, .
\ee
Again, there is no interesting phase structure, though we note the existence of an extremal solution of finite horizon radius when
\be 
q^2_{\rm ext} = \frac{D-1}{D-3} \frac{r_+^{2D-4}}{L^2} \, .
\ee
Similar considerations apply as well to the theories of nonlinear electrodynamics we have studied. We do not report further on this here, as it does not provide any new insights.

\subsection{Thermodynamics of spherical black holes}
\label{sec:bhthermo}

Let us now consider in more detail the thermodynamics of regular black holes with $ k = + 1$. To illustrate some of the main points, we will begin by considering the Hayward-AdS case (model I) in detail and then make more general remarks about some of the other models.

\begin{figure}
    \centering
    \includegraphics[width=0.75\linewidth]{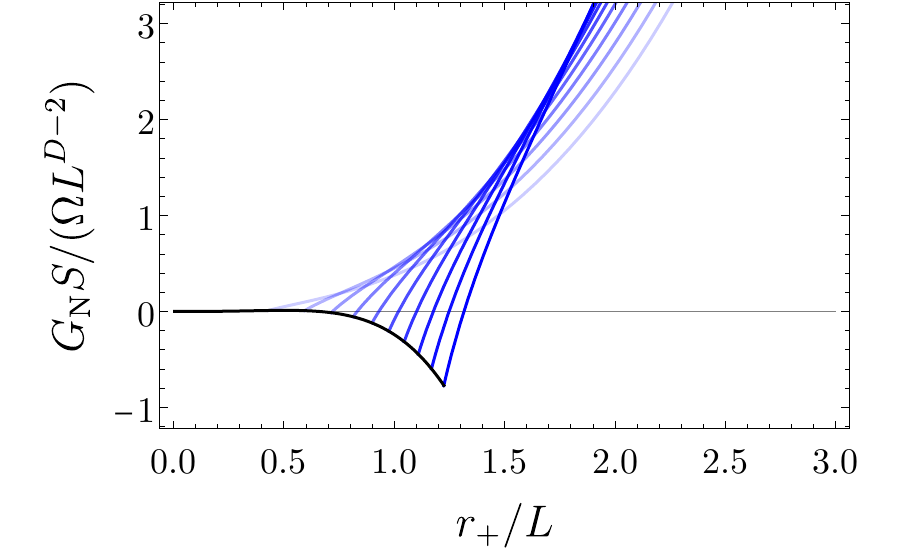}
    \caption{{\bf Wald entropy: Hayward-AdS model.} The Wald entropy is plotted as a function of the horizon radius for the Hayward model in $D = 5$. The blue curves show the entropy for different values of $a=\alpha/L^2$, ranging from $a= 0$ to the maximum allowed value of $a= 1$ from the lightest to the darkest hue. The black curve shows the entropy at extremality, with all blue curves originating along this line. Plots in higher dimensions are qualitatively similar.}
    \label{fig:Hayward_entropy}
\end{figure}

The mass, temperature, and entropy of the Hayward-AdS model read
\begin{align}\label{MTSHaywardAdS}
    M &= \frac{(D-2)\Omega_{D-2} r_+^{D-1}}{16 \pi L^2 G_{\rm N}} \left[1 + \frac{L^2}{r_+^2 - \alpha} \right] \, ,
    \\
    T &= \frac{1}{4 \pi r_+^3} \left[(D-3) r_+^2 - \alpha (D-1) + \frac{(D-1) (r_+^2-\alpha)^2}{L^2} \right] \, ,
    \\
    S &= \frac{\Omega_{D-2} r_+^{D-2}}{4 G_{\rm N}} \,\, {}_2F_1 \!\Big(2, 1 - \frac{D}{2}; 2 - \frac{D}{2}; \frac{\alpha}{r_+^2} \Big) \, .
\end{align}
The expression for the entropy is not particularly transparent, so let us consider it in more detail. Restricting to AdS asymptotics, it is convenient to work with the dimensionless variables $a = \alpha/L^2$ [Eq.~\eqref{aaa}] and $x = r_+/L$. Then, it is easy to see that when $a \ll 1$, the entropy reproduces the area law from general relativity,
\be 
S = \frac{\Omega x^{D-2}}{4 G} \left[1 + \frac{2 (D-2) a}{(D-4) x^2} + \cdots \right] \, .
\ee
On the other hand, when the horizon radius becomes comparable to the coupling constant, $x \approx \sqrt{a}$, the behavior can differ dramatically due to the fact that higher-order terms with a potentially negative denominator (for certain values of $D$) are then no longer suppressed. We show a representative plot in Figure~\ref{fig:Hayward_entropy}, which illustrates the Wald entropy of the Hayward black hole as a function of the horizon radius. In fact, for black holes sufficiently close to extremality, the Wald entropy can become negative. It has long been known that the Wald entropy can in some circumstances become negative for higher-curvature black holes~\cite{Cvetic:2001bk, Clunan:2004tb, Bueno:2017qce}, though a general consensus on what this means does not exist.
From a pragmatic perspective, one must bear in mind that the Wald entropy is sensitive to \textit{all} higher-curvature terms added to the action, irrespective of whether they produce dynamics or not. Therefore, by adding suitable total derivative terms to the action, it is always possible to shift the entropy by an overall constant. Thus, it is not clear whether any robust physical meaning can be ascribed to the absolute value of the Wald entropy, as opposed to the differences of Wald entropy between two states. In the present case, the solutions with negative Wald entropy occur close to extremality --- at very low Hawking temperature. In this regime, we do not expect that the semiclassical computation of the entropy will be reliable anyway. For example, it has recently been understood that quantum gravitational fluctuations of the metric become significant for near-extremal black holes and actually dominate over the Bekenstein--Hawking entropy at low temperatures~\cite{Iliesiu:2020qvm}. Similar corrections will be relevant for the black holes considered here in the regimes where the entropy takes on negative values. However, for the vacuum case the precise resolution of this issue is not important, since the solutions with negative entropy never dominate the canonical ensemble --- they exist only below the Hawking--Page transition temperature, which will be discussed below. 

Using the definition of the pressure from~\eqref{eq:thermo_eqns}, we can rearrange the equation for the temperature to obtain the equation of state,
\be 
 P = \frac{(D-2)r_+^3 T}{4 G_{\rm N} \left(r_+^2 - \alpha \right)^2}  - \frac{(D-2) \left[(D-3)r_+^2 - \alpha (D-1) \right]}{16 \pi G_{\rm N} \left(r_+^2 - \alpha \right)^2} \, .
\ee
The factors of $r_+^2 - \alpha$ appearing in the denominator indicate that $\alpha$ plays the role of an effective ``molecular volume'' for the black holes, analogous to a van der Waals fluid.\footnote{Recall that the {\em van der Waals equation of state} takes the form
\be \label{VdW}
P = \frac{T}{v-b} - \frac{a}{v^2}\,,
\ee 
where $b$ represents the finite size of the molecule and $a$ characterizes the strength of intermolecular forces. This equation admits a critical point, characterized by $P_c, v_c, T_c$, with a critical ratio $P_c v_c/T_c=3/8.$ 
}\ The equations of motion naturally enforce that $r_+^2 > \alpha$, since there is a minimum permissible horizon size that corresponds to the extremal limit,
\be 
r_+^2 \ge r_{\rm ext}^2 = \alpha - \frac{(D-3) L^2}{2(D-1)} +    \frac{L\sqrt{8 \alpha (D-1)+ L^2 (D-3)^2}}{2 (D-1)}  > \alpha \, .
\ee
 In analogy to the van der Waals gas \eqref{VdW}, we can introduce a {\em molecular volume parameter} $b$ and {\em specific volume} $v$ [cf.\ expression for the thermodynamic volume $V$ in \eqref{eq:thermo_eqns}] as 
\be 
b = \frac{4 G_{\rm N} \sqrt{\alpha}}{D-2}  \, , \quad v = \frac{4 G_{\rm N}}{D-2}r_+  \, ,
\ee
in terms of which the above equation of state reads:
\be 
P=\frac{Tv^3}{(v^2-b^2)^2}
-\frac{G_{\rm N}\Big((D-3)v^2-b^2(D-1)\Big)}{(D-2)\pi(v^2-b^2)^2}\,,
\ee 
and the large volume behavior is governed by an ideal gas law:
\be 
P = \frac{T}{v} + \mathcal{O}(1/v^2) \, .
\ee

We assess the system for critical points in the standard manner, by checking under what conditions the equation of state has a critical (inflection) point,\footnote{In general, one should compute the derivatives with respect to the thermodynamic volume. However, since here the thermodynamic volume is a function only of $r_+$, this simpler method produces the same results.}
\be \label{PPP}
\frac{\partial P}{\partial v} = \frac{\partial^2 P}{\partial v^2} = 0 \, .
\ee
The Hayward-AdS black hole has a single critical point with critical parameters 
\begin{align}
    P_c &= \frac{\left(2D-3 - \sqrt{3D(D-2)} \right) G_{\rm N} }{4 \pi (D-2) b^2} \, , \quad v_c = b \sqrt{\frac{3(D-1)}{D-3} + \frac{2 \sqrt{3D(D-2)}}{D-3}}  \, ,
    \nonumber
    \\
    T_c &= \frac{2 G_{\rm N} \left(2 D- 3-\sqrt{3 D(D-2)}\right) \sqrt{3 (D-1) +2
   \sqrt{3 D(D-2)}}}{3 \pi  b (D-2) \sqrt{D-3}} \, .
\end{align}
The ratio of these parameters matches identically the van der Waals ratio in {\em any dimension},
\be \label{rhoCHayward}
\frac{P_c v_c}{T_c} = \frac{3}{8}  \, .
\ee
As can be inferred from the reciprocal dependence of $P_c$ and $T_c$ on the coupling parameter, the critical point is introduced by the higher-curvature terms in the action. In pure Einstein-AdS gravity, there is no critical behavior for spherical black holes \cite{Hawking:1982dh}. Straightforward computations --- following, for example, Ref.~\cite{Kubiznak:2012wp} --- confirm that the critical point is governed by standard mean field theory critical exponents.

\begin{figure}
    \centering
    \includegraphics[width=\linewidth]{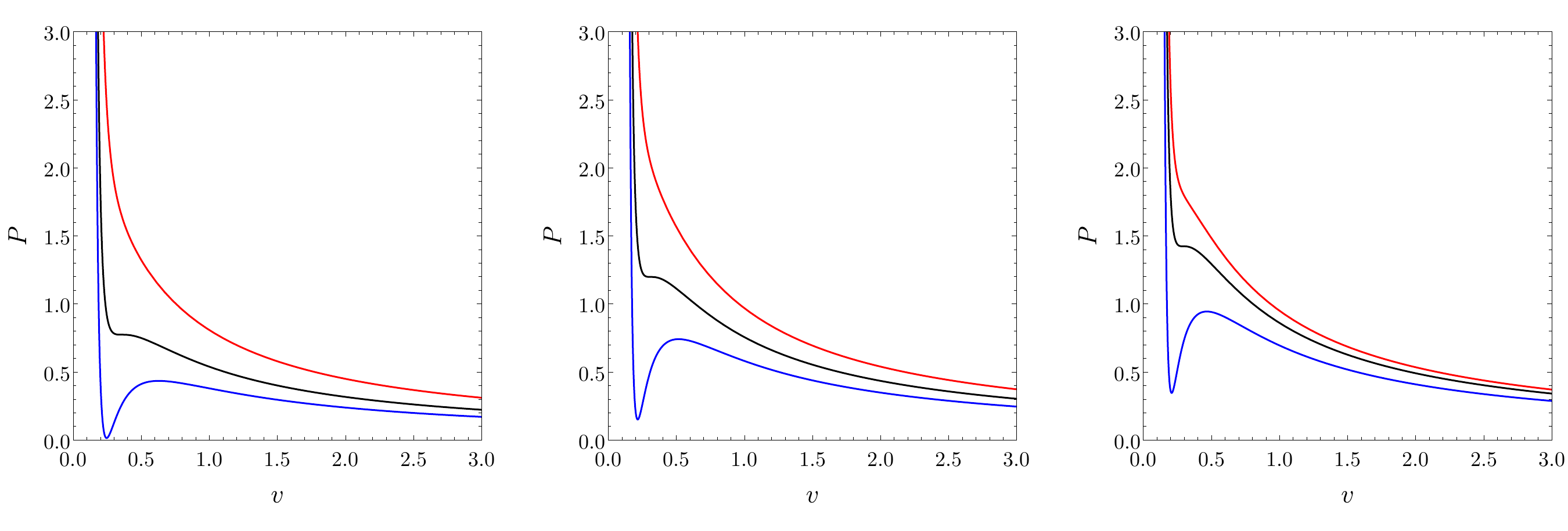}
    \caption{{\bf $\boldsymbol{P-v}$ diagram: Hayward-AdS black hole.} Isotherms for the Hayward-AdS black hole (model I) for $D = 5, 7, 9$ (from left to right). In each case, the lower blue curve has $T < T_c$, the middle black curve has $T = T_c$, and the upper red curve has $T > T_c$. In each plot, we have set the effective molecular size parameter $b = 0.1$. The units are such that $G_{\rm N} = 1$. }
    \label{fig:hayward-pv}
\end{figure}

In Figure~\ref{fig:hayward-pv}, we show representative isotherms for five, seven, and nine dimensions. The blue curves have $T < T_c$ and display standard `oscillatory' van der Waals behavior --- characteristic of {\em small black hole/large black hole (SBH/LBH)} first-order phase transition {\`a} la  \cite{Kubiznak:2012wp}. The black curves have $T = T_c$ and show the expected inflection point associated with a critical point, where the potential phase transition would be of the {\em second-order}. The red curves have $T > T_c$ and feature ideal gas-like behavior. The results are qualitatively identical in any odd dimension.

To assess whether or not the critical point and the associated SBH/LBH
phase transition are  physical, we need to examine the free energy

\be 
F=M-TS\,.
\ee 
Due to the appearance of the hypergeometric function in the entropy, the $D$-dimensional form of the free energy is not illuminating.
Focusing on the case of the critical point itself, we find that the free energy is positive (larger than that of thermal AdS) in $D = 5$ and negative (less than that of thermal AdS) in all higher odd dimensions. This indicates that the critical points and the associated phase behavior are physical in all odd dimensions $D \ge 7$. 

\begin{figure}[t]
    \centering \includegraphics[width=\linewidth]{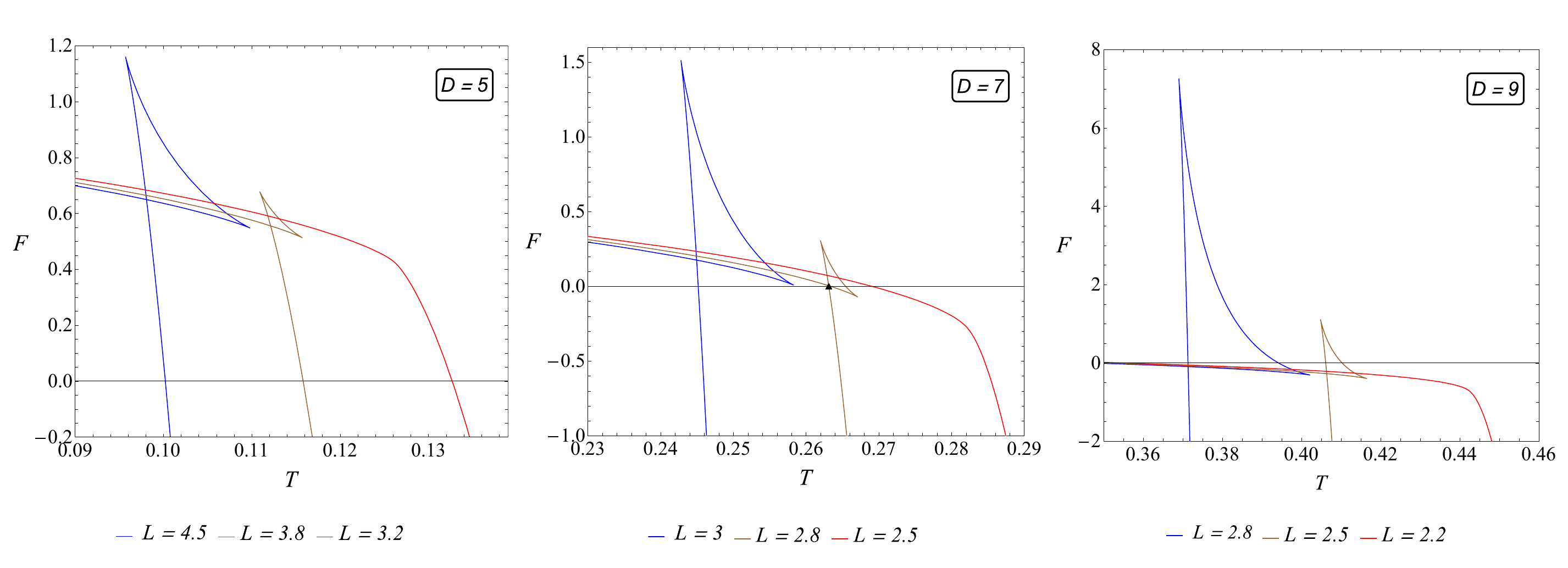}
    \caption{{\bf Free energy: Hayward-AdS model.} The free energy $F$ is plotted parametrically as a function of temperature $T$, using the horizon radius as a parameter, for the Hayward-AdS black hole (model I) in various dimensions. The coupling $\alpha$ remains fixed at the value $\alpha=0.2$, and we allow the cosmological constant parameter $L$ to vary. In the $D=7$ case, the triangle indicates the triple point where the small black hole, large black hole, and radiation phases coexist (see also Figure~\ref{fig:coexistence}). The temperature range was chosen to zoom in on the regions with interesting behavior. The units are such that $G_{\rm N} = 1$.}
    \label{fig:PT-Hayward}
\end{figure}

In Figure~\ref{fig:PT-Hayward}, we show plots of the free energy against temperature for the Hayward-AdS black hole in five, seven, and nine dimensions. The temperature range was chosen to make the interesting phase structure more easily visible. We see that in five dimensions, the critical point and the associated SBH/LBH phase behavior occurs for positive free energy, and hence is unphysical as thermal AdS dominates in this range of temperatures. 
Therefore, in this dimension, we only observe the standard Hawking--Page thermal AdS/LBH first-order phase transition, which occurs at a temperature where the free energy vanishes \cite{Hawking:1982dh}.\footnote{Here, we have set the free energy of thermal AdS to zero and measured the black hole free energy with respect to this value. However, since we are in odd dimensions and consider spherical black holes, there is a non-trivial energy shift corresponding to the Casimir energy of the dual CFT on a sphere. While it would be interesting to derive the corresponding Casimir energy for black holes in the resummed quasi-topological gravity, such a shift (which shifts both the black hole and thermal AdS free energies) does not play any role for the phase transitions discussed in this paper.}  

\begin{figure}[t]
    \centering \includegraphics[width=0.75\linewidth]{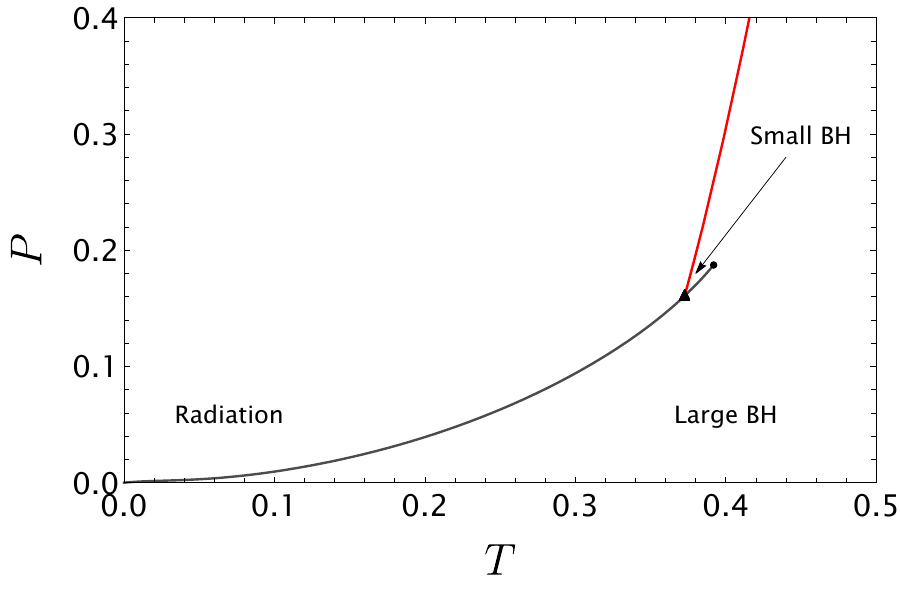}
    \caption{{\bf Phase diagram: Hayward-AdS model.} The pressure $P$ is plotted as a function of temperature $T$ for the $D=7$ case with a fixed coupling parameter $\alpha = 0.1$. The coexistence lines separate regions corresponding to the small black hole, large black hole, and radiation phases. As the temperature increases from low values, the black line follows the coexistence between the radiation and large black hole phases, ending at the triangle, which marks the triple point where all three phases coexist. Beyond this point, the black line represents the coexistence of small and large black hole phases, ending at the dot that denotes the critical point. The red line represents the coexistence between the radiation and large black hole phases at pressures above the triple point, ending at the maximum pressure $P_{\rm max} = (D-1)(D-2)/(16\pi G_{\rm N}\alpha)$, imposed by the condition $L^2 > \alpha$. For the chosen value of $\alpha$, this corresponds to $P_{\rm max} \approx 5.97$. This point is not visible in the figure, as the plot is zoomed in on the region where the most interesting features of the phase diagram appear. The units are such that $G_{\rm N} = 1$.}
    \label{fig:coexistence}
\end{figure}

However, in higher dimensions the situation is markedly different: The critical point always occurs for negative free energy, and hence is physical. In a window of pressures near to but below the critical pressure, the first-order SBH/LBH phase transition occurs also at negative free energy and is also physical. Thus, for these black holes the phase structure is such that at very low temperature thermal AdS dominates, then as the temperature is increased the Hawking--Page transition takes place between thermal AdS and a {\em small} regular black hole. As the temperature is further increased, a first-order SBH/LBH phase transition potentially occurs. We display the corresponding phase diagram for the $D=7$ Hayward-AdS black hole in Figure~\ref{fig:coexistence}. In particular, we observe a triple point where the three phases of thermal AdS, SBH, and LBH meet. Such a phase transition between solid, liquid, and gaseous states has been previously observed for higher-dimensional Lovelock black holes with exotic horizon geometries and 4D Gauss--Bonnet-de Sitter black holes~\cite{HullMann:2021,MarksSimovicMann:2021,HullSimovic:2023}. This is, to the best of our knowledge, the first time such a triple point has been observed for spherical black holes.

The Hawking--Page transition is physical in any odd dimension. Unfortunately, an expression for the temperature at which the Hawking--Page transitions occurs cannot be obtained in a simple closed form. However, it is straightforward to obtain a perturbative expansion in the limit of small coupling parameter $\alpha$,
\be 
T_{\rm HP} = \frac{(D-2)}{2 \pi L} \left[ 1 - \frac{(3D-4) \alpha}{2 (D-4) L^2} +\mathcal{O}(\alpha^2) \right] \, .
\ee
We see that for the Hayward-AdS model, the temperature at which the transition occurs is \textit{decreased} relative to the Einstein gravity value, at least in the limit of small $\alpha/L^2$. 

It is natural to wonder whether a type of generalized Hawking--Page transition is possible between the black holes and the regular solutions without horizons, i.e., the solitonic configurations. Being completely regular, one can construct thermal versions of these solutions by Wick rotating to Euclidean signature and identifying the imaginary time coordinate with any desired periodicity. However, these ``thermal solitons'' never dominate the canonical ensemble. This is simply because their free energy always exceed thermal AdS: the thermal solitons have nonzero, positive mass and therefore positive free energy relative to thermal AdS. 

\subsubsection{Generic equation of state: finite molecular volume}
\label{sec:gen_eos_k1}

Let us make some more general considerations. We will focus here on vacuum $k = +1$ black holes with cosmological constant, but allow for any resummation. The equation of state can be easily found, and it reads (setting $G_{\rm N} = 1$) 
\be 
P = \frac{(D-2) h_{\rm f}'(\psi_+) T}{4 r_+} - \frac{(D-2) \left[(D-1) r_+^2 h_{\rm f}(\psi_+) - 2 h_{\rm f}'(\psi_+)\right]}{16 \pi r_+^2} \, .
\ee

In any model that regularizes the singularity, $h_{\rm f}(x)$ necessarily diverges  at $x = \psi_0$ where $f(r) = 1 + \psi_0 r^2 +\mathcal{O}(r^3)$ --- see Section~\ref{sec:regularity conditions}. Hence, in any such model the pressure will diverge at finite volume. This means that the feature we observed for the Hayward-AdS model --- that the coupling parameter introduces an effective molecular volume, analogous to a van der Waals gas --- holds for \textit{any} model that regularizes the singularity.\footnote{It is natural to wonder whether one can choose a resummation $h_{\rm f}(x)$ such that the equation of state matches exactly the van der Waals fluid. This does not appear to be possible. For example, if one chooses the resummation such that the temperature-dependent part of the equation of state matches exactly that of the van der Waals fluid, then one finds that the temperature-independent part gives an infinite series of corrections (a type of virial expansion) and does not match the van der Waals case. More importantly, it appears that the corresponding resummation does not actually yield sensible black hole solutions when chosen in this manner.}\ 

In fact, it is a necessary condition for the equation of state: if the equation of state describes a regular black hole, then it must have a term analogous to finite molecular volume. In all of the models we have studied in Table~\ref{tab:metrics}, the parameter $\psi_0 = -1/\alpha$. This means (since $\psi_+ = 1/r_+^2$) that the pressure exhibits a divergence for all such models when $r_+^2 = \alpha$. Thus, the same specific volume and molecular volume parameters hold for all such models,
\be 
b = \frac{4 G_{\rm N} \sqrt{\alpha}}{D-2}  \, , \quad v = \frac{4 G_{\rm N}}{D-2}r_+  \, .
\ee

\begin{figure}
    \centering
    \includegraphics[width=\linewidth]{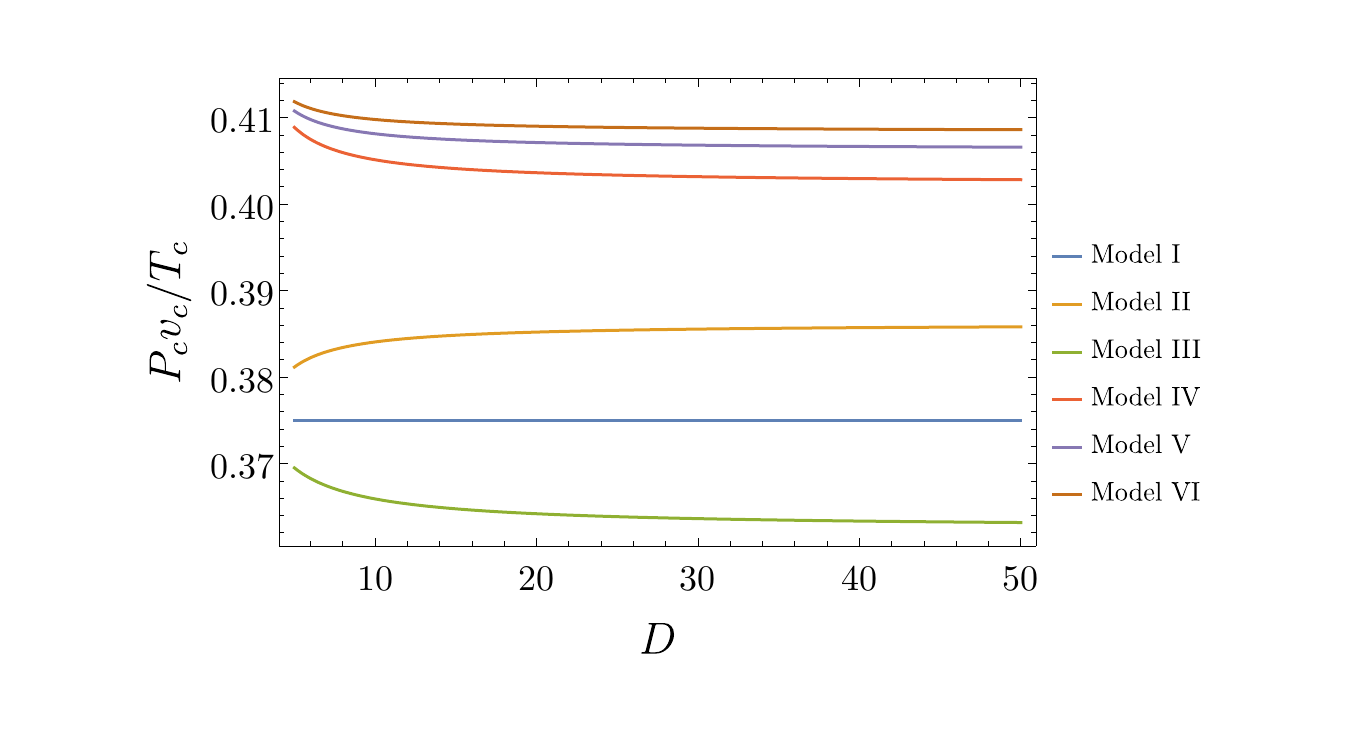}
    \caption{{{\bf Critical ratio $P_c v_c/T_c$: various AdS models.} We display a  dimensional dependence of the critical ratio for the first six models in Table~\ref{tab:metrics} in vacuum-AdS.} The Hayward-AdS case (model I) is the unique one for which the critical ratio matches that of the van der Waals fluid, and is also the unique one for which the critical ratio is independent of dimension. The plot illustrates that, while the other models exhibit dimensional dependence, it is very weak. Note that we have interpolated between integer dimensions.}
    \label{fig:critical_ratios}
\end{figure}

 \begingroup
\setlength{\tabcolsep}{5pt} 
\renewcommand{\arraystretch}{3}
\begin{table*}[t!]
	\footnotesize
	\centering
	\begin{tabular}{|c||c|c|c|c|c|c|}
	\hline
        Model & I & II & III & IV & V & VI
        \\
        \hline
        $\displaystyle h(\psi)$ & $\displaystyle \frac{\psi}{1-\alpha \psi}  $ & $\displaystyle - \frac{\log \left(1-\alpha \psi \right)}{\alpha} $ & $\displaystyle \frac{\psi}{\left(1-\alpha \psi \right)^2}$ & $\displaystyle \frac{\psi}{1-\alpha^2\psi^2}$ & $\displaystyle \frac{\psi}{\sqrt{1-\alpha^2\psi^2}}$ & $\displaystyle \frac{\arctanh(\alpha \psi)}{\alpha}$
        \\
        \hline
        $\displaystyle \lim_{D\to\infty} \frac{P_c v_c}{T_c}$ & $\displaystyle \frac{3}{8}\,=\,0.375$ & $\displaystyle 0.386193$ & $\displaystyle 0.362512$ & $\displaystyle 0.402311$ & $\displaystyle 0.406247$ & $\displaystyle 0.408348$
        \\ 
        \hline
	\end{tabular}
	\caption{{\bf Ultimate critical ratio: limit of infinite dimensions.} We tabulate the critical ratios for the vacuum-AdS regular black holes for the different models in the strict $D \to \infty$ limit. The result for model I is exact, and independent of dimension. All other models exhibit  weak dependence on dimension, and the $D \to \infty$ value provides a good approximation for general dimensions. Despite the very different functional forms for the different models, the critical ratios of all models fall within  $\approx 0.04$ of each other.}
	\label{tab:crit_ratios_large_d}
\end{table*}
\endgroup
    
Following the procedure outlined above for the Hayward-AdS case [Eqs.~\eqref{PPP}--\eqref{rhoCHayward}], we find that all of the models presented in Table~\ref{tab:metrics} have exactly one critical point. Moreover, the ratio of critical values is model-dependent. As we saw above, the ratio $P_c v_c/T_c = 3/8$ for the Hayward-AdS model in any dimension, matching precisely the van der Waals gas. The other models have different critical ratios that vary depending on the spacetime dimension. Thus, the matching of the Hayward-AdS model and the van der Waals fluid is a curious coincidence. We illustrate this fact in Figure~\ref{fig:critical_ratios}, where we plot the critical ratios for the first six models of Table~\ref{tab:metrics} in different dimensions. Note that the dimensional dependence of the critical ratios for models II--VI is very weak. Thus, while the general ratios are very complicated functions of $D$, they can be well-approximated by their $D \to \infty$ limit --- see Table~\ref{tab:crit_ratios_large_d}. 

In the case of Model VII, the analysis is more complicated because of the additional parameter ${\rm N}$. However, with some amount of tedious computation, it is possible to show that 
\be 
\frac{3}{8} \le \frac{P_c v_c}{T_c} \le \frac{1}{2}  \,  \quad \forall \, {\rm N} \ge 1 \quad  \forall \,  D \ge 5 \, .
\ee
The lower bound is obtained for ${\rm N} = 1$, since model VII reduces to model I for ${\rm N} =1$. The upper limit of $1/2$ is obtained in the combined limit ${\rm N} \to \infty$ and $D \to \infty$. Based on this and the  observations for the other models, it is tempting to speculate that $1/2$ may be a universal upper bound for the critical ratio of spherically symmetric vacuum regular black holes in any resummation. 

Remarkably, despite the very different forms of the resummations, the critical ratios are very close for all models, differing only by about $0.04$ in absolute terms between all models, or about $10\%$.  This observation seems nontrivial. One has essentially complete freedom to specify the resummation, and each resummation describes a unique equation of state. In principle, then, it would seem that the critical ratios compared between different models should have very little to do with one another. However, instead we find that for all resummations studied, the critical ratios falls within a narrow range (and, moreover, appear to be bounded from above). This is analogous to the {\em  principle of corresponding states} for ordinary fluids~\cite{correspondingStates}. As described there, the \textit{empirically measured} critical ratios for various different gases all fall within a narrow window.\footnote{This should not be confused with van der Waal's \textit{law} of corresponding states: that, when expressed as fractions of the critical values, the van der Waals equation takes the same form for any fluid. The van der Waals equation is one equation of state, and here we are making a comparison across several different equations of state which all make very similar predictions.}\ We therefore interpret our observation as the regular black hole analog of the principle of corresponding states.

\subsubsection{Phase structure with electric charge}

Allowing for minimally coupled matter with $T_t^t = T_r^r$ as we have been considering throughout, the equation of state reads 
\begin{align}\label{eq:eos_min_mat}
P &= \frac{(D-2) h_{\rm f}'(\psi_+) T}{4 r_+} - \frac{(D-2) \left[(D-1) r_+^2 h_{\rm f}(\psi_+) - 2 h_{\rm f}'(\psi_+)\right]}{16 \pi r_+^2}-T_t^t(r_+) \, ,
\end{align}
where by $T_t^t(r_+)$ we indicate that the EMT component is to be evaluated at the horizon. By noting that $\rho_{\rm E} = -T_t^t$, we see that it is the energy density of the matter that contributes directly to the thermodynamic equation of state. 

Focusing first on the Maxwell theory, we can write the equation of state more explicitly, namely:
\be 
P = \frac{(D-2) h_{\rm f}'(\psi_+) T}{4 r_+} - \frac{(D-2) \left[(D-1) r_+^2 h_{\rm f}(\psi_+) - 2 h_{\rm f}'(\psi_+)\right]}{16 \pi r_+^2} + \frac{(D-3)(D-2) q^2}{16 \pi r_+^{2(D-2)}} \, .
\ee
When the gravitational sector is general relativity (corresponding to $h_{\rm f}(x) = x$), the equation of state reduces to the one studied in Refs.~\cite{Kubiznak:2012wp, Gunasekaran:2012dq}. Recall that in the Einstein--Maxwell theory, there exists a single critical point in a general dimension with critical values 
\be 
v_c = \frac{4\left[q^2 (D-2)(2D-5) \right]^\frac{1}{2(D-3)} }{D-2}\, , \quad T_c = \frac{4 (D-3)^2}{\pi (D-2)(2D-5) v_c} \, ,\quad P_c = \frac{(D-3)^2}{\pi (D-2)^2 v_c^2} \, ,
\ee
and critical ratio
\be\label{eq:ein_max_crit} 
\frac{P_c v_c}{T_c} = \frac{2D-5}{4D-8} \, .
\ee

\begin{figure}
    \centering
    \includegraphics[width=\linewidth]{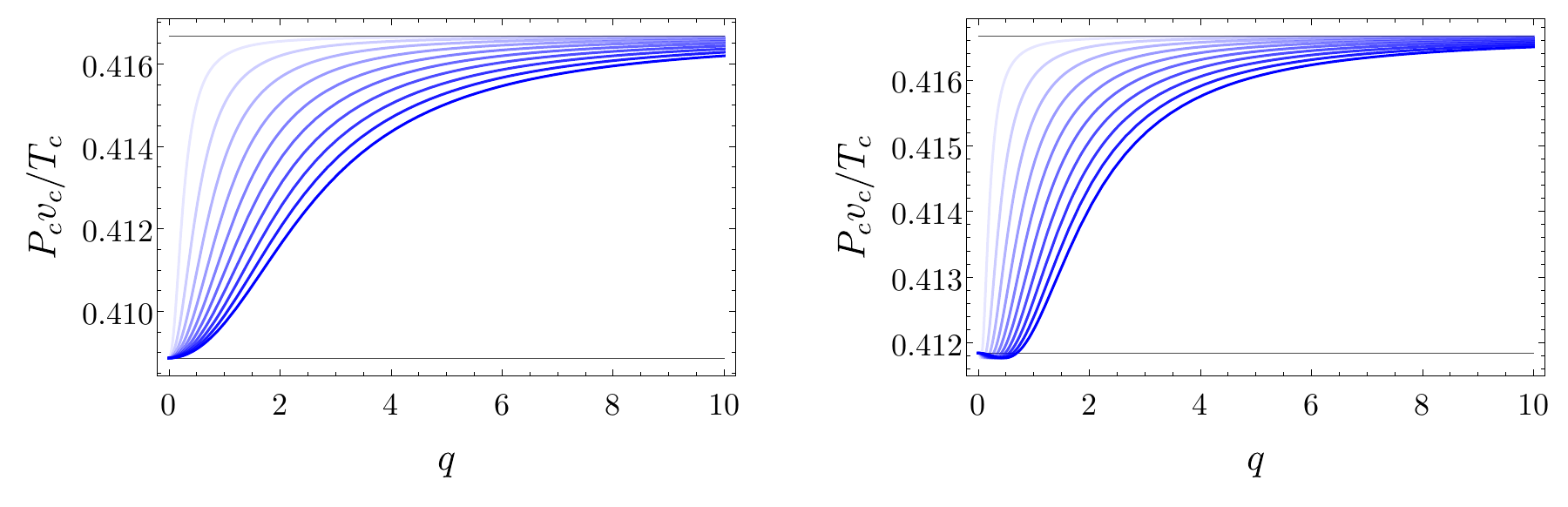}
    \caption{{\bf Critical ratio: charge dependence.} Ratio of critical parameters for Maxwell charged black holes in model IV (left) and model VI (right) in the five-dimensional case. The blue curves correspond to different values of $\alpha$, with the lighter (upper-leftmost) curves corresponding to smaller values, and the darker curves corresponding to larger values. The lower horizontal line is the $q = 0$ critical ratio, while the upper horizontal line is the critical ratio in the Einstein--Maxwell theory. The units are such that $G_{\rm N} = 1$.}
    \label{fig:crit_ratio_maxwell}
\end{figure}

When, instead of general relativity, we consider one of the resummations from Table~\ref{tab:metrics}, it is no longer possible to study the criticality analytically. Upon scanning the parameter space for the different models numerically, we find that the electric charge introduces no additional critical points. All models exhibit a single critical point as before, and the $P-v$ diagrams are qualitatively identical to those presented in Figure~\ref{fig:hayward-pv}. Because there are now two dimensionful parameters in the thermodynamic description, the critical ratio is no longer (dimension-dependent) constant, but a function of the dimensionless ratio $q^{2/(D-3)}/\alpha$. We illustrate this in Figure~\ref{fig:crit_ratio_maxwell}, where we show the critical ratio as a function of $q$ for different values of $\alpha$ for models IV and VI. In both cases, as the charge increases, the critical ratio interpolates between its neutral value and the Einstein--Maxwell result~\eqref{eq:ein_max_crit}. As the rightmost plot for model VI illustrates, the critical ratio need not change monotonically. However, we find that in all cases the Einstein--Maxwell ratio~\eqref{eq:ein_max_crit} is approached from below, consistent with the idea that $P_cv_c/T_c \le 1/2$.

\begin{figure}[t]
    \centering \includegraphics[width=\linewidth]{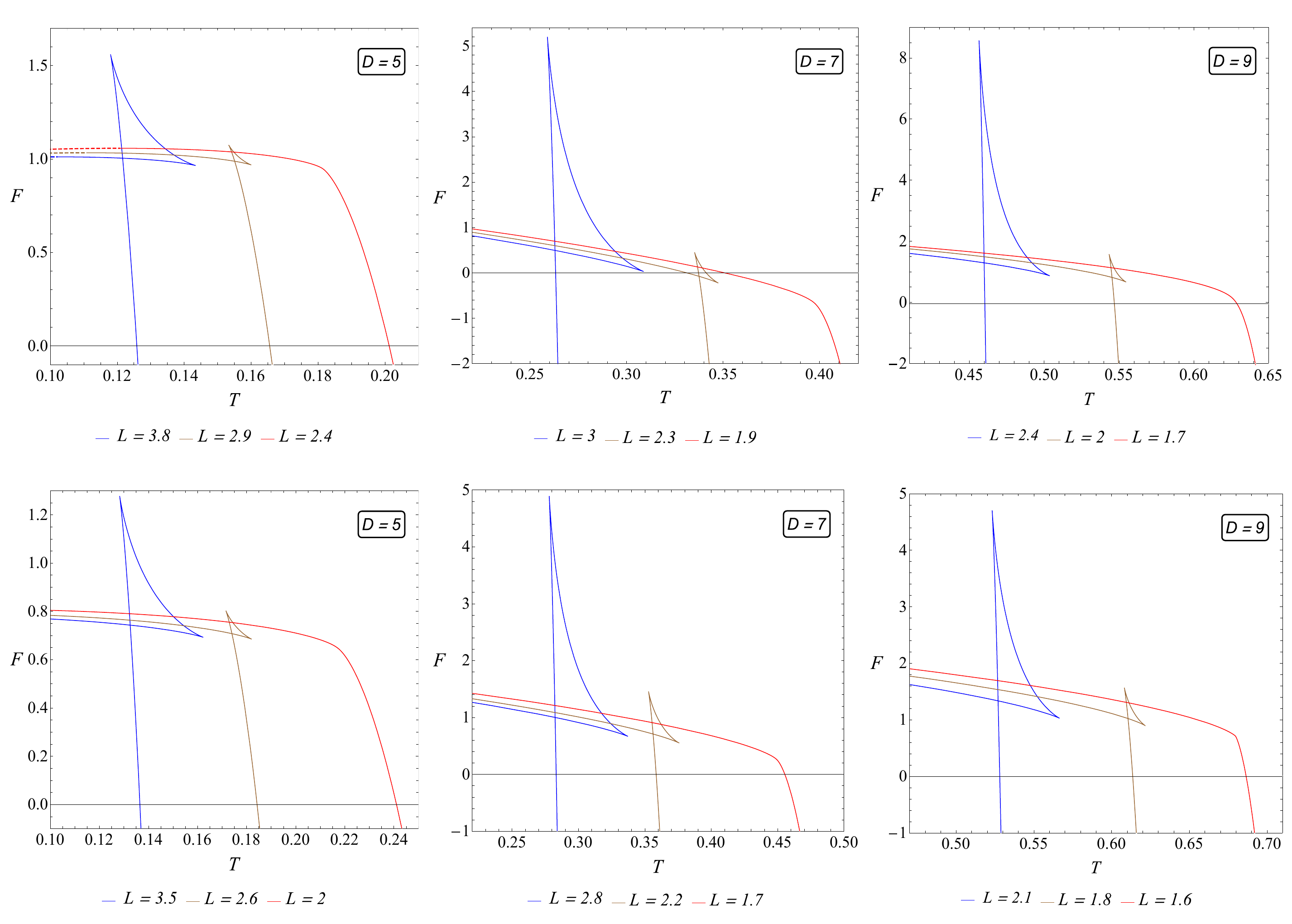}
    \caption{{\bf Free energy: charged Maxwell case.} The free energy $F$ is plotted parametrically as a function of temperature $T$, using the horizon radius as a parameter, for model IV (top) and model VI (bottom), both minimally coupled to the Maxwell field, across various dimensions. The coupling $\alpha$ and charge parameter $q$ remain fixed at the values $\alpha=0.2$ and $q=0.3$, respectively, and we allow the cosmological constant parameter $L$ to vary. The dashed lines indicate branches of black holes with negative Wald entropy. However, in the plots above these are only visible for $D=5$ in the top left as the temperature range has been chosen to zoom in on the regions with interesting behavior. The units are such that $G_{\rm N} = 1$.} 
    \label{fig:PT-Maxwell}
\end{figure}

In Figure~\ref{fig:PT-Maxwell}, we include representative plots of the free energy for models IV and VI in five, seven, and nine dimensions (for comparison with the Hayward case considered previously). The free energy is qualitatively the same as what we saw earlier, exhibiting swallowtails with associated first-order phase transitions below the critical pressure and monotonic behavior above. The key difference compared to earlier is that in this case there is no Hawking--Page transition,\footnote{This is because one no longer has a charged thermal AdS in the canonical (fixed charge) ensemble.}\ and therefore all critical and phase behavior is physical. Qualitatively similar behavior is seen in all models. Note that now, because of the conserved charge, the negative entropy black holes play a role in the thermodynamics in principle. Here, we take the pragmatic view that these black holes are still physical since, as described earlier, it is always possible to shift the Wald entropy by a constant. It is of course possible that this is not the case (e.g., due to the semiclassical approximation breaking down near the extremal solutions), which remains to be assessed.

In the case of Born--Infeld electrodynamics, the equation of state becomes
\begin{align} 
P &= \frac{(D-2) h_{\rm f}'(\psi_+) T}{4 r_+} - \frac{(D-2) \left[(D-1) r_+^2 h_{\rm f}(\psi_+) - 2 h_{\rm f}'(\psi_+)\right]}{16 \pi r_+^2} \nonumber
\\
&\quad+\frac{b \sqrt{4 b^2 +2 (D-3)(D-2) q^2 r_+^{4-2D}}}{8 \pi
   }-\frac{b^2}{4 \pi } \, .
\end{align}
In four dimensions, this equation of state was studied in Ref.~\cite{Gunasekaran:2012dq} and later in Ref.~\cite{ZouZhangWang:2014} for higher dimensions. The main findings of that work indicate that Born--Infeld electrodynamics can introduce novel zeroth-order phase transitions absent in the Maxwell case, but in general does not introduce any further (physical) critical points. 

To see how these features translate to more general gravitational theories, let us make a few remarks about this equation of state. First, in the limit $b \to \infty$, the Born--Infeld term contributions reduce to the Maxwell ones. For large $b$, we therefore expect a single critical point with $P_cv_c/T_c$ behaving like those presented in Figure~\ref{fig:crit_ratio_maxwell}. On the other hand, in the limit $b \to 0$, the charge contribution vanishes and the equation of state reduces to the vacuum one. This is the situation we considered earlier. For very small $b$, we therefore expect a single critical point with critical ratio given approximately by those from Table~\ref{tab:crit_ratios_large_d}. General values of $b$ interpolate between these two possibilities.

If we consider the RegMax theory instead of Born--Infeld electrodynamics, our discussion above remains qualitatively unchanged. The equation of state takes a functionally different form~\eqref{eq:eos_min_mat} with the EMT and electric field given by~\eqref{eq:Ttt_regmax} and 
\eqref{eq:RM_Efield}, 
respectively. (The general expressions are rather complicated so we refrain from writing them explicitly.) Despite the different functional forms, we find again exactly one critical point irrespective of the value of the RegMax coupling parameter $\beta$. Just as in Born--Infeld, variations of this parameter interpolate between the Maxwell cases and the vacuum case. 

\begin{figure}[H]
    \centering
    \includegraphics[width=\linewidth]{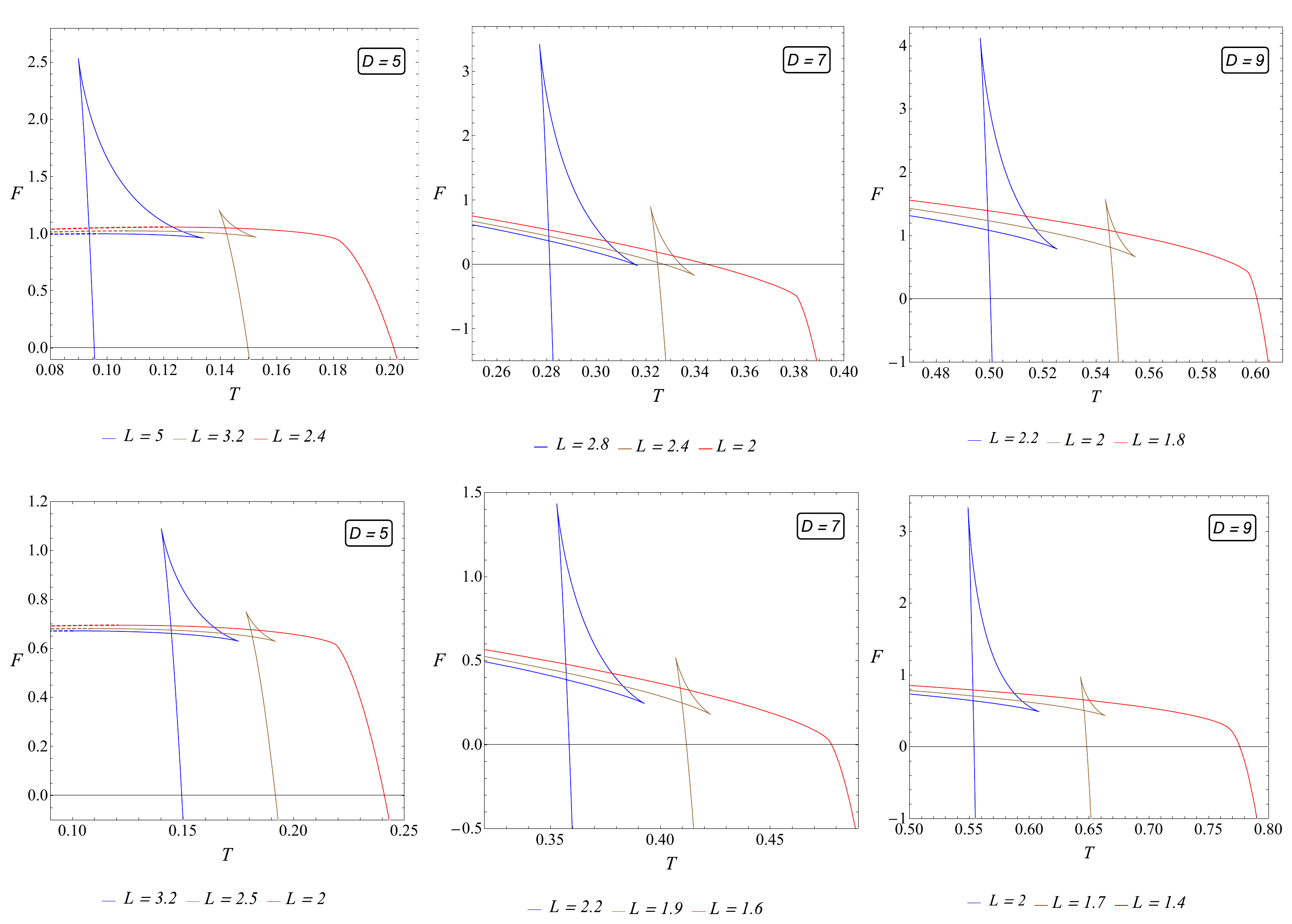}
    \caption{{\bf Free energy: Born--Infeld and RegMax cases.} The free energy $F$ is plotted parametrically as a function of temperature $T$, using the horizon radius as a parameter across various dimensions. The coupling $\alpha$ and charge parameter $q$ remain fixed at the values $\alpha = 0.2$ and $q = 0.3$, respectively, and we allow the cosmological constant parameter $L$ to vary. \textbf{Top:} Model IV, minimally coupled to the Born--Infeld field, with the electric field regularization parameter fixed at $b = 10$. \textbf{Bottom:} Model VI, minimally coupled to the RegMax field, with the electric field regularization parameter fixed at $\beta = 4$. The dashed lines indicate branches of black holes with negative Wald entropy. However, in the plots above these are only visible for $D=5$ (left column) as the temperature range has been chosen to zoom in on regions displaying interesting behavior. The units are such that $G_{\rm N} = 1$.}
    \label{fig:PT-NLE}
\end{figure}

In Figure~\ref{fig:PT-NLE}, we display representative plots of the free energy for regular black holes with Born--Infeld and RegMax nonlinear electrodynamics. The introduction of these terms results in no qualitative differences as far as the phase structure is concerned, consisting of swallowtails and first-order phase transitions. The same qualitative behavior occurs in all dimensions and in all of the resummations we have studied.

\subsection{Thermodynamics of topological black holes}
\label{sec:topothermo}

Finally, we consider the thermodynamics of topological ($k = -1$) black holes. For general models, the vacuum equation of state reads (setting $G_{\rm N} = 1$) 
\be 
P = \frac{(D-2) h_{\rm f}'(\psi_+) T}{4 r_+} - \frac{(D-2) \left[(D-1) r_+^2 h_{\rm f}(\psi_+) - 2k h_{\rm f}'(\psi_+)\right]}{16 \pi r_+^2} \, ,
\ee
where $\psi_+ = k/r_+^2$. Having $k = -1$ results in crucial differences relative to the spherical $k = +1$ case, depending on the details of the resummation. Namely, we observe two completely different kinds of behavior --- depending on {\em whether or not} the function $h_{\rm f}(x)$ remains invariant under the substitution $\alpha \to - \alpha$. Let us discuss these cases separately, starting with the $\alpha\to -\alpha$ symmetric case.

\subsubsection[Models with $\alpha\to -\alpha$ invariant resummation]{Models with $\boldsymbol{\alpha\to -\alpha}$ invariant resummation}

If the function $h_{\rm f}(x)$ is such that it remains invariant under the substitution $\alpha \to - \alpha$, then the previously observed feature of the higher-curvature terms introducing finite molecular volume persists. This is the case, for instance, in models IV, V, VI, and VII with ${\rm N}$ even. However, the general features of these black holes are quite exotic --- even more so than their Einstein gravity counterparts. We illustrate  the thermodynamics by way of example, focusing on model V of Table~\ref{tab:metrics} in five dimensions for concreteness. The essential features we shall describe here apply to any model with $\alpha \to - \alpha$ symmetry from Table~\ref{tab:metrics}, and in all dimensions.

The mass, entropy, and temperature of model V in five dimensions read
\begin{align}
    M &= \frac{3 \Omega_{3} L^2 x^4}{16 \pi G_{\rm N}} \left[1- \frac{1}{\sqrt{x^4-\mathsf{a}^4}} \right] \, ,
    \\
    S &= \frac{3 \Omega_{3} L^3}{4 G_{\rm N}} \int \frac{x^8}{\left(x^4 - \mathsf{a}^4 \right)^{3/2}} {\rm d} x \, ,
    \\
    T &= \frac{-2 \mathsf{a}^4 \left(-1 + \sqrt{x^4-\mathsf{a}^4} \right) + x^4 \left(-1 + 2 \sqrt{x^4-\mathsf{a}^4} \right)}{2 \pi L x^5} \, ,
\end{align}
where we introduced $\mathsf{a} = \sqrt{\alpha}/L$, $x = r_+/L$, and it is to be understood that we have specialized to $k = -1$. The expression for the entropy can be obtained in a closed form in terms of hypergeometric functions, but since the integral form is much simpler, we have opted to present it instead. 

We observe that the mass and the entropy are divergent as $x \to \mathsf{a}$. Explicitly, we have
\begin{align}
    M &= - \frac{3 \Omega_3 L^2 \mathsf{a}^{5/2}}{32 \pi G_{\rm N} \sqrt{x-\mathsf{a}}} + \frac{3 \Omega_3 L^2 \mathsf{a}^4}{16 \pi G_{\rm N}} + \mathcal{O}\left(\sqrt{x-\mathsf{a} }\right) \, ,
    \\
    S &= - \frac{3 \Omega_3 L^3 \mathsf{a}^{7/2}}{16 G_{\rm N} \sqrt{x-\mathsf{a}}} - \frac{5\Omega_3 L^3 \mathsf{a}^3 \sqrt{\pi} \Gamma[5/4]}{8 G_{\rm N} \Gamma[3/4]} + \mathcal{O}\left(\sqrt{x-\mathsf{a} }\right) \, .
\end{align}
The fact that negative-mass hyperbolic black holes exist is not so surprising --- the same thing happens in Einstein gravity. However, in that case, the negative mass is bounded from below by a function of the cosmological constant --- see Eq.~\eqref{eqn:topo-mass-bound}. The key difference here is that the mass can be \textit{arbitrarily} negative, with $M \to -\infty$ for a black hole with finite horizon radius. On the other hand, the temperature is perfectly finite in this limit,
\be 
T = \frac{1}{2 \pi L \mathsf{a}} - \frac{9 (x - \mathsf{a})}{2 \pi L \mathsf{a}^2} + \mathcal{O}\left(x-\mathsf{a}\right)^2 \, .
\ee
It is interesting to note that the divergences of the mass and entropy are such that they combine to \textit{cancel} in the free energy,
\be \label{badfreeenergy}
F = M - TS = \frac{\mathsf{a}^2 \Omega_3  L^2 }{16 \pi G_{\rm N}} \left[3 \mathsf{a}^2 + \frac{5 \sqrt{\pi} \Gamma[5/4]}{\Gamma[3/4]} \right]  + \mathcal{O}\left(\sqrt{x-\mathsf{a} }\right)\, .
\ee
This fact will be manifest in the free energy plots we consider below.

To gain some further insight, let us consider the metric in the limit $x \to \mathsf{a}$. A simple computation reveals that
\be 
\lim_{r_+ \to \sqrt{\alpha}} f(r) = -1 + \frac{r^2}{\alpha} \, .
\ee
This is simply the AdS-Rindler metric, albeit with a cosmological length scale set by the coupling constant. 
This point deserves further elaboration. In Einstein gravity, the AdS-Rindler metric is obtained only when the mass parameter $m = 0$, and this solution is unique. However, here the situation is more subtle and, at fixed coupling, there are actually \textit{two} distinct AdS-Rindler solutions. These are obtained when
\begin{align} 
x_1 = \left(1 + \mathsf{a}^4 \right)^{1/4} \quad  \Rightarrow \quad f(r) &= -1 + \frac{r^2}{L^2 \sqrt{1 + \alpha^2/L^4}} \, ,
\\
\text{or} \quad x_2 = \mathsf{a} \,  \quad \Rightarrow \quad  f(r) &= -1 + \frac{r^2}{\alpha} \, .
\end{align}
The first case is the natural deformation of the Einstein gravity result and limits to it as $\alpha \to 0$. In this case, the mass of the solution is identically zero and the entropy, at least for small values of the coupling, is positive. The second possibility is the problematic case identified above, and is achieved only in an asymptotic sense, as certain physical quantities approach infinity. This branch is nonperturbative in the coupling, and has no analog in Einstein gravity.

\begin{figure}[htp]
    \centering
    \includegraphics[width=\linewidth]{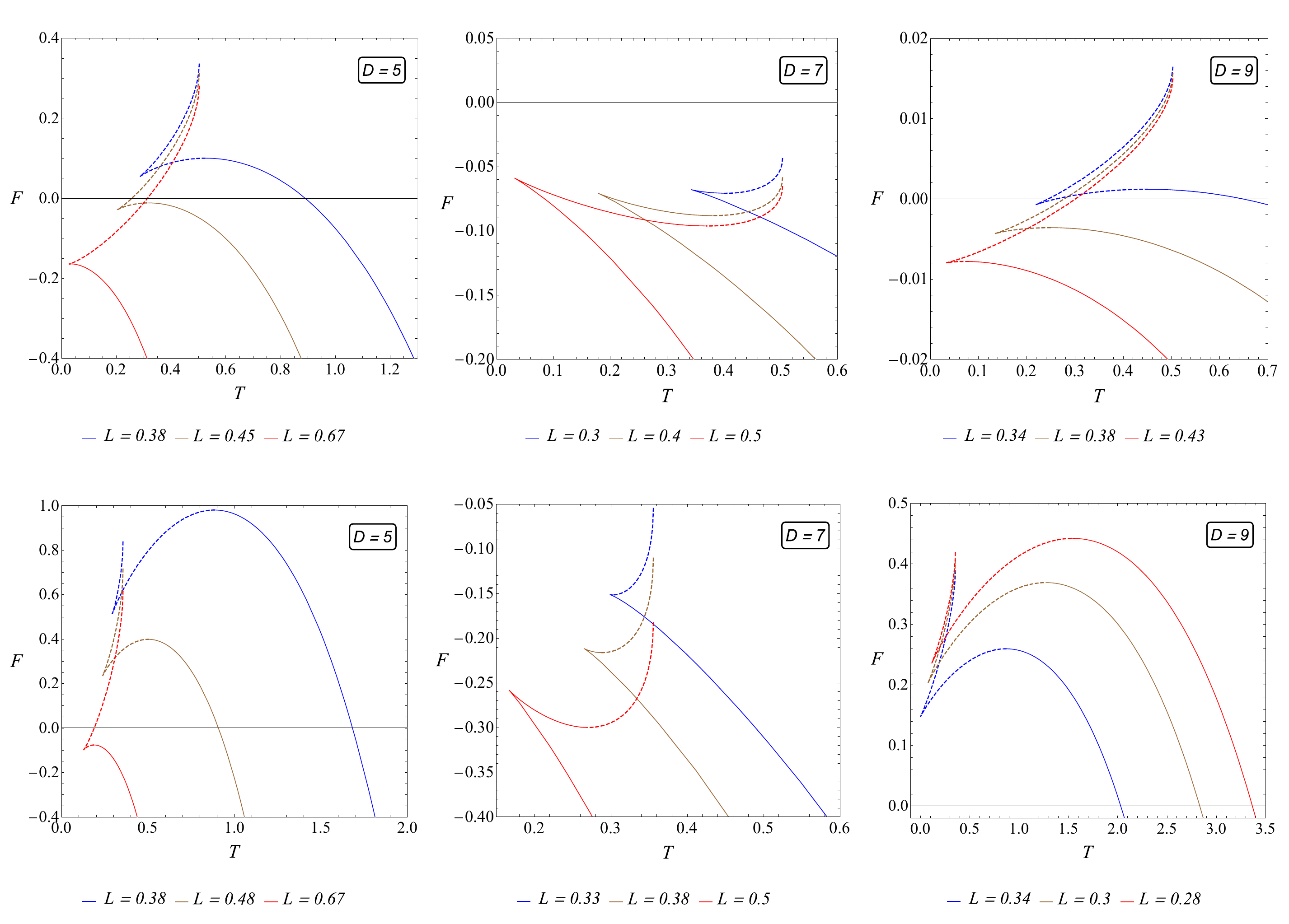}
    \caption{{\bf Free energy: topological black holes with $\boldsymbol{\alpha\to -\alpha}$ symmetry.} The free energy $F$ is plotted parametrically as a function of temperature $T$, using the horizon radius as a parameter, across various dimensions for the topological black holes with $k=-1$ of model V. \textbf{Top:} No matter fields are present, and and the coupling is fixed at $\alpha=0.1$. \textbf{Bottom:} Maxwell-charged case with coupling $\alpha=0.2$ and charge parameter $q=0.02$. The temperature range has been chosen to zoom in on regions displaying interesting behavior. In all cases, a dashed line indicates a branch of black holes with negative Wald entropy. Note that in all cases the upper branch (which corresponds to negative mass and entropy) is thermally unstable, as indicated by the upward-curving free energy curves. The units are such that $G_{\rm N} = 1$. }
    \label{fig:PT-hyp}
\end{figure}

The above raises a puzzle: Typically, in black hole thermodynamics, one would like to consider the maximally symmetric solution as the ground state. Here, however, we have \textit{two} such solutions, one with rather pathological properties. Which, then, is the proper ground state? If we resort to thermodynamic arguments, then, happily, it seems the natural ground state is the one with \textit{vanishing} mass and finite entropy. The free energy of this solution is always less than that of the pathological one, indicating that it is thermodynamically favored. This can be seen explicitly, expanding the free energy for small $\mathsf{a}$,
\be 
F_1 = - \frac{\Omega_3 L^2}{8 \pi G_{\rm N}} \left[1 - 4 \mathsf{a}^2 + \cdots \right] \, .
\ee
This can be compared with Eq.~\eqref{badfreeenergy} above which, recall, is always positive. We show this in the free energy plots of Figure~\ref{fig:PT-hyp}, where we see that the pathological AdS-Rindler is a \textit{global maximum} of the free energy in all cases.

On the other hand, one may interpret the negative mass and entropy as indicative of an instability. Similar arguments were made in Ref.~\cite{Emparan:1999pm} in relation to Taub-NUT spacetimes, for which the thermodynamics is similarly problematic. An instability is plausible since, as can been seen in Figure~\ref{fig:PT-hyp}, the heat capacity becomes \textit{negative} as $x \to \mathsf{a}$, and actually diverges in that limit. We find that all of these thermally unstable solutions do not actually dominate the ensemble, and hence this potential pathology is avoided here. However, this may still be problematic for the $\alpha \to -\alpha$ symmetric resummations, as it could suggest the lack of a stable ground state in, for example, the microcanonical ensemble.

As we mentioned at the beginning, while the precise calculations above have been carried out for model V in five dimensions, our conclusions are qualitatively general. For each model from Table~\ref{tab:metrics} with $\alpha \to -\alpha$ symmetry, the mass and entropy diverge to negative infinity as $x \to \mathsf{a}$. At the same time, the temperature approaches a finite number in this limit, as does the free energy. In all cases, the limiting spacetime geometry is AdS-Rindler. None of these models exhibit critical phenomena or phase behaviour of any kind, and the negative entropy, thermally unstable black holes never dominate the canonical ensemble.

As in the $k=+1$ case, the equation of state for topological black holes with $k=-1$ can be extended to include the appropriate matter contribution. When considering minimal coupling to the Maxwell field, the resulting phase structure does not differ significantly from the uncharged case. The main features persist, most notably the absence of a critical point in the resummations characterized by the symmetry $\alpha \rightarrow -\alpha$. The phase structure of these models, and in particular model V, is shown in the bottom row of Figure~\ref{fig:PT-hyp}. As before, regions of negative entropy are indicated with dashed lines. 

\subsubsection[Models breaking $\alpha\to-\alpha$ symmetry]{Models breaking $\boldsymbol{\alpha\to-\alpha}$ symmetry}

The thermodynamics is completely different for models whose $h_{\rm f}(x)$ is \textit{not} invariant under $\alpha \to - \alpha$ substitution. 
This is the case for models I, II, III, and VII with ${\rm N}$ odd. To parallel the discussion of the previous subsection, let us begin by examining in detail the thermodynamics of the models that do not have $\alpha \to -\alpha$ symmetry. As a representative example, we focus on the Hayward metric (model I).

The mass, entropy, and temperature of the Hayward solution with $k = -1$ read
\begin{align}
 M &= \frac{3 \Omega_3 L^2 x^4}{16 \pi G_{\rm N}} \left[1- \frac{1}{x^2 + \mathsf{a}^2} \right] \, ,
 \\
 S &= \frac{\Omega_3 L^3}{8 G_{\rm N}} \left[2 x^3 - 12 \mathsf{a}^2 x - \frac{15 \pi \mathsf{a}^3}{2} - \frac{3 x \mathsf{a}^4}{x^2+\mathsf{a}^2} + 15 \mathsf{a}^3 \arctan \frac{x}{\mathsf{a}} \right] \, ,
 \\
 T &= \frac{2 \mathsf{a}^4+\mathsf{a}^2 \left(4 x^2-2\right)+x^2 \left(2 x^2-1\right)}{2 \pi  L x^3} \, .
\end{align}
The notation we have introduced is the same as before, $x = r_+/L$ and $\mathsf{a} = \sqrt{\alpha}/L$, and we have presented the expression for the entropy explicitly since the integral has a simple evaluation in this case. Crucially, we see directly that all thermodynamic quantities are finite and without divergences at finite $x$.

For large black holes, the mass and entropy reduce appropriately to their values in general relativity. The behavior is different for small black holes. As one decreases the horizon radius at fixed coupling, eventually the extremal solution is reached, which occurs when
\be 
x_{\rm ext} = \frac{1}{2} \sqrt{1-4\mathsf{a}^2 + \sqrt{1+8\mathsf{a}^2}} \, .
\ee
The mass of the extremal solution is in all cases negative, but uniformly bounded below by the Einstein gravity result,
\be 
M_{\rm ext} \ge - \frac{3 \Omega_3 L^2}{64 \pi G_{\rm N}} \, .
\ee
On the other hand, for $\mathsf{a} \lessapprox 0.261862$ the entropy at extremality (and the full solution space) is always positive. For $\mathsf{a}$ exceeding this value, the extremal entropy is negative, but again uniformly bounded, 
\be 
S_{\rm ext} \ge - \frac{15 \pi \Omega_3 L^3}{16 G_{\rm N}} \, .
\ee
Hence a constant shift to the entropy can eliminate all negative entropy solutions. As we described earlier, such a constant shift is always achievable within the Wald formalism.

These features are representative, at least qualitatively, of all models in Table~\ref{tab:metrics} that lack the $\alpha \to -\alpha$ symmetry. These resummations are therefore considerably better behaved for the topological black holes.

\begin{figure}[H]
    \centering
    \includegraphics[width=\linewidth]{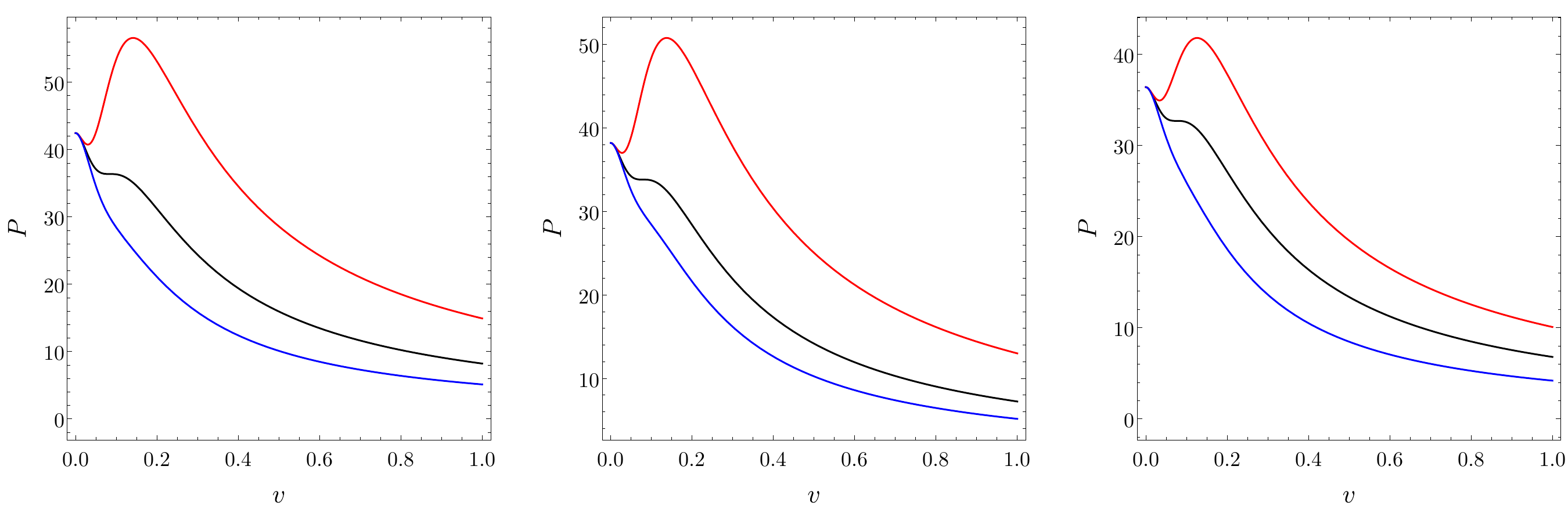}
    \caption{{\bf $\boldsymbol{P-v}$ diagram: topological  Hayward-AdS black hole.} Isotherms for the $k=-1$ Hayward-AdS black hole (model I) for $D = 5, 7, 9$ (left to right) feature a `reverse van der Waals' behavior. In each case, the lower blue curve has $T < T_c$, the middle black curve has $T = T_c$, and the upper red curve has $T > T_c$. In each plot, we have set the effective molecular size parameter $b = 0.1$. The units are such that $G_{\rm N} = 1$.}
    \label{fig:Hayward-pv-km1}
\end{figure}

Let us now explore the equation of state, which reads
\be 
P = \frac{(D-2) r_+^3 T}{4 \left(r_+^2+\alpha\right)^2} + \frac{(D-2) \left(\alpha  (D-1)+(D-3) r_+^2\right)}{16 \pi  \left(r_+^2+\alpha\right)^2} \, .
\ee
The key difference from the $k=+1$ case is that the denominator now contains the term $r_+^2+\alpha$ compared to $r_+^2-\alpha$ for $k = +1$. This results in the pressure remaining uniformly bounded, for all choices of volume, see Figure~\ref{fig:Hayward-pv-km1}. Physically, this is more difficult to interpret in terms of an ordinary fluid, but the most natural interpretation would seem to be that when $k = -1$ in these models, the short-range interactions involving large numbers of molecules are strongly attractive, rather than the repulsive interactions that would be generated if the molecules were hard spheres.

Contrary to the `symmetric models' (which do not admit any critical points), these models admit exactly one critical point. For the Hayward model, the characteristic swallowtail behavior is observed, but it occurs only within a narrowly confined region of the parameter space. Consequently, rather than illustrating it with a figure, we provide the critical values, which are:
\begin{align}
    P_c &= \frac{2D-3 + \sqrt{3 D (D-2)}}{4 \pi (D-2) b^2} \, , \qquad v_c = b \sqrt{ \frac{2 \sqrt{3} \sqrt{(D-2) D}}{D-3} - \frac{3(D-1)}{D-3}} \, ,
    \nonumber
    \\
    T_c &= \frac{2 \left[2D-3 + \sqrt{3 D(D-2)} \right]}{3 \pi (D-2) b^2 v_c} \, .
\end{align}
Interestingly, the ratio of these quantities is again given by 
\be 
\frac{P_c v_c}{T_c} = \frac{3}{8} \, ,
\ee
the same as in the spherical case and the same as for the van der Waals fluid. The ratios for the other models differ from the van der Waals value and exhibit weak dependence on the dimension. The critical ratio in model II monotonically decreases from $P_c v_c/T_c \approx 0.315701$ in $D = 5$ to $P_c v_c/T_c = 0.311755$ as $D \to \infty$, while the critical ratio in model III monotonically decreases from $P_c v_c/T_c \approx 0.00600468$ in $D = 5$ to $P_c v_c/T_c \approx 0.00318023$ when $D \to \infty$. The critical ratio in model VII ({\rm N} odd) can be worked out exactly, but the expressions are not illuminating. However, in this case, the single critical point is characterized by the critical ratio, which can become arbitrarily large as the dimension and ${\rm N}$ increase. 

To uncover the nature of the above critical point, let us study the isotherms in the $P-v$ diagram. Upon doing so, we discover the `reverse van der Waals behavior', observed previously in Ref.~\cite{Mo:2014qsa} for Lovelock black holes. That is, for temperatures above the critical temperature, phase structure exists, while for lower temperatures the behavior of the pressure is monotonic. We illustrate this `reverse' van der Waals behavior in Figure~\ref{fig:Hayward-pv-km1} for the Hayward model across several different dimensions.
This is to be compared to the `normal' van der Waals behavior, where the isotherms below the critical temperature display the characteristic oscillation, see Figure~\ref{fig:hayward-pv}.
The isotherms in models II, III, and VII with ${\rm N}$ odd are qualitatively identical to what is shown here. 

\begin{figure}[htp]
    \centering
    \includegraphics[width=\linewidth]{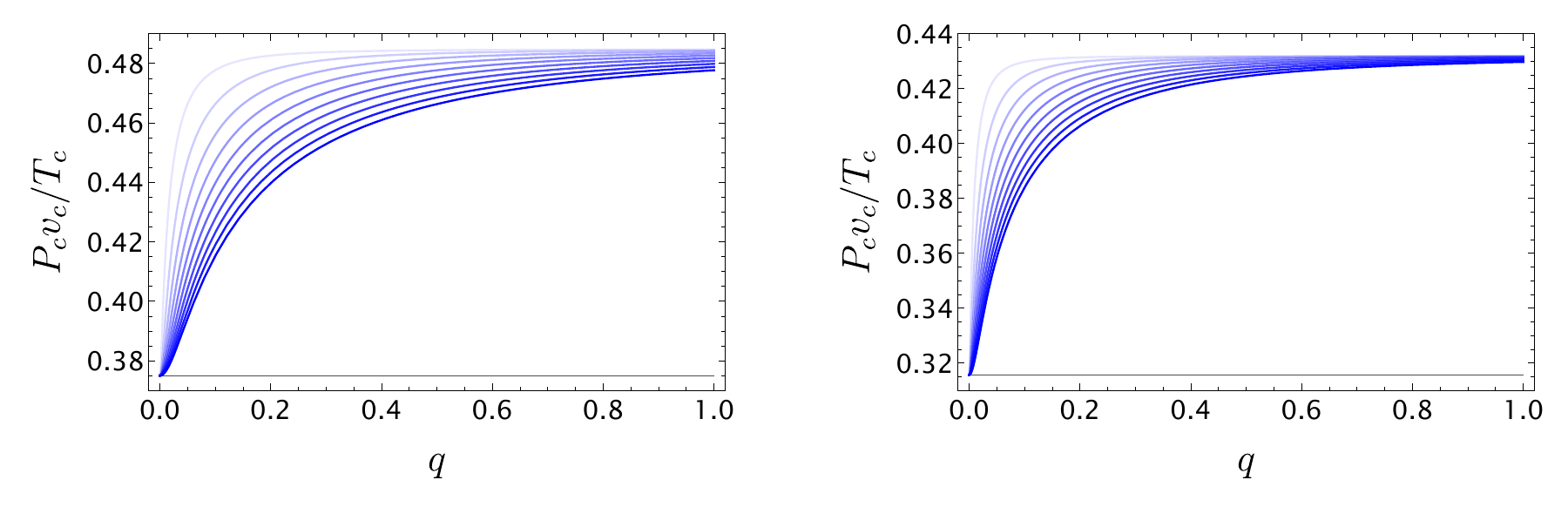}
    \caption{{\bf Critical ratio for topological black holes: charge dependence}. Ratio of critical parameters for Maxwell charged topological black holes in model I (left) and model II (right) in the five-dimensional case. The blue curves correspond to different values of $\alpha$, with the lighter (upper-leftmost) curves corresponding to smaller values, and the darker curves corresponding to larger values. The lower horizontal line is the uncharged critical ratio. The units are such that $G_{\rm N}=1$.}
    \label{fig:ratio-hyp}
\end{figure}

As in the uncharged case, for models that do \textit{not} exhibit the symmetry $\alpha \rightarrow -\alpha$, the swallowtail structure appears only within a very narrow region of the parameter space. The expressions for the critical pressure, reduced volume, and temperature are lengthy and offer little additional insight, so we omit them here. The critical ratio now depends on the spacetime dimensionality, the coupling parameter $\alpha$, and the charge parameter $q$. This dependence is illustrated in Figure~\ref{fig:ratio-hyp}, where the critical ratio is shown as a function of the charge parameter $q$ for various values of $\alpha$ in both the Hayward model (model I) and model II. In each case, the ratio increases monotonically with $q$ until it reaches a maximum value. From these plots, we observe that the ratio remains consistently below the conjectured bound $P_c v_c / T_c \leq 1/2$, contrary to the behavior of model VII described above.

\section{Discussion}
\label{sec:discussion}

We have performed a detailed study of black holes with anti-de Sitter asymptotics in theories that incorporate infinite towers of quasi-topological corrections to the action. We examined the basic properties of the AdS solutions, including how the higher-curvature terms `renormalize' the bare cosmological and Newton constants. We presented a careful analysis of the theory minimally coupled to matter which satisfies $T_t^t = T_r^r$, deriving sufficient conditions to ensure the resulting solutions are regular. As examples of such matter, we analyzed the theories coupled to Maxwell and nonlinear electrodynamics, in particular emphasizing how finite self-energy in the case of nonlinear electrodynamics results in intricate behavior for the curvature of the regular core of the metrics. Finally, we performed a comprehensive investigation of the black hole thermodynamics, studying phase transitions and critical behavior. 

Let us summarize some of the most interesting findings of our study:

\paragraph{Universality of regular black branes}\mbox{}\\ 
We have performed a preliminary investigation of regular black branes and topological black holes in resummed quasi-topological gravities. The regular black branes revealed particularly interesting features. These objects \textit{necessarily} have extremal inner horizons, i.e., inner horizons of vanishing surface gravity, without any fine tuning of parameters. As such, it is possible that such regular black branes avoid altogether the issues associated with mass inflation instabilities of the inner horizon. It would be particularly interesting to explore this further in future work. We have also shown that the thermodynamics of these objects is completely universal and insensitive to the particular model/resummation. In fact, the thermodynamics of regular black branes is completely identical to that of black branes in Einstein gravity. 

\paragraph{The nature of the regular black hole core}\mbox{}\\ 
In the case of vacuum regular black holes, the metric exhibits a de Sitter ``core'' at small radius. Here, we have shown that the nature of the core, i.e., whether it is de Sitter or anti-de Sitter, is governed by the sign of $\mathcal{S}(r)$ --- or, equivalently, the sign of the Misner--Sharp mass --- in the vicinity of $r = 0$. If this quantity is positive, then the black hole has a de Sitter core, while the core will be anti-de Sitter in the case where it is negative. Along these lines, we demonstrated that any black hole in the Maxwell theory must necessarily have an anti-de Sitter core. On the other hand, we found a relationship between the nature of the core and the electrostatic self-energy $M_\star$ for theories of nonlinear electrodynamics. Specifically, if $M > M_\star$, then the black holes have a de Sitter core analogous to the vacuum case, while if $M < M_\star$, then the core is anti-de Sitter, like the Maxwell case.

\paragraph{Regular black holes and molecular volume}\mbox{}\\ 
We have proven that in any resummed quasi-topological gravity that admits regular spherical black holes, the corresponding equation of state will exhibit a divergence at finite volume. This feature is exactly what would be expected for an equation of state that incorporates the finite volume effects of the constituent molecules. Of course, our proof only shows that regularity is \textit{sufficient} for a gravitational equation of state to describe `molecules' of finite volume. It would be interesting to delve into this further, assessing to what extent it is also necessary. If finite molecular volume could be rigorously tied to the existence of a regular core, then this would provide an interesting thermodynamic probe of the interior.

\paragraph{The law of corresponding states at the critical point}\mbox{}\\ 
The ratio of critical values $P_c v_c/T_c$ provides a dimensionless characterization of the fluid under consideration. We have studied this ratio for a large number of resummations (actually infinitely many)  that lead to regular black holes and also in the presence of various forms of charged matter. Intriguingly, the Hayward black hole was the unique one among the models we have studied for which the critical ratio is independent of dimension and also matches exactly the van der Waals ratio $P_c v_c/T_c = 3/8$ for both spherical and hyperbolic topologies. For {\em spherical} black holes, we have found a remarkably small variance between the possible values, with all cases, across all dimensions and all matter models, falling into the approximate range $0.35$ to $0.5$. We have interpreted this universality as a manifestation of the principle of corresponding states for regular black holes.  In fact, we have observed in all cases that the ratio of critical values appears to satisfy a universal bound 
\be 
\frac{P_c v_c}{T_c} \le 1/2\,.
\ee 
Saturation of this bound emerges in the $D \to \infty$ limit of some of the resummations. This bound appears to be consistent with the broader literature as well, included charged matter systems in Einstein gravity, rotating black holes, and isolated critical points~\cite{Gunasekaran:2012dq, Altamirano:2014tva, Hu:2024ldp}. It is therefore tempting to speculate that this bound applies not only to regular black holes, but is perhaps universal. Certainly, if true, this bound would apply only in the spherical sector (for example, isolated critical points in Lovelock theory have a critical ratio that exceeds $1/2$~\cite{Dolan:2014vba}). Of course, the literature on the subject of black hole criticality is quite extensive, so there may exist known counterexamples. In any case, it would be interesting to better understand if this is really the case, or to establish the necessary and sufficient conditions under which this bound holds. 

Finally, let us discuss a few possibilities for future work that would directly build off what we have done here.

\paragraph{Other forms of matter}\mbox{}\\ 
We have have developed the necessary formalism to study minimal matter coupled to infinite towers of higher-curvature corrections, provided that the matter satisfies $T_t^t = T_r^r$. One obvious candidate for future work would be to perform a study of minimal matter where this condition on the EMT is relaxed. However, there is considerable scope for further work within the same context we have studied here. For example, any model of nonlinear electrodynamics built as a functional of the Maxwell invariant will satisfy the $T_t^t = T_r^r$ condition. Moreover, one could consider gravity coupled to generalized conformal scalars as in Refs.~\cite{Oliva:2011np,Hennigar:2016xwd, Dykaar:2017mba}, which also meet this criteria.  

\paragraph{Non-minimally coupled matter}\mbox{}\\ 
We have performed a quite extensive analysis of infinite towers of higher-curvature corrections coupled minimally to matter. In particular, we have derived sufficient conditions under which the resulting solutions remain regular. However, it must be emphasized that it is somewhat unnatural to simultaneously consider minimal matter (e.g., a two-derivative Maxwell theory) while taking into account an infinite tower of corrections in the gravitational sector. To partially alleviate this, we have considered coupling to theories of nonlinear electrodynamics, which effectively incorporate a tower of corrections in the matter sector as well. Going forward, it would be interesting to consider simultaneously infinite towers of corrections in the gravitational and matter sectors, allowing for nonminimal couplings. This would more effectively capture what would arise in general effective field theory constructions or top-down models. One class of candidate theories in which this could be pursued would be electromagnetic quasi-topological gravities~\cite{Cano:2020qhy}; another would be the towers of Horndeski theories considered in Refs.~\cite{Fernandes:2025fnz, Fernandes:2025eoc}. 

\paragraph{Stability}\mbox{}\\ 
A pressing issue in any formulation of regular black holes is the problem of stability. On general grounds, one expects that the inner horizons of regular black holes will be susceptible to mass inflation instabilities. Moreover, here one may worry whether the resummation of quasi-topological theories produces stable dynamics. 

Regarding mass inflation, although this phenomenon has a kinematical origin, it is not obvious that in a theory capable of resolving singularities the corresponding build up of energy will backreact in a singular manner. It would be particularly interesting to investigate this issue for the black branes we have studied here. Since these metrics are automatically inner extremal, they have the potential to completely evade this classical instability. Of course, inner horizons are only a problem for predictability if they are \textit{Cauchy} horizons. It is not obvious that the regular black hole inner horizons actually will be Cauchy horizons in a fully dynamical set up~\cite{Bueno:2024eig}. Dynamical stability is a very difficult problem to address within a bottom-up framework. Certainly, if the series of corrections is truncated at finite order, it will be sensible within the EFT regime, but not beyond~\cite{Bueno:2023jtc}. However, in resumming the full infinite tower of corrections, there is the possibility that stable dynamics emerges. This is because, since the construction is a bottom-up one, one can add to the action all those terms which do not contribute to the equations of motion in spherical symmetry. Such terms, however, can and will contribute to the equations of motion of the perturbations. Thus, a careful assessment of the stability of the full tower of corrections must carefully and thoroughly take all such corrections into account. It is possible that, in this way, the quasi-topological theories can be `completed' to an infinite derivative theory of gravity which evades completely the Ostrogradsky-type instabilities. 

\paragraph{Mass inflation toy models}\mbox{}\\ 
We have provided sufficient conditions a resummation must satisfy to ensure that the corresponding black hole solutions are regular with minimal matter. In particular, what this means is that some resummations that are regular in vacuum become singular when matter is included. Here, our focus has been on the regular configurations, but the alternative presents an interesting possibility. In the resummations that are singular with charged matter, the singularity no longer occurs at $r = 0$, but instead at intermediate values of $r$. One could imagine tuning the position of the singularity such that it occurs at or in the vicinity of the inner horizon. As such, it may be possible to use these models as toy models to capture some of the effects expected for the backreaction of mass inflation instabilities in general relativity.

\section*{Acknowledgements}

We would like to thank Pablo Bueno, Pablo Cano, {\'A}ngel Murcia, Javier Moreno, Julio Oliva, and Simone Speziale for useful discussions on this manuscript. 
DK is grateful for support from GAČR 23-07457S grant of the Czech Science Foundation and  the Charles University Research Center Grant No.\ UNCE24/SCI/016. He would also like to thank the Perimeter Institute for Theoretical Physics for hospitality, where part of this work was completed. SM is supported by the Quantum Gravity Unit of the Okinawa Institute of Science and Technology (OIST). IS is supported by an International Macquarie University Research Excellence Scholarship (IMQRES). IS would also like to thank DK for hosting him as a visitor at the Institute of Theoretical Physics at Charles University, where parts of this project were completed. Research at Perimeter Institute is supported by the Government of Canada through the Department of Innovation, Science and Economic Development and by the
Province of Ontario through the Ministry of Colleges and Universities.

\appendix

\section{Quasi-topological theories to quintic order}
\label{app:lags}

Here, we collect explicit expressions for quasi-topological theories up to fifth order in curvature. From these `seed' densities quasi-topological theories of \textit{any} order can be constructed utilizing the recurrence formula~\eqref{recursive}.  

In the following, we denote by $R$ the Ricci scalar, by $Z_{ab} \equiv R_{ab} - 1/D g_{ab} R$ the traceless part of the Ricci tensor, and by $W_{abcd}$ the Weyl tensor. The first five densities then read 
\begin{subequations}\label{eq:Znexplicit}
\begin{align}
 \mathcal{Z}_{(1)}&=R\,, \\
 \mathcal{Z}_{(2)}&=\frac{\mathcal{Z}_{(1)}^2}{D(D-1)}-\frac{1}{(D-2)} \left [\frac{4 Z_{ab}Z^{ab}}{D-2}-\frac{W_{abcd} W^{abcd}}{D-3}\right]\,, \\
\nonumber \mathcal{Z}_{(3)}&=\frac{3\mathcal{Z}_{(1)}\mathcal{Z}_{(2)}}{D(D-1)}-\frac{2 \mathcal{Z}_{(1)}^3}{D^2(D-1)^2}+\frac{24}{(D-2)^2(D-3)} \left[\frac{ W\indices{_a_c^b^d} Z^a_b Z^c_d}{(D-2)}-
   \frac{   W_{a c d e}W^{bcde}Z^a_b}{(D-4)} \right]\\&+\frac{16
   Z^a_b Z^b_cZ_a^c}{(D-2)^4}   +\frac{2(2 D-3) W\indices{^a^b_c_d}W\indices{^c^d_e_f}W\indices{^e^f_a_b}}{ (D-2)(D-3) (D ((D-9)
   D+26)-22)} \,,   \\
\nonumber
\mathcal{Z}_{(4)}&=\frac{4\mathcal{Z}_{(1)}\mathcal{Z}_{(3)}-3 \mathcal{Z}_{(2)}^2}{D(D-1)}+\frac{96}{(D-2)^2(D-3)} \left[\frac{(D-1)\left ( W_{abcd} W^{abcd} \right)^2}{8D(D-2)^2(D-3)}-\frac{4Z_{a c} Z_{d e} W^{bdce} Z^{a}_b}{(D-2)^2(D-4)}\right. \\ \nonumber &-\frac{(2D-3) Z_e^f Z^e_f W_{abcd} W^{abcd}}{4(D-1)(D-2)^2}-
\frac{2 W_{acbd} W^{c efg} W^d{}_{efg} Z^{ab} }{D(D-3)(D-4)}  +\frac{(D^2-3D+3) \left (Z_a^b Z_b^a\right )^2}{D(D-1)(D-2)^3}\\& \left. -\frac{Z_a^b Z_b^c Z_c^d Z_d^a}{(D-2)^3}+\frac{(2D-1)W_{abcd} W^{aecf} Z^{bd} Z_{ef}}{D(D-2)(D-3)}\right]\,,\\ \nonumber
\mathcal{Z}_{(5)}&=\frac{5\mathcal{Z}_{(1)}\mathcal{Z}_{(4)}-2\mathcal{Z}_{(2)}\mathcal{Z}_{(3)}}{D(D-1)}+\frac{6 \mathcal{Z}_{(1)}\mathcal{Z}_{(2)}^2-8 \mathcal{Z}_{(1)}^2\mathcal{Z}_{(3)}}{D^2(D-1)^2}+\frac{768 Z_a^b Z_b^c Z_c^d Z_d^e Z_e^a}{5(D-2)^6(D-3)(D-4)} \\ \nonumber & + \frac{24(D-1)W_{ghij} W^{ghij}W\indices{_a_b^c^d}W\indices{_c_d^e^f}W\indices{_e_f^a^b} }{D(D-2)^3 (D-3)^2(D^3-9 D^2+26D-22)}-\frac{96(3D
-1)W^{ghij} W_{ghij}  W_{a c d e}W^{bcde} Z^a_b}{10D(D-1)(D-2)^4(D-3)^2(D-4)} \\ \nonumber & -\frac{768(D^2-2D+2)Z_a^b Z_b^a Z_c^d Z_d^e Z_e^c}{D(D-1)^2(D-2)^6(D-3)(D-4)}  -\frac{96(3D-1)(D^2+2D-4)W^{ghij} W_{ghij} Z_c^d Z_d^e Z_e^c}{D(D-1)(D+1)(D-2)^6(D-3)(D-4)}\\ \nonumber & +\frac{24(15D^5-148 D^4+527 D^3-800 D^2+472D-88)W\indices{_a_b^c^d}W\indices{_c_d^e^f}W\indices{_e_f^a^b} Z_{g}^h Z_h^g}{D(D-1)(D-2)^3(D-3)(D-4)(D^5-15D^4+91 D^3-277 D^2+418D-242)} \\\nonumber & +\frac{960 (D-1)}{(D-2)^4(D-3)^2} \left[\frac{(5D^2-7D+6)Z_g^h Z_h^g W_{abcd} Z^{ac} Z^{bd}}{10D(D-1)^2(D-2)}-\frac{Z_{a}^b Z_{b}^{c} Z_{cd} Z_{ef} W^{eafd}}{(D-1)(D-2)} \right. \\ \nonumber &- \frac{2(3D-1)Z^{ab} W_{acbd} Z^{ef}  W\indices{_e^c_f^g} Z^d_g}{D(D^2-1)(D-4)} +\frac{(D-3)W_{a c d e}W^{bcde} Z^a_b Z_f^g Z_g^f}{5D(D-1)^2(D-4)}\\  &\left. -\frac{(D-2)(D-3)(3D-2) Z^a_b Z^b_c W_{daef} W^{efgh} W_{gh}{}^{dc}}{4(D-1)^2(D-4)(D^2-6D+11)}+\frac{W_{ghij}W^{ghij} Z^{ac}Z^{bd}W_{abcd}}{20D(D-1)^2}\right]\,.
\end{align}
\end{subequations}
Note that the second-order quasi-topological density is simply the Gauss--Bonnet term, just written in a non-standard form. 

Quasi-topological gravities are not unique. These theories are defined in a bottom-up way based on their properties on spherically symmetric spacetimes. Hence, if there exists a density $\mathcal{T}_{(n)}$ which has the property that $\mathcal{T}_{(n)}$ does not contribute to the equations of motion in spherical symmetry, then $\tilde{\mathcal{Z}}_{(n)} = \mathcal{Z}_{(n)} + \beta_n \mathcal{T}_{(n)}$ will also be a quasi-topological theory with exactly the same equations of motion on spherical backgrounds. Therefore, the above densities are simply \textit{representatives} at each order. While this degeneracy is unimportant in spherical symmetry (and hence for us), it can have important consequences beyond spherical symmetry. For example, such terms would contribute in the presence of angular momentum. They could also contribute to perturbations of spherically symmetric spacetimes.

\section{Energy Conditions}\label{app:EnergyConditions}

\subsection{Null energy condition and strong energy condition}

Regular black holes are nonsingular ultracompact objects that may form through the gravitational collapse of a star. To prevent the formation of a singularity in dust collapse models, some repulsive effect must counteract the collapse. This repulsion might arise from the breakdown of general relativity in the final stages of the collapse and/or from a violation of classical energy conditions. Once the collapse is complete, an idealized static configuration remains. In four-dimensional spacetimes, most uncharged regular black holes feature a de Sitter core, as seen in the models proposed by Bardeen \cite{1968qtr..conf...87B} and Hayward \cite{Hayward:2005gi}, among others. This core structure satisfies the {\em null energy condition (NEC)} near the center while violating the {\em strong energy condition (SEC)} \cite{Maeda:2022}. Although matter is typically required to support such geometries in four-dimensional cases, in quasi-topological gravity, these regular geometries emerge as vacuum solutions.

Here, we study the two energy conditions by treating the higher-order curvature terms of the quasi-topological theories as an {\em effective EMT}, defined by the Einstein tensor $G_{\mu\nu}$ in general relativity as follows:
\be 
T_{\mu\nu}\equiv \frac{1}{{8\pi G_{\rm N}}} G_{\mu\nu}\,.
\ee 
We restrict our considerations to regular black holes with horizons of spherical topology in $D=5$ spacetime dimensions. The spacetime geometry is then described by a line element of the form 
\begin{align}
	{\rm d}s^2=g_{\mu\nu}{\rm d}x^{\mu}{\rm d}x^{\nu}=-f(r){\rm d}t^2+f(r)^{-1}{\rm d}r^2+r^2{\rm d}\theta^2+r^2\sin^2\!{\theta}{\rm d}\phi^2+r^2\sin^2\!{\theta}\cos^2\!{\phi} {\rm d}\psi^2\,.\label{eq:metric}
\end{align}
We assume a general vector of the form 
\begin{align}
	k^{\mu}=\left(\frac{c_{t}}{\sqrt{\sigma f(r)}},c_{r}\sqrt{\sigma f(r)},\frac{c_{\theta}}{r},\frac{c_{\phi}}{r \sin{\theta}},\frac{c_{\psi}}{r\sin{\theta}\cos{\phi}}\right)\,,
\end{align}
where $\sigma=\pm 1$. One can pick the coefficients $c_{t}$, $c_{r}$, $c_{\theta}$, and $c_{\phi}$ to satisfy specific constraints that will enforce the vector $k^{\mu}$ to be null or timelike. The inner product is given by
\begin{align}
	g_{\mu\nu}k^{\mu}k^{\nu}=\sigma \left(-c^2_{t}+c^2_{r}+\sigma\left(c^2_{\theta}+c^2_{\phi}+c^2_{\psi}\right)\right)\,,
\end{align}
and thus the conditions for the vector $k^{\mu}$ to be null and timelike, respectively, are 
\begin{eqnarray}
{\rm null}: c^2_{t}&=&c^2_{r}+\sigma\left(c^2_{\theta}+c^2_{\phi}+c^2_{\psi}\right)\,,
    \label{app:eq:null.condition}\\
    {\rm timelike}: c^2_{t}&=&c^2_{r}+\sigma\left(c^2_{\theta}+c^2_{\phi}+c^2_{\psi}\right)+\sigma\,.
    \label{app:eq:timelike.condition}
\end{eqnarray}
To evaluate the NEC, we use the null constraint \eqref{app:eq:null.condition}. Contraction with the EMT yields
\begin{align}
	T_{\mu\nu}k^{\mu}k^{\nu}=\frac{c^2_{\theta}+c^2_{\phi}+c^2_{\psi}}{2r^2}\left(4\left(1-f(r)\right)+rf'(r)+r^2f''(r)\right) \,.
\end{align}
Similarly, to evaluate the SEC, we have 
\begin{align}
	\Big(T_{\mu\nu}&-\frac{1}{D-2}g_{\mu\nu}\Big)t^{\mu}t^{\nu}\nonumber
    \\
    & =\frac{4 (c_{\phi}^2 + c_{\psi}^2 + c_{\theta}^2)(1-f(r)) + (1 + c_{\phi}^2 + c_{\psi}^2 + c_{\theta}^2)\left(rf'(r)+r^2 f''(r)\right)+ 2rf'(r)}{2 r^2},
\end{align}
where $D=5$ has been used in calculating the right-hand side and $t^{\mu}$ is a timelike vector obtained by applying the constraint \eqref{app:eq:timelike.condition} 
on $k^{\mu}$.

\begin{figure}[t]
    \centering
    \includegraphics[width=\linewidth]{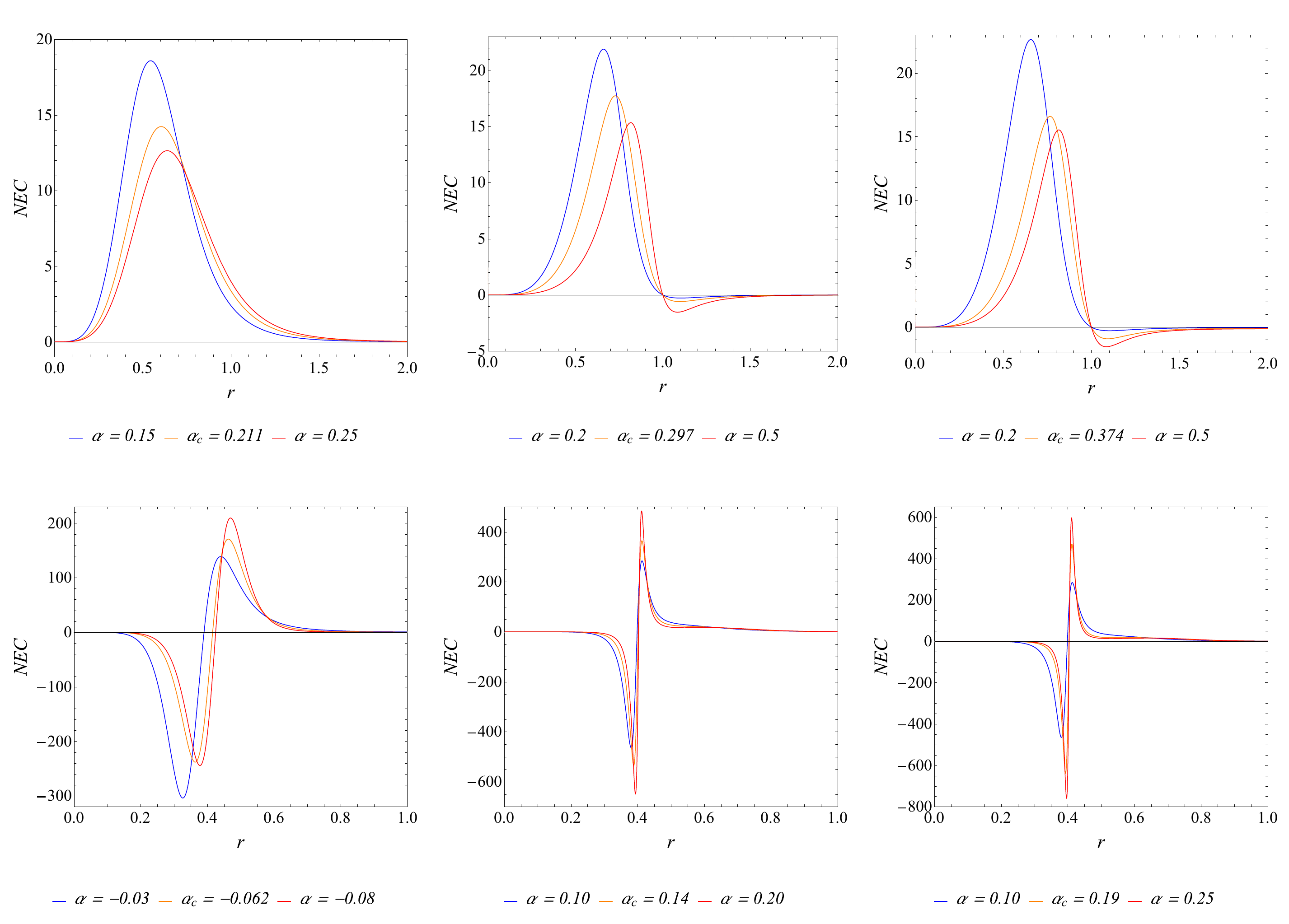}
    \caption{{\bf NEC: models I, IV, and V}. The quantity $T_{\mu\nu}k^{\mu}k^{\nu}/(c^2_{\theta}+c^2_{\phi}+c^2_{\psi})$, referred to as NEC for simplicity, 
    as a function of the radial coordinate $r$ for various spherical regular black hole models, including solitonic configurations, with $k=+1$ in five dimensions. The coupling parameter $\alpha$ is varied, while the cosmological constant parameter is fixed at $L = 1$. \textbf{Top:} From left to right, the uncharged models I, IV, and V are shown. \textbf{Bottom:} From left to right, the Maxwell-charged models I, IV, and V are displayed, with the charge parameter fixed at $q = 0.4$.}
    \label{fig:NEC}
\end{figure}

In what follows, we evaluate the NEC and SEC for models I, IV, and V. The corresponding behavior is displayed in Figures~\ref{fig:NEC} and \ref{fig:SEC}, and summarized in Table~\ref{tab:energy-conditions}. 

\subsection{Model I: Hayward model}
Explicit evaluation of the NEC for the metric function of the 5D Hayward model in the uncharged and charged cases leads to 
\begin{align}
	T_{\mu\nu}k^{\mu}k^{\nu}\big|_{q=0}=\frac{16\alpha L^6m^2r^4}{\big((L^2-\alpha)r^4+\alpha L^2 m\big)^3}\left(c^2_{\theta}+c^2_{\phi}+c^2_{\psi}\right)\ge 0, \label{eq:NEC}
\end{align}
which is satisfied in the entire spacetime provided that we have proper AdS asymptotics, which require $L^2>\alpha$ or equivalently $a<1$, and 
\begin{align}
	T_{\mu\nu}k^{\mu}k^{\nu}\big|_{q\neq0}=\frac{2r^6\left(3\alpha L^4 q^2r^6-3L^6q^2r^6+\alpha L^6(-15q^4+21mq^2r^2-8m^2r^4)\right)}{\left[-L^2r^6+\alpha \big(r^6+L^2(q^2-mr^2)\big)\right]^3}\left(c^2_{\theta}+c^2_{\phi}+c^2_{\psi}\right).
\end{align}
Although the NEC is marginally satisfied everywhere for the charged metric when only radial null vectors are considered, i.e., $c_{\theta}=c_{\phi}=c_{\psi}=0$, this does not hold in general. Expanding around the center, we find
\begin{align}
	T_{\mu\nu}k^{\mu}k^{\nu}\big|_{q\neq0}=-\frac{30r^6}{\alpha^2 q^2}\left(c^2_{\theta}+c^2_{\phi}+c^2_{\psi}\right)+\mathcal{O}(r^8),
\end{align} 
which demonstrates that the NEC is violated near the center in the charged case.

\begin{figure}[t]
    \centering
    \includegraphics[width=\linewidth]{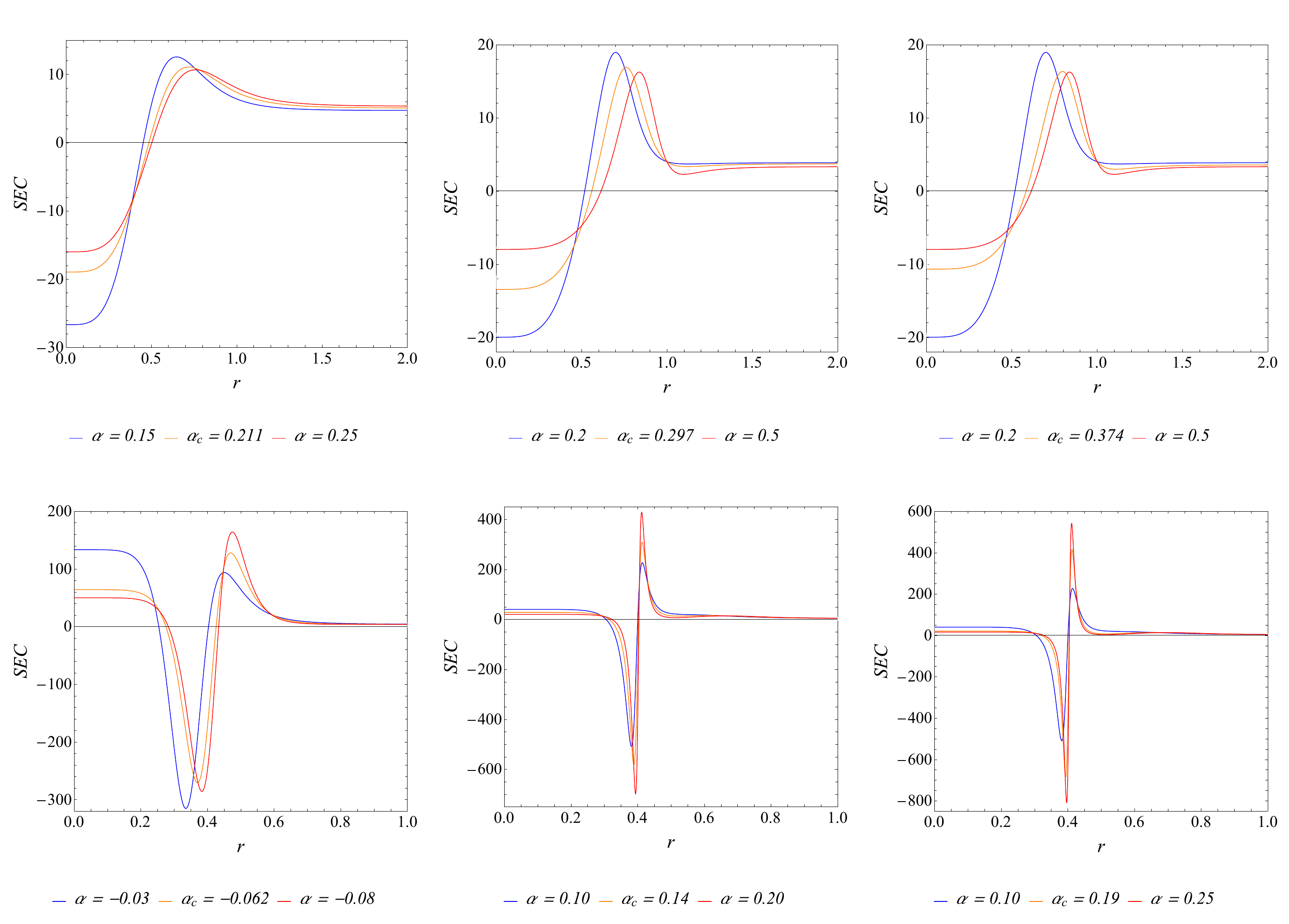}
    \caption{{\bf SEC: models I, IV, and V.} The SEC is plotted as a function of the radial coordinate $r$ for various spherical regular black hole models, including solitonic configurations, with $k=+1$ in five dimensions. The coupling parameter $\alpha$ is varied, while the cosmological constant paramter is fixed at $L=1$. For simplicity, we consider radial timelike vectors with $c_{\theta} = c_{\phi} = c_{\psi} = 0$. \textbf{Top:} From left to right, the uncharged models I, IV, and V are shown. \textbf{Bottom:} From left to right, the Maxwell-charged models I, IV, and V are displayed, with the charge parameter fixed at $q = 0.4$.}
    \label{fig:SEC}
\end{figure}

The expressions for the SEC, which encodes the attractive nature of gravity for timelike observers, are more complex than those for the NEC and do not provide further insight. We thus omit them here. However, we briefly comment on the series expansion near the center:
\begin{align}
	\left(T_{\mu\nu}-\frac{1}{D-2}g_{\mu\nu}T\right)t^{\mu}t^{\nu}\bigg|_{q=0, D=5}=-\frac{4}{\alpha} + \frac{8 (3 + 2 c_{\phi}^2 + 2 c_{\psi}^2 + 2 c_{\theta}^2) r^4}{m \alpha^2}+\cdots.
\end{align}
Since $\alpha>0$ in the uncharged case, we note that the SEC is violated close to the center as expected due to the presence of a dS core. This signals the repulsive nature of gravity in this region, taking under consideration the effective EMT. Nonetheless, the higher-order corrections of the theory's action, which does not require any matter, suffice to ensure regularity. In the charged case on the other hand, where regularity is achieved through an AdS core, the situation is entirely different. The SEC's expansion near the center is given by
\begin{align}
	\left(T_{\mu\nu}-\frac{1}{D-2}g_{\mu\nu}T\right)t^{\mu}t^{\nu}\bigg|_{q\neq 0, D=5}=-\frac{4}{\alpha} - \frac{10 (4 + 3 c_{\phi}^2 + 3 c_{\psi}^2 + 3 c_{\theta}^2) }{q^2 \alpha^2}r^6+\cdots ,
\end{align}
which implies that it is satisfied since in this case regularity of the spacetime requires $\alpha<0$.\footnote{Note that models IV and V (discussed in what follows) are symmetric with respect to $\alpha \to - \alpha$ and thus can admit charged regular black hole geometries for $\alpha>0$.}\ 

Typically, situations where the NEC is violated and the SEC is satisfied arise in phantom scalar field theories or modified gravity theories with non-minimally coupled scalar fields \cite{Kontou:2020}. We note that, while the SEC is satisfied in the charged case near the center, there exists a region where it is violated, as can be seen in Figure~\ref{fig:SEC}, indicating the presence of a region with repulsive gravity.

\subsection{Model IV}
We continue with the analysis of the model in the fourth row of Table~\ref{tab:metrics}. For this model, as well as the one in the fifth row (discussed in the next section), we expect properties distinct from the Hayward model due to the aforementioned $\alpha \leftrightarrow -\alpha$ symmetry. As the explicit expressions for the NEC are quite lengthy, we provide only their expansions near the center, given by
\begin{align}
		T_{\mu\nu}k^{\mu}k^{\nu}\big|_{q=0}=\frac{8 (c_{\phi}^2 + c_{\psi}^2 + c_{\theta}^2) r^4}{m \alpha^2}+\cdots
\end{align}
for the uncharged case, and by
\begin{align}
	T_{\mu\nu}k^{\mu}k^{\nu}\big|_{q\neq 0}=-\frac{15 (c_{\phi}^2 + c_{\psi}^2 + c_{\theta}^2) r^6}{q^2 \alpha^2}+\cdots
\end{align}
for the charged case. Analogous to the behavior in the Hayward model (model I in Table~\ref{tab:metrics}), we observe that the NEC is satisfied near the center in the uncharged case and violated in the charged case.
 
However, a notable difference arises: Here, there are regions where the NEC is violated even in the uncharged case, a property that is absent in the previous model.

On the other hand, the SEC exhibits the same behavior as in the Hayward model and model II. The expansions close to the center are given by
\begin{align}
	\left(T_{\mu\nu}-\frac{1}{D-2}g_{\mu\nu}T\right)t^{\mu}t^{\nu}\bigg|_{q=0, D=5}=-\frac{4}{|\alpha|} + \frac{4 \left(3 + 2 c_{\phi}^2 + 2 c_{\psi}^2 + 2 c_{\theta}^2\right) r^4}{m \alpha^2}+\cdots
\end{align}
for the uncharged case, and 
\begin{align}
\left(T_{\mu\nu}-\frac{1}{D-2}g_{\mu\nu}T\right)t^{\mu}t^{\nu}\bigg|_{q\neq 0,D=5}=\frac{4}{|\alpha|} - \frac{5 \left(4 + 3 c_{\phi}^2 + 3 c_{\psi}^2 + 3 c_{\theta}^2\right) r^6}{q^2 \alpha^2},	
\end{align}
for the charged case. We note that the behavior of the energy conditions is independent of the sign of $\alpha$, showing a dS core with SEC violation in the uncharged case, and an AdS core with the SEC satisfied near the center in the charged case. However, in both cases, there are regions where the SEC is violated, as illustrated in Figure~\ref{fig:SEC}. 
  
\subsection{Model V}
Lastly, we consider model V described by the fifth row of Table~\ref{tab:metrics}. We expect features similar to those of model IV, and specifically a violation of the NEC for the uncharged case since the metric functions of these two models behave in a similar manner. Once again, the explicit evaluation of the NEC leads to rather lengthy expressions that we omit for the sake of brevity. The relevant expansions near the center mimic those of model IV described in the previous subsection, with differences only in their coefficients. In the uncharged case, we obtain  
\begin{align}
	T_{\mu\nu}k^{\mu}k^{\nu}\big|_{q=0}=\frac{9 (c_{\phi}^2 + c_{\psi}^2 + c_{\theta}^2) r^4}{m \alpha^2}+\cdots,
\end{align}
with the NEC satisfied near the center, while for the charged case, we find 
\begin{align}
		T_{\mu\nu}k^{\mu}k^{\nu}\big|_{q\neq 0}=-\frac{16 (c_{\phi}^2 + c_{\psi}^2 + c_{\theta}^2) r^6}{q^2 \alpha^2}+\cdots,
\end{align}
and the NEC is violated. Analogous to model IV, NEC-violating regions exist even in the uncharged case. Additionally, the expansions of the SEC near the center have the same form as those of model IV, at least up to $\mathcal{O}(r^0)$, which indicates a violation of the SEC and a dS core for the uncharged case, while in the charged case the SEC is satisfied with an AdS core.

The expansions of the SEC in the uncharged and charged case, respectively, are given by
\begin{align}
	\left(T_{\mu\nu}-\frac{1}{D-2}g_{\mu\nu}T\right)t^{\mu}t^{\nu}\bigg|_{q=0, D=5}=-\frac{4}{|\alpha|}+\cdots,
\end{align}
and
\begin{align}
	\left(T_{\mu\nu}-\frac{1}{D-2}g_{\mu\nu}T\right)t^{\mu}t^{\nu}\bigg|_{q\neq 0, D=5}=\frac{4}{|\alpha|}+\cdots.
\end{align}
Similar to model IV, we note that SEC-violating regions exist in both cases. Table~\ref{tab:energy-conditions} summarizes the results of our analysis of the NEC and SEC for models I, IV, and V of Table~\ref{tab:metrics}.

\begin{table}[h!]
	\centering
	\begin{tabular}{|c|c|c|c|c|}
		\hline
		& \multicolumn{2}{c|}{$q = 0$} & \multicolumn{2}{c|}{$q \neq 0$} \\
		\hline
		& NEC & SEC & NEC & SEC \\
		\hline
		Model I & \cmark (\cmark) & \xmark (\xmark) & \xmark (\xmark) & \xmark (\cmark) \\
		Model IV & \xmark (\cmark) & \xmark (\xmark) & \xmark (\xmark) & \xmark (\cmark) \\
		Model V & \xmark (\cmark) & \xmark (\xmark) & \xmark (\xmark) & \xmark (\cmark) \\ 
		\hline
	\end{tabular}
	\caption{{\bf NEC and SEC: summary.} Status of the NEC and SEC for the cosmological extensions of models I (Hayward), IV, and V of Table~\ref{tab:metrics} with and without charge. The behavior near the center is indicated in parentheses.}
	\label{tab:energy-conditions}
\end{table}

\bibliography{bib}

\providecommand{\href}[2]{#2}\begingroup\raggedright\begin{thebibliography}{100}

\bibitem{Hawking:1973uf}
S.~W. Hawking and G.~F.~R. Ellis, \emph{{The Large Scale Structure of Space-Time}}.
\newblock Cambridge Monographs on Mathematical Physics. Cambridge University Press, 2, 2023, \href{http://dx.doi.org/10.1017/9781009253161}{10.1017/9781009253161}.

\bibitem{hoffmann1937choice}
B.~Hoffmann and L.~Infeld, \emph{On the choice of the action function in the new field theory}, \href{http://dx.doi.org/10.1103/PhysRev.51.765}{\emph{Phys. Rev.} {\bfseries 51} (1937) 765}.

\bibitem{Ayon-Beato:1998hmi}
E.~Ayon-Beato and A.~Garcia, \emph{{Regular black hole in general relativity coupled to nonlinear electrodynamics}}, \href{http://dx.doi.org/10.1103/PhysRevLett.80.5056}{\emph{Phys. Rev. Lett.} {\bfseries 80} (1998) 5056--5059}, [\href{https://arxiv.org/abs/gr-qc/9911046}{{\ttfamily gr-qc/9911046}}].

\bibitem{1968qtr..conf...87B}
J.~{Bardeen}, \emph{{Non-singular general relativistic gravitational collapse}},  in \emph{Proceedings of the 5th International Conference on Gravitation and the Theory of Relativity}, p.~87, Sept., 1968.

\bibitem{Hayward:2005gi}
S.~A. Hayward, \emph{{Formation and evaporation of regular black holes}}, \href{http://dx.doi.org/10.1103/PhysRevLett.96.031103}{\emph{Phys. Rev. Lett.} {\bfseries 96} (2006) 031103}, [\href{https://arxiv.org/abs/gr-qc/0506126}{{\ttfamily gr-qc/0506126}}].

\bibitem{Dymnikova:1992ux}
I.~Dymnikova, \emph{{Vacuum nonsingular black hole}}, \href{http://dx.doi.org/10.1007/BF00760226}{\emph{Gen. Rel. Grav.} {\bfseries 24} (1992) 235--242}.

\bibitem{Sakharov:1966aja}
A.~D. Sakharov, \emph{{Nachal'naia stadija rasshirenija Vselennoj i vozniknovenije neodnorodnosti raspredelenija veshchestva}}, {\emph{Sov. Phys. JETP} {\bfseries 22} (1966) 241}.

\bibitem{Borde:1994ai}
A.~Borde, \emph{{Open and closed universes, initial singularities and inflation}}, \href{http://dx.doi.org/10.1103/PhysRevD.50.3692}{\emph{Phys. Rev. D} {\bfseries 50} (1994) 3692--3702}, [\href{https://arxiv.org/abs/gr-qc/9403049}{{\ttfamily gr-qc/9403049}}].

\bibitem{Lemos:2011dq}
J.~P.~S. Lemos and V.~T. Zanchin, \emph{{Regular black holes: Electrically charged solutions, Reissner-Nordstr\"om outside a de Sitter core}}, \href{http://dx.doi.org/10.1103/PhysRevD.83.124005}{\emph{Phys. Rev. D} {\bfseries 83} (2011) 124005}, [\href{https://arxiv.org/abs/1104.4790}{{\ttfamily 1104.4790}}].

\bibitem{Bambi:2013ufa}
C.~Bambi and L.~Modesto, \emph{{Rotating regular black holes}}, \href{http://dx.doi.org/10.1016/j.physletb.2013.03.025}{\emph{Phys. Lett. B} {\bfseries 721} (2013) 329--334}, [\href{https://arxiv.org/abs/1302.6075}{{\ttfamily 1302.6075}}].

\bibitem{Simpson:2018tsi}
A.~Simpson and M.~Visser, \emph{{Black-bounce to traversable wormhole}}, \href{http://dx.doi.org/10.1088/1475-7516/2019/02/042}{\emph{JCAP} {\bfseries 02} (2019) 042}, [\href{https://arxiv.org/abs/1812.07114}{{\ttfamily 1812.07114}}].

\bibitem{Rodrigues:2018bdc}
M.~E. Rodrigues and M.~V. de~Sousa~Silva, \emph{{Bardeen Regular Black Hole With an Electric Source}}, \href{http://dx.doi.org/10.1088/1475-7516/2018/06/025}{\emph{JCAP} {\bfseries 06} (2018) 025}, [\href{https://arxiv.org/abs/1802.05095}{{\ttfamily 1802.05095}}].

\bibitem{Carballo-Rubio:2018pmi}
R.~Carballo-Rubio, F.~Di~Filippo, S.~Liberati, C.~Pacilio and M.~Visser, \emph{{On the viability of regular black holes}}, \href{http://dx.doi.org/10.1007/JHEP07(2018)023}{\emph{JHEP} {\bfseries 07} (2018) 023}, [\href{https://arxiv.org/abs/1805.02675}{{\ttfamily 1805.02675}}].

\bibitem{Carballo-Rubio:2018jzw}
R.~Carballo-Rubio, F.~Di~Filippo, S.~Liberati and M.~Visser, \emph{{Phenomenological aspects of black holes beyond general relativity}}, \href{http://dx.doi.org/10.1103/PhysRevD.98.124009}{\emph{Phys. Rev. D} {\bfseries 98} (2018) 124009}, [\href{https://arxiv.org/abs/1809.08238}{{\ttfamily 1809.08238}}].

\bibitem{Carballo-Rubio:2019nel}
R.~Carballo-Rubio, F.~Di~Filippo, S.~Liberati and M.~Visser, \emph{{Opening the Pandora\textquoteright{}s box at the core of black holes}}, \href{http://dx.doi.org/10.1088/1361-6382/ab8141}{\emph{Class. Quant. Grav.} {\bfseries 37} (2020) 14}, [\href{https://arxiv.org/abs/1908.03261}{{\ttfamily 1908.03261}}].

\bibitem{Carballo-Rubio:2019fnb}
R.~Carballo-Rubio, F.~Di~Filippo, S.~Liberati and M.~Visser, \emph{{Geodesically complete black holes}}, \href{http://dx.doi.org/10.1103/PhysRevD.101.084047}{\emph{Phys. Rev. D} {\bfseries 101} (2020) 084047}, [\href{https://arxiv.org/abs/1911.11200}{{\ttfamily 1911.11200}}].

\bibitem{DiFilippo:2022qkl}
F.~Di~Filippo, R.~Carballo-Rubio, S.~Liberati, C.~Pacilio and M.~Visser, \emph{{On the Inner Horizon Instability of Non-Singular Black Holes}}, \href{http://dx.doi.org/10.3390/universe8040204}{\emph{Universe} {\bfseries 8} (2022) 204}, [\href{https://arxiv.org/abs/2203.14516}{{\ttfamily 2203.14516}}].

\bibitem{Carballo-Rubio:2022kad}
R.~Carballo-Rubio, F.~Di~Filippo, S.~Liberati, C.~Pacilio and M.~Visser, \emph{{Regular black holes without mass inflation instability}}, \href{http://dx.doi.org/10.1007/JHEP09(2022)118}{\emph{JHEP} {\bfseries 09} (2022) 118}, [\href{https://arxiv.org/abs/2205.13556}{{\ttfamily 2205.13556}}].

\bibitem{Carballo-Rubio:2025fnc}
R.~Carballo-Rubio et~al., \emph{{Towards a Non-singular Paradigm of Black Hole Physics}},  \href{https://arxiv.org/abs/2501.05505}{{\ttfamily 2501.05505}}.

\bibitem{Bronnikov:2000vy}
K.~A. Bronnikov, \emph{{Regular magnetic black holes and monopoles from nonlinear electrodynamics}}, \href{http://dx.doi.org/10.1103/PhysRevD.63.044005}{\emph{Phys. Rev. D} {\bfseries 63} (2001) 044005}, [\href{https://arxiv.org/abs/gr-qc/0006014}{{\ttfamily gr-qc/0006014}}].

\bibitem{Ayon-Beato:2000mjt}
E.~Ayon-Beato and A.~Garcia, \emph{{The Bardeen model as a nonlinear magnetic monopole}}, \href{http://dx.doi.org/10.1016/S0370-2693(00)01125-4}{\emph{Phys. Lett. B} {\bfseries 493} (2000) 149--152}, [\href{https://arxiv.org/abs/gr-qc/0009077}{{\ttfamily gr-qc/0009077}}].

\bibitem{Bronnikov:2000yz}
K.~A. Bronnikov, \emph{{Comment on `Regular black hole in general relativity coupled to nonlinear electrodynamics'}}, \href{http://dx.doi.org/10.1103/PhysRevLett.85.4641}{\emph{Phys. Rev. Lett.} {\bfseries 85} (2000) 4641}.

\bibitem{Ayon-Beato:2004ywd}
E.~Ayon-Beato and A.~Garcia, \emph{{Four parametric regular black hole solution}}, \href{http://dx.doi.org/10.1007/s10714-005-0050-y}{\emph{Gen. Rel. Grav.} {\bfseries 37} (2005) 635}, [\href{https://arxiv.org/abs/hep-th/0403229}{{\ttfamily hep-th/0403229}}].

\bibitem{Dymnikova:2004zc}
I.~Dymnikova, \emph{{Regular electrically charged structures in nonlinear electrodynamics coupled to general relativity}}, \href{http://dx.doi.org/10.1088/0264-9381/21/18/009}{\emph{Class. Quant. Grav.} {\bfseries 21} (2004) 4417--4429}, [\href{https://arxiv.org/abs/gr-qc/0407072}{{\ttfamily gr-qc/0407072}}].

\bibitem{Berej:2006cc}
W.~Berej, J.~Matyjasek, D.~Tryniecki and M.~Woronowicz, \emph{{Regular black holes in quadratic gravity}}, \href{http://dx.doi.org/10.1007/s10714-006-0270-9}{\emph{Gen. Rel. Grav.} {\bfseries 38} (2006) 885--906}, [\href{https://arxiv.org/abs/hep-th/0606185}{{\ttfamily hep-th/0606185}}].

\bibitem{Balart:2014jia}
L.~Balart and E.~C. Vagenas, \emph{{Regular black hole metrics and the weak energy condition}}, \href{http://dx.doi.org/10.1016/j.physletb.2014.01.024}{\emph{Phys. Lett. B} {\bfseries 730} (2014) 14--17}, [\href{https://arxiv.org/abs/1401.2136}{{\ttfamily 1401.2136}}].

\bibitem{Fan:2016rih}
Z.-Y. Fan, \emph{{Critical phenomena of regular black holes in anti-de Sitter space-time}}, \href{http://dx.doi.org/10.1140/epjc/s10052-017-4830-9}{\emph{Eur. Phys. J. C} {\bfseries 77} (2017) 266}, [\href{https://arxiv.org/abs/1609.04489}{{\ttfamily 1609.04489}}].

\bibitem{Bronnikov:2017sgg}
K.~A. Bronnikov, \emph{{Nonlinear electrodynamics, regular black holes and wormholes}}, \href{http://dx.doi.org/10.1142/S0218271818410055}{\emph{Int. J. Mod. Phys. D} {\bfseries 27} (2018) 1841005}, [\href{https://arxiv.org/abs/1711.00087}{{\ttfamily 1711.00087}}].

\bibitem{Junior:2023ixh}
J.~T. S.~S. Junior, F.~S.~N. Lobo and M.~E. Rodrigues, \emph{{(Regular) Black holes in conformal Killing gravity coupled to nonlinear electrodynamics and scalar fields}}, \href{http://dx.doi.org/10.1088/1361-6382/ad210e}{\emph{Class. Quant. Grav.} {\bfseries 41} (2024) 055012}, [\href{https://arxiv.org/abs/2310.19508}{{\ttfamily 2310.19508}}].

\bibitem{Murk:2024nod}
S.~Murk and I.~Soranidis, \emph{{Light rings and causality for nonsingular ultracompact objects sourced by nonlinear electrodynamics}}, \href{http://dx.doi.org/10.1103/PhysRevD.110.044064}{\emph{Phys. Rev. D} {\bfseries 110} (2024) 044064}, [\href{https://arxiv.org/abs/2406.07957}{{\ttfamily 2406.07957}}].

\bibitem{Cano:2020ezi}
P.~A. Cano and A.~Murcia, \emph{{Resolution of Reissner-Nordstr\"om singularities by higher-derivative corrections}}, \href{http://dx.doi.org/10.1088/1361-6382/abd923}{\emph{Class. Quant. Grav.} {\bfseries 38} (2021) 075014}, [\href{https://arxiv.org/abs/2006.15149}{{\ttfamily 2006.15149}}].

\bibitem{Li:2024rbw}
Z.-C. Li and H.~Lu, \emph{{Regular Electric Black Holes from EMS Gravity}},  \href{https://arxiv.org/abs/2407.07952}{{\ttfamily 2407.07952}}.

\bibitem{Bueno:2024dgm}
P.~Bueno, P.~A. Cano and R.~A. Hennigar, \emph{{Regular black holes from pure gravity}}, \href{http://dx.doi.org/10.1016/j.physletb.2025.139260}{\emph{Phys. Lett. B} {\bfseries 861} (2025) 139260}, [\href{https://arxiv.org/abs/2403.04827}{{\ttfamily 2403.04827}}].

\bibitem{Oliva:2010eb}
J.~Oliva and S.~Ray, \emph{{A new cubic theory of gravity in five dimensions: Black hole, Birkhoff's theorem and C-function}}, \href{http://dx.doi.org/10.1088/0264-9381/27/22/225002}{\emph{Class. Quant. Grav.} {\bfseries 27} (2010) 225002}, [\href{https://arxiv.org/abs/1003.4773}{{\ttfamily 1003.4773}}].

\bibitem{Quasi}
R.~C. Myers and B.~Robinson, \emph{{Black Holes in Quasi-topological Gravity}}, \href{http://dx.doi.org/10.1007/JHEP08(2010)067}{\emph{JHEP} {\bfseries 08} (2010) 067}, [\href{https://arxiv.org/abs/1003.5357}{{\ttfamily 1003.5357}}].

\bibitem{Dehghani:2011vu}
M.~H. Dehghani, A.~Bazrafshan, R.~B. Mann, M.~R. Mehdizadeh, M.~Ghanaatian and M.~H. Vahidinia, \emph{{Black Holes in Quartic Quasitopological Gravity}}, \href{http://dx.doi.org/10.1103/PhysRevD.85.104009}{\emph{Phys. Rev. D} {\bfseries 85} (2012) 104009}, [\href{https://arxiv.org/abs/1109.4708}{{\ttfamily 1109.4708}}].

\bibitem{Ahmed:2017jod}
J.~Ahmed, R.~A. Hennigar, R.~B. Mann and M.~Mir, \emph{{Quintessential Quartic Quasi-topological Quartet}}, \href{http://dx.doi.org/10.1007/JHEP05(2017)134}{\emph{JHEP} {\bfseries 05} (2017) 134}, [\href{https://arxiv.org/abs/1703.11007}{{\ttfamily 1703.11007}}].

\bibitem{Cisterna:2017umf}
A.~Cisterna, L.~Guajardo, M.~Hassaine and J.~Oliva, \emph{{Quintic quasi-topological gravity}}, \href{http://dx.doi.org/10.1007/JHEP04(2017)066}{\emph{JHEP} {\bfseries 04} (2017) 066}, [\href{https://arxiv.org/abs/1702.04676}{{\ttfamily 1702.04676}}].

\bibitem{Bueno:2019ycr}
P.~Bueno, P.~A. Cano and R.~A. Hennigar, \emph{{(Generalized) quasi-topological gravities at all orders}}, \href{http://dx.doi.org/10.1088/1361-6382/ab5410}{\emph{Class. Quant. Grav.} {\bfseries 37} (2020) 015002}, [\href{https://arxiv.org/abs/1909.07983}{{\ttfamily 1909.07983}}].

\bibitem{Bueno:2022res}
P.~Bueno, P.~A. Cano, R.~A. Hennigar, M.~Lu and J.~Moreno, \emph{{Generalized quasi-topological gravities: the whole shebang}}, \href{http://dx.doi.org/10.1088/1361-6382/aca236}{\emph{Class. Quant. Grav.} {\bfseries 40} (2023) 015004}, [\href{https://arxiv.org/abs/2203.05589}{{\ttfamily 2203.05589}}].

\bibitem{Moreno:2023rfl}
J.~Moreno and A.~J. Murcia, \emph{{Classification of generalized quasitopological gravities}}, \href{http://dx.doi.org/10.1103/PhysRevD.108.044016}{\emph{Phys. Rev. D} {\bfseries 108} (2023) 044016}, [\href{https://arxiv.org/abs/2304.08510}{{\ttfamily 2304.08510}}].

\bibitem{Bueno:2024zsx}
P.~Bueno, P.~A. Cano, R.~A. Hennigar and A.~J. Murcia, \emph{{Regular black holes from thin-shell collapse}}, \href{http://dx.doi.org/10.1103/PhysRevD.111.104009}{\emph{Phys. Rev. D} {\bfseries 111} (2025) 104009}, [\href{https://arxiv.org/abs/2412.02740}{{\ttfamily 2412.02740}}].

\bibitem{Bueno:2024eig}
P.~Bueno, P.~A. Cano, R.~A. Hennigar and A.~J. Murcia, \emph{{Dynamical Formation of Regular Black Holes}}, \href{http://dx.doi.org/10.1103/PhysRevLett.134.181401}{\emph{Phys. Rev. Lett.} {\bfseries 134} (2025) 181401}, [\href{https://arxiv.org/abs/2412.02742}{{\ttfamily 2412.02742}}].

\bibitem{Bueno:2025gjg}
P.~Bueno, P.~A. Cano, R.~A. Hennigar, A.~J. Murcia and A.~Vicente-Cano, \emph{{Regular black holes from Oppenheimer-Snyder collapse}},  \href{https://arxiv.org/abs/2505.09680}{{\ttfamily 2505.09680}}.

\bibitem{DiFilippo:2024mwm}
F.~Di~Filippo, I.~Kol\'a\v{r} and D.~Kubizňák, \emph{{Inner-extremal regular black holes from pure gravity}}, \href{http://dx.doi.org/10.1103/PhysRevD.111.L041505}{\emph{Phys. Rev. D} {\bfseries 111} (2025) L041505}, [\href{https://arxiv.org/abs/2404.07058}{{\ttfamily 2404.07058}}].

\bibitem{Konoplya:2024hfg}
R.~A. Konoplya and A.~Zhidenko, \emph{{Infinite tower of higher-curvature corrections: Quasinormal modes and late-time behavior of D-dimensional regular black holes}}, \href{http://dx.doi.org/10.1103/PhysRevD.109.104005}{\emph{Phys. Rev. D} {\bfseries 109} (2024) 104005}, [\href{https://arxiv.org/abs/2403.07848}{{\ttfamily 2403.07848}}].

\bibitem{Konoplya:2024kih}
R.~A. Konoplya and A.~Zhidenko, \emph{{Dymnikova black hole from an infinite tower of higher-curvature corrections}}, \href{http://dx.doi.org/10.1016/j.physletb.2024.138945}{\emph{Phys. Lett. B} {\bfseries 856} (2024) 138945}, [\href{https://arxiv.org/abs/2404.09063}{{\ttfamily 2404.09063}}].

\bibitem{Ma:2024olw}
T.-X. Ma and Y.-Q. Wang, \emph{{Frozen boson stars in an infinite tower of higher-derivative gravity}},  \href{https://arxiv.org/abs/2406.08813}{{\ttfamily 2406.08813}}.

\bibitem{Frolov:2024hhe}
V.~P. Frolov, A.~Koek, J.~P. Soto and A.~Zelnikov, \emph{{Regular black holes inspired by quasi-topological gravity}},  \href{https://arxiv.org/abs/2411.16050}{{\ttfamily 2411.16050}}.

\bibitem{Wang:2024zlq}
S.-W. Wang, S.-P. Wu and S.-W. Wei, \emph{{Are regular black holes from pure gravity classified within the same thermodynamical topology?}}, \href{http://dx.doi.org/10.1016/j.physletb.2025.139402}{\emph{Phys. Lett. B} {\bfseries 864} (2025) 139402}, [\href{https://arxiv.org/abs/2412.05811}{{\ttfamily 2412.05811}}].

\bibitem{Fernandes:2025fnz}
P.~G.~S. Fernandes, \emph{{Singularity resolution and inflation from an infinite tower of regularized curvature corrections}},  \href{https://arxiv.org/abs/2504.07692}{{\ttfamily 2504.07692}}.

\bibitem{Fernandes:2025eoc}
P.~G.~S. Fernandes, \emph{{Regular BTZ black holes from an infinite tower of corrections}},  \href{https://arxiv.org/abs/2504.08565}{{\ttfamily 2504.08565}}.

\bibitem{Simovic:2023yuv}
F.~Simovic and I.~Soranidis, \emph{{Euclidean and Hamiltonian thermodynamics for regular black holes}}, \href{http://dx.doi.org/10.1103/PhysRevD.109.044029}{\emph{Phys. Rev. D} {\bfseries 109} (2024) 044029}, [\href{https://arxiv.org/abs/2309.09439}{{\ttfamily 2309.09439}}].

\bibitem{Soranidis:2024}
I.~Soranidis, \emph{{Euclidean methods and phase transitions for the strongest deformations compatible with Schwarzschild asymptotics}}, \href{http://dx.doi.org/10.1103/PhysRevD.109.044041}{\emph{Phys. Rev. D} {\bfseries 109} (2024) 044041}, [\href{https://arxiv.org/abs/2310.07228}{{\ttfamily 2310.07228}}].

\bibitem{Julio}
M.~Aguayo, L.~Gajardo, N.~Grandi, J.~Moreno, J.~Oliva and M.~Reyes, \emph{{Holographic explorations of regular black holes in pure gravity}},  \href{https://arxiv.org/abs/2505.XXXXX}{{\ttfamily 2505.XXXXX}}.

\bibitem{Lovelock1}
D.~Lovelock, \emph{Divergence-free tensorial concomitants}, \href{http://dx.doi.org/10.1007/BF01817753}{\emph{Aequ. Math.} {\bfseries 4} (1970) 127--138}.

\bibitem{Lovelock2}
D.~Lovelock, \emph{{The Einstein tensor and its generalizations}}, \href{http://dx.doi.org/10.1063/1.1665613}{\emph{J. Math. Phys.} {\bfseries 12} (1971) 498--501}.

\bibitem{Fels:2001rv}
M.~E. Fels and C.~G. Torre, \emph{{The Principle of symmetric criticality in general relativity}}, \href{http://dx.doi.org/10.1088/0264-9381/19/4/303}{\emph{Class. Quant. Grav.} {\bfseries 19} (2002) 641--676}, [\href{https://arxiv.org/abs/gr-qc/0108033}{{\ttfamily gr-qc/0108033}}].

\bibitem{Deser:2003up}
S.~Deser and B.~Tekin, \emph{{Shortcuts to high symmetry solutions in gravitational theories}}, \href{http://dx.doi.org/10.1088/0264-9381/20/22/011}{\emph{Class. Quant. Grav.} {\bfseries 20} (2003) 4877--4884}, [\href{https://arxiv.org/abs/gr-qc/0306114}{{\ttfamily gr-qc/0306114}}].

\bibitem{Frausto:2024egp}
G.~Frausto, I.~Kol\'a\v{r}, T.~M\'alek and C.~Torre, \emph{{Symmetry reduction of gravitational Lagrangians}},  \href{https://arxiv.org/abs/2410.11036}{{\ttfamily 2410.11036}}.

\bibitem{Mann:1997iz}
R.~B. Mann, \emph{{Topological black holes: Outside looking in}}, {\emph{Annals Israel Phys. Soc.} {\bfseries 13} (1997) 311}, [\href{https://arxiv.org/abs/gr-qc/9709039}{{\ttfamily gr-qc/9709039}}].

\bibitem{Bueno:2017sui}
P.~Bueno and P.~A. Cano, \emph{{On black holes in higher-derivative gravities}}, \href{http://dx.doi.org/10.1088/1361-6382/aa8056}{\emph{Class. Quant. Grav.} {\bfseries 34} (2017) 175008}, [\href{https://arxiv.org/abs/1703.04625}{{\ttfamily 1703.04625}}].

\bibitem{Zhou:2022yio}
T.~Zhou and L.~Modesto, \emph{{Geodesic incompleteness of some popular regular black holes}}, \href{http://dx.doi.org/10.1103/PhysRevD.107.044016}{\emph{Phys. Rev. D} {\bfseries 107} (2023) 044016}, [\href{https://arxiv.org/abs/2208.02557}{{\ttfamily 2208.02557}}].

\bibitem{Bueno:2025jgc}
P.~Bueno, O.~Lasso~Andino, J.~Moreno and G.~van~der Velde, \emph{{On regular charged black holes in three dimensions}},  \href{https://arxiv.org/abs/2503.02930}{{\ttfamily 2503.02930}}.

\bibitem{Birmingham:1998nr}
D.~Birmingham, \emph{{Topological black holes in Anti-de Sitter space}}, \href{http://dx.doi.org/10.1088/0264-9381/16/4/009}{\emph{Class. Quant. Grav.} {\bfseries 16} (1999) 1197--1205}, [\href{https://arxiv.org/abs/hep-th/9808032}{{\ttfamily hep-th/9808032}}].

\bibitem{Emparan:1999gf}
R.~Emparan, \emph{{AdS / CFT duals of topological black holes and the entropy of zero energy states}}, \href{http://dx.doi.org/10.1088/1126-6708/1999/06/036}{\emph{JHEP} {\bfseries 06} (1999) 036}, [\href{https://arxiv.org/abs/hep-th/9906040}{{\ttfamily hep-th/9906040}}].

\bibitem{Eriksen:1995ws}
E.~Eriksen and O.~Groen, \emph{{The de Sitter universe models}}, \href{http://dx.doi.org/10.1142/S0218271895000090}{\emph{Int. J. Mod. Phys. D} {\bfseries 4} (1995) 115--159}.

\bibitem{DeLorenzo:2017tgx}
T.~De~Lorenzo and A.~Perez, \emph{{Light Cone Thermodynamics}}, \href{http://dx.doi.org/10.1103/PhysRevD.97.044052}{\emph{Phys. Rev. D} {\bfseries 97} (2018) 044052}, [\href{https://arxiv.org/abs/1707.00479}{{\ttfamily 1707.00479}}].

\bibitem{Banados:1992gq}
M.~Banados, M.~Henneaux, C.~Teitelboim and J.~Zanelli, \emph{{Geometry of the (2+1) black hole}}, \href{http://dx.doi.org/10.1103/PhysRevD.48.1506}{\emph{Phys. Rev. D} {\bfseries 48} (1993) 1506--1525}, [\href{https://arxiv.org/abs/gr-qc/9302012}{{\ttfamily gr-qc/9302012}}].

\bibitem{born1933electromagnetic}
M.~Born and L.~Infeld, \emph{Electromagnetic mass}, {\emph{Nature} {\bfseries 132} (1933) 970--970}.

\bibitem{Born:1934gh}
M.~Born and L.~Infeld, \emph{{Foundations of the new field theory}}, \href{http://dx.doi.org/10.1098/rspa.1934.0059}{\emph{Proc. Roy. Soc. Lond. A} {\bfseries 144} (1934) 425--451}.

\bibitem{Gibbons-Rasheed:1995}
G.~Gibbons and D.~Rasheed, \emph{Electric-magnetic duality rotations in non-linear electrodynamics}, \href{http://dx.doi.org/10.1016/0550-3213(95)00409-L}{\emph{Nucl. Phys. B} {\bfseries 454} (1995) 185--206}, [\href{https://arxiv.org/abs/9506035}{{\ttfamily 9506035}}].

\bibitem{plebanski1970lectures}
J.~Pleba{\'n}ski, \emph{Lectures on non-linear electrodynamics: an extended version of lectures given at the Niels Bohr Institute and NORDITA, Copenhagen, in October 1968}, vol.~25.
\newblock Nordita, 1970.

\bibitem{Russo:2022qvz}
J.~G. Russo and P.~K. Townsend, \emph{{Nonlinear electrodynamics without birefringence}}, \href{http://dx.doi.org/10.1007/JHEP01(2023)039}{\emph{JHEP} {\bfseries 01} (2023) 039}, [\href{https://arxiv.org/abs/2211.10689}{{\ttfamily 2211.10689}}].

\bibitem{Tahamtan:2021}
T.~Tahamtan, \emph{Compatibility of nonlinear electrodynamics models with robinson-trautman geometry}, \href{http://dx.doi.org/10.1103/PhysRevD.109.044041}{\emph{Phys. Rev. D} {\bfseries 103} (2021) 064052}, [\href{https://arxiv.org/abs/2010.01689}{{\ttfamily 2010.01689}}].

\bibitem{Hale-Kubizňák-Svítek-Tahamtan:2023}
T.~Hale, D.~Kubiz{\v{n}}{\'a}k, O.~Sv{\'\i}tek and T.~Tahamtan, \emph{Solutions and basic properties of regularized maxwell theory}, \href{http://dx.doi.org/10.1103/PhysRevD.107.124031}{\emph{Phys. Rev. D} {\bfseries 107} (2023) 124031}, [\href{https://arxiv.org/abs/2303.16928}{{\ttfamily 2303.16928}}].

\bibitem{Kubiznak-Tahamtan-Svitek:2022}
D.~Kubiz{\v{n}}{\'a}k, T.~Tahamtan and O.~Svitek, \emph{Slowly rotating black holes in nonlinear electrodynamics}, \href{http://dx.doi.org/10.1103/PhysRevD.105.104064}{\emph{Phys. Rev. D} {\bfseries 105} (2022) 104064}, [\href{https://arxiv.org/abs/2203.01919}{{\ttfamily 2203.01919}}].

\bibitem{Hale:2025veb}
T.~Hale, D.~Kubiz\v{n}\'ak, J.~Men\v{s}\'\i{}kov\'a, R.~B. Mann and J.~Yang, \emph{{On thermodynamics of charged and accelerating black holes}},  \href{https://arxiv.org/abs/2501.13679}{{\ttfamily 2501.13679}}.

\bibitem{Dey:2004}
T.~K. Dey, \emph{Born--infeld black holes in the presence of a cosmological constant}, \href{http://dx.doi.org/10.1016/j.physletb.2004.06.047}{\emph{Phys. Lett. B} {\bfseries 595} (2004) 484--490}, [\href{https://arxiv.org/abs/0406169}{{\ttfamily 0406169}}].

\bibitem{Li:2016nll}
S.~Li, H.~Lu and H.~Wei, \emph{{Dyonic (A)dS Black Holes in Einstein-Born-Infeld Theory in Diverse Dimensions}}, \href{http://dx.doi.org/10.1007/JHEP07(2016)004}{\emph{JHEP} {\bfseries 07} (2016) 004}, [\href{https://arxiv.org/abs/1606.02733}{{\ttfamily 1606.02733}}].

\bibitem{Kubizňák-Svítek-Tahamtan:2024}
D.~Kubiz{\v{n}}{\'a}k, O.~Sv{\'\i}tek and T.~Tahamtan, \emph{Regularized conformal electrodynamics: novel c metric in 2+ 1 dimensions}, \href{http://dx.doi.org/10.1103/PhysRevD.110.064054}{\emph{Phys, Rev. D} {\bfseries 110} (2024) 064054}, [\href{https://arxiv.org/abs/2404.14335}{{\ttfamily 2404.14335}}].

\bibitem{Kastor:2011qp}
D.~Kastor, S.~Ray and J.~Traschen, \emph{{Mass and Free Energy of Lovelock Black Holes}}, \href{http://dx.doi.org/10.1088/0264-9381/28/19/195022}{\emph{Class. Quant. Grav.} {\bfseries 28} (2011) 195022}, [\href{https://arxiv.org/abs/1106.2764}{{\ttfamily 1106.2764}}].

\bibitem{Hennigar:2015esa}
R.~A. Hennigar, W.~G. Brenna and R.~B. Mann, \emph{{{$P-v$} criticality in quasitopological gravity}}, \href{http://dx.doi.org/10.1007/JHEP07(2015)077}{\emph{JHEP} {\bfseries 07} (2015) 077}, [\href{https://arxiv.org/abs/1505.05517}{{\ttfamily 1505.05517}}].

\bibitem{Gunasekaran:2012dq}
S.~Gunasekaran, R.~B. Mann and D.~Kubizňák, \emph{{Extended phase space thermodynamics for charged and rotating black holes and Born-Infeld vacuum polarization}}, \href{http://dx.doi.org/10.1007/JHEP11(2012)110}{\emph{JHEP} {\bfseries 11} (2012) 110}, [\href{https://arxiv.org/abs/1208.6251}{{\ttfamily 1208.6251}}].

\bibitem{Cadoni:2016hhd}
M.~Cadoni, A.~M. Frassino and M.~Tuveri, \emph{{On the universality of thermodynamics and $\eta/s$ ratio for the charged Lovelock black branes}}, \href{http://dx.doi.org/10.1007/JHEP05(2016)101}{\emph{JHEP} {\bfseries 05} (2016) 101}, [\href{https://arxiv.org/abs/1602.05593}{{\ttfamily 1602.05593}}].

\bibitem{Hennigar:2017umz}
R.~A. Hennigar, \emph{{Criticality for charged black branes}}, \href{http://dx.doi.org/10.1007/JHEP09(2017)082}{\emph{JHEP} {\bfseries 09} (2017) 082}, [\href{https://arxiv.org/abs/1705.07094}{{\ttfamily 1705.07094}}].

\bibitem{Hennigar:2020fkv}
R.~A. Hennigar, D.~Kubiznak, R.~B. Mann and C.~Pollack, \emph{{Lower-dimensional Gauss\textendash{}Bonnet gravity and BTZ black holes}}, \href{http://dx.doi.org/10.1016/j.physletb.2020.135657}{\emph{Phys. Lett. B} {\bfseries 808} (2020) 135657}, [\href{https://arxiv.org/abs/2004.12995}{{\ttfamily 2004.12995}}].

\bibitem{Cvetic:2001bk}
M.~Cvetic, S.~Nojiri and S.~D. Odintsov, \emph{{Black hole thermodynamics and negative entropy in de Sitter and anti-de Sitter Einstein-Gauss-Bonnet gravity}}, \href{http://dx.doi.org/10.1016/S0550-3213(02)00075-5}{\emph{Nucl. Phys. B} {\bfseries 628} (2002) 295--330}, [\href{https://arxiv.org/abs/hep-th/0112045}{{\ttfamily hep-th/0112045}}].

\bibitem{Clunan:2004tb}
T.~Clunan, S.~F. Ross and D.~J. Smith, \emph{{On Gauss-Bonnet black hole entropy}}, \href{http://dx.doi.org/10.1088/0264-9381/21/14/009}{\emph{Class. Quant. Grav.} {\bfseries 21} (2004) 3447--3458}, [\href{https://arxiv.org/abs/gr-qc/0402044}{{\ttfamily gr-qc/0402044}}].

\bibitem{Bueno:2017qce}
P.~Bueno and P.~A. Cano, \emph{{Universal black hole stability in four dimensions}}, \href{http://dx.doi.org/10.1103/PhysRevD.96.024034}{\emph{Phys. Rev. D} {\bfseries 96} (2017) 024034}, [\href{https://arxiv.org/abs/1704.02967}{{\ttfamily 1704.02967}}].

\bibitem{Iliesiu:2020qvm}
L.~V. Iliesiu and G.~J. Turiaci, \emph{{The statistical mechanics of near-extremal black holes}}, \href{http://dx.doi.org/10.1007/JHEP05(2021)145}{\emph{JHEP} {\bfseries 05} (2021) 145}, [\href{https://arxiv.org/abs/2003.02860}{{\ttfamily 2003.02860}}].

\bibitem{Hawking:1982dh}
S.~W. Hawking and D.~N. Page, \emph{{Thermodynamics of Black Holes in anti-De Sitter Space}}, \href{http://dx.doi.org/10.1007/BF01208266}{\emph{Commun. Math. Phys.} {\bfseries 87} (1983) 577}.

\bibitem{Kubiznak:2012wp}
D.~Kubizňák and R.~B. Mann, \emph{{P-V criticality of charged AdS black holes}}, \href{http://dx.doi.org/10.1007/JHEP07(2012)033}{\emph{JHEP} {\bfseries 07} (2012) 033}, [\href{https://arxiv.org/abs/1205.0559}{{\ttfamily 1205.0559}}].

\bibitem{HullMann:2021}
B.~Hull and R.~B. Mann, \emph{Thermodynamics of exotic black holes in lovelock gravity}, \href{http://dx.doi.org/10.1103/PhysRevD.104.084032}{\emph{Phys. Rev. D} {\bfseries 104} (2021) 084032}, [\href{https://arxiv.org/abs/2102.05282}{{\ttfamily 2102.05282}}].

\bibitem{MarksSimovicMann:2021}
G.~A. Marks, F.~Simovic and R.~B. Mann, \emph{Phase transitions in {4D Gauss--Bonnet--de Sitter} black holes}, \href{http://dx.doi.org/10.1103/PhysRevD.104.104056}{\emph{Phys.\ Rev.\ D.} {\bfseries 104} (2021) 104056}, [\href{https://arxiv.org/abs/2107.11352}{{\ttfamily 2107.11352}}].

\bibitem{HullSimovic:2023}
B.~Hull and F.~Simovic, \emph{Exotic black hole thermodynamics in third-order lovelock gravity}, \href{http://dx.doi.org/10.1088/1361-6382/acdb3d}{\emph{Class. Quant. Grav.} {\bfseries 40} (2023) 145016}, [\href{https://arxiv.org/abs/2208.05500}{{\ttfamily 2208.05500}}].

\bibitem{correspondingStates}
E.~A. Guggenheim, \emph{The principle of corresponding states}, \href{http://dx.doi.org/10.1063/1.1724033}{\emph{The Journal of Chemical Physics} {\bfseries 13} (07, 1945) 253--261}, [\href{https://arxiv.org/abs/https://pubs.aip.org/aip/jcp/article-pdf/13/7/253/18793644/253\_1\_online.pdf}{{\ttfamily https://pubs.aip.org/aip/jcp/article-pdf/13/7/253/18793644/253\_1\_online.pdf}}].

\bibitem{ZouZhangWang:2014}
D.-C. Zou, S.-J. Zhang and B.~Wang, \emph{Critical behavior of born-infeld ads black holes in the extended phase space thermodynamics}, \href{http://dx.doi.org/10.1103/PhysRevD.89.044002}{\emph{Phys, Rev. D} {\bfseries 89} (2014) 044002}, [\href{https://arxiv.org/abs/1311.7299}{{\ttfamily 1311.7299}}].

\bibitem{Emparan:1999pm}
R.~Emparan, C.~V. Johnson and R.~C. Myers, \emph{{Surface terms as counterterms in the AdS / CFT correspondence}}, \href{http://dx.doi.org/10.1103/PhysRevD.60.104001}{\emph{Phys. Rev. D} {\bfseries 60} (1999) 104001}, [\href{https://arxiv.org/abs/hep-th/9903238}{{\ttfamily hep-th/9903238}}].

\bibitem{Mo:2014qsa}
J.-X. Mo and W.-B. Liu, \emph{{$P-V$ criticality of topological black holes in Lovelock-Born-Infeld gravity}}, \href{http://dx.doi.org/10.1140/epjc/s10052-014-2836-0}{\emph{Eur. Phys. J. C} {\bfseries 74} (2014) 2836}, [\href{https://arxiv.org/abs/1401.0785}{{\ttfamily 1401.0785}}].

\bibitem{Altamirano:2014tva}
N.~Altamirano, D.~Kubizňák, R.~B. Mann and Z.~Sherkatghanad, \emph{{Thermodynamics of rotating black holes and black rings: phase transitions and thermodynamic volume}}, \href{http://dx.doi.org/10.3390/galaxies2010089}{\emph{Galaxies} {\bfseries 2} (2014) 89--159}, [\href{https://arxiv.org/abs/1401.2586}{{\ttfamily 1401.2586}}].

\bibitem{Hu:2024ldp}
Y.-P. Hu, H.~Zhang, Y.-S. An, G.-Y. Sun, W.-L. You, D.-N. Shi et~al., \emph{{Quantum anomaly triggers the violation of scaling laws in gravitational system}},  \href{https://arxiv.org/abs/2410.23783}{{\ttfamily 2410.23783}}.

\bibitem{Dolan:2014vba}
B.~P. Dolan, A.~Kostouki, D.~Kubizňák and R.~B. Mann, \emph{{Isolated critical point from Lovelock gravity}}, \href{http://dx.doi.org/10.1088/0264-9381/31/24/242001}{\emph{Class. Quant. Grav.} {\bfseries 31} (2014) 242001}, [\href{https://arxiv.org/abs/1407.4783}{{\ttfamily 1407.4783}}].

\bibitem{Oliva:2011np}
J.~Oliva and S.~Ray, \emph{{Conformal couplings of a scalar field to higher curvature terms}}, \href{http://dx.doi.org/10.1088/0264-9381/29/20/205008}{\emph{Class. Quant. Grav.} {\bfseries 29} (2012) 205008}, [\href{https://arxiv.org/abs/1112.4112}{{\ttfamily 1112.4112}}].

\bibitem{Hennigar:2016xwd}
R.~A. Hennigar, R.~B. Mann and E.~Tjoa, \emph{{Superfluid Black Holes}}, \href{http://dx.doi.org/10.1103/PhysRevLett.118.021301}{\emph{Phys. Rev. Lett.} {\bfseries 118} (2017) 021301}, [\href{https://arxiv.org/abs/1609.02564}{{\ttfamily 1609.02564}}].

\bibitem{Dykaar:2017mba}
H.~Dykaar, R.~A. Hennigar and R.~B. Mann, \emph{{Hairy black holes in cubic quasi-topological gravity}}, \href{http://dx.doi.org/10.1007/JHEP05(2017)045}{\emph{JHEP} {\bfseries 05} (2017) 045}, [\href{https://arxiv.org/abs/1703.01633}{{\ttfamily 1703.01633}}].

\bibitem{Cano:2020qhy}
P.~A. Cano and A.~Murcia, \emph{{Electromagnetic Quasitopological Gravities}}, \href{http://dx.doi.org/10.1007/JHEP10(2020)125}{\emph{JHEP} {\bfseries 10} (2020) 125}, [\href{https://arxiv.org/abs/2007.04331}{{\ttfamily 2007.04331}}].

\bibitem{Bueno:2023jtc}
P.~Bueno, P.~A. Cano and R.~A. Hennigar, \emph{{On the stability of Einsteinian cubic gravity black holes in EFT}}, \href{http://dx.doi.org/10.1088/1361-6382/ad4f41}{\emph{Class. Quant. Grav.} {\bfseries 41} (2024) 137001}, [\href{https://arxiv.org/abs/2306.02924}{{\ttfamily 2306.02924}}].

\bibitem{Maeda:2022}
H.~Maeda, \emph{{Quest for realistic non-singular black-hole geometries: regular-center type}}, \href{http://dx.doi.org/10.1007/JHEP11(2022)108}{\emph{JHEP} {\bfseries 11} (2022) 108}, [\href{https://arxiv.org/abs/2107.04791}{{\ttfamily 2107.04791}}].

\bibitem{Kontou:2020}
E.-A. Kontou and K.~Sanders, \emph{Energy conditions in general relativity and quantum field theory}, \href{http://dx.doi.org/10.1088/1361-6382/ab8fcf}{\emph{Class. Quant. Grav.} {\bfseries 37} (2020) 193001}, [\href{https://arxiv.org/abs/2003.01815}{{\ttfamily 2003.01815}}].

\end{thebibliography}\endgroup

\end{document}